\documentclass[a4paper, amsfonts, amssymb, amsmath, reprint, showkeys, nofootinbib, twoside]{revtex4-1}
\usepackage[english]{babel}
\usepackage[utf8]{inputenc}
\usepackage[colorinlistoftodos, color=green!40, prependcaption]{todonotes}
\usepackage[pdftex, pdftitle={Article}, pdfauthor={Author}]{hyperref} 
\usepackage{float}
\usepackage{multirow} 
\usepackage{graphicx}
\usepackage{wrapfig}
\begin{document}
\title{Entropic organization of topologically modified ring polymers in spherical confinement}

\author{Kingkini Roychoudhury$^{[1,2]}$, Shreerang Pande$^{[1]}$, Indrakanty S. Shashank$^{[1]}$, Debarshi Mitra$^{[1]}$,  and Apratim Chatterji$^{[1]}$}
    \email[Correspondence email address: ]{apratim@iiserpune.ac.in}
    \affiliation{1. Dept. of Physics, Indian Institute of Science Education and Research, Pune, India-411008.\\
    2. Dept. of Physics, Indian Institute of Science Education and Research, Berhampur, India-760010}

\date{\today} 

\begin{abstract}
It has been shown that under high cylindrical confinement, two ring polymers with excluded volume interactions 
between monomers, segregate to two halves of the cylinder to maximize their entropy. 
In contrast, two ring polymers remain mixed within a sphere, as there is no 
symmetry breaking direction [{\em Nat Rev Microbiol}, {\bf 8}, 600-607 (2010)].
Therefore, in order to observe emergent organization of ring polymers in a sphere,
we can introduce an asymmetric topological modification to the polymer architecture by creating a small loop 
and a big loop within the ring polymer. We consider the bead-spring model of polymers where there are
only repulsive excluded volume interactions between the monomers ensuring that the organization we observe
is purely entropy-driven. We find that for a single topologically modified polymer within a sphere, the monomers of the
bigger loop are statistically more probable to be found closer to the periphery.
However, the situation is reversed when we have multiple such topologically modified polymers in a sphere.
The monomers of the small loops are found closer to the walls of the sphere.  We can increase this 
localization and radial organization  of polymer segments 
by increasing the number of  small loops in each ring polymer. We study how these loops interact with 
each other within a polymer, as well as with loops of other polymers in spherical confinement. 
We compare contact maps of multiple 
such topologically  modified polymers in a sphere.  Finally, we discuss the plausible relevance 
of our studies to eukaryotic  chromosomes that are confined within a spherical nucleus. 

\end{abstract}


\maketitle

\section{Introduction}

The organization of topologically modified polymers, with multiple internal loops along chain contour, 
is now a growing topic of investigation in polymer physics  
\cite{dna2,knot_review,Polovnikov2023,Mithun2,Haddad2017, Tubiana2024}. 
This is due to the realisation that understanding the emergent properties of such polymers
is relevant for the understanding of chromosome-organization and segregation within cells \cite{Cook2009, Mithun, Nicodemi2014, Gilbert2017,  Michieletto2020,  Kadam2023, Dutta2023, DiStefano2021, Abdulla2023, Schiessel2023, Klempahn2024, Joyeux2020,Conforto2024, Dixon2016}. 

Ring-polymer dynamics has continued to  be of interest to polymer physicists since the last four decades,
as ring polymers do not have any free ends. Consequently, polymer relaxation by reptation  is not 
possible in melts of ring polymers \cite{Smrek2021,Zhou2019,Vettorel2009,Halverson2011_1,Halverson2011_2,Rosa2019,Schram2019,Pachong2020,Smrek2015}.
There has been renewed interest in the physics of ring polymers in the last two  decades, when it was
realized that bacterial chromosomes having cyclic topology can be modelled as ring polymers 
\cite{Shin2014,Shin2015,Joyeux2021,Joyeux2023}. 
It is established that the segregation of daughter chromosomes in rod-shaped bacterial cells, such 
as the {\em E.coli}  \cite{Badrinarayanan2015,woldy,Trueba869,WOLDRINGH2006273}, relies on the entropic 
repulsion between different daughter chromosomes  \cite{Jun2006, Jun2010, Jun2012, ha2_2012, ha3_2012,Harju2024,Woldringh_r_2024}.
We have recently shown by simulation studies that if one assumes that 
DNA-ring-polymers adopt specific modified polymer topologies while  remaining confined in a (sphero-)
cylindrical cell, then the emergent organization of polymer segments along the cell long axis
is quantitatively to the organization of {\em E.coli} chromosomes as seen {\em in-vivo} \cite{dna1,dna2,dna3}. 
However, deciphering the mechanisms underlying the organization of chromosomes in the cells
of higher-order organisms is significantly more complex as compared to that in bacterial 
cells \cite{Jun2007}.  The chromosomes  of eukaryotes are confined in a spherical nucleus, and moreover there are 
multiple chromosomes  within  the nucleus. Moreover, gene regulation,  carried out by linker
proteins, bring different (enhancer-promoter) DNA-segments  in spatial proximity and thereby create loops \cite{Junier2010,dieterloop}. 
This  study is one of  the first steps to investigate the physical properties of `{\em specifically designed}' 
topologically modified coarse grained models pf polymers with loops within  spherical confinement, 
with the expectation that the principles  elucidated  from this study might be useful in the future to
comprehend  chromosome organization within a nucleus.

In eukaryotic cells,  multiple chromosomes are compacted at various length scales within a spherical 
nucleus of diameter $\approx 10 \mu$. The compaction of the double-helical DNA strands to form nucleosomes 
is achieved by histone proteins and leads to  the  formation of chromatin fiber. 
Moreover, there are linker proteins which connect different DNA-segments which are located far apart
along the chain contour. This  results in the formation of loops of different sizes
along the chain contour, and may also lead to compaction and organization 
of the chromosome at higher length scales \cite{Mondal2015,Agarwal2018,Agarwal2019,Agarwal2019_2,Mondal2020,Salari2022, Salari2024, Barbieri2012, Brackley2016, sbsmodel,Virnau2}. The formation of loops along the chain contour 
leads to the observation of topologically associated domains (TADs)
\cite{jost,Lioy2018,cmap_caul,Mirny2019,Dekker2016}.

Euchromatin, which is  transcriptionally more active regions of the chromosome, is less compacted than the transcriptionally 
less active  heterochromation. Euchromatin segments are found towards the center of the nucleus. 
On the other hand, the more compacted heterochromatin is found towards the peripheral regions. 
Moreover, the multiple chromosomal chains within the nucleus  also remain  spatially  segregated from 
each other and do not get mixed with each other \cite{Rosa2008, Rosa2010, Halverson2014}. We aim to explore 
possibilities to develop a mechanistic  understanding of the emergence of some of these observations using basic 
principles of polymer physics, avoiding much of  the complexities inherent to chromosomes inside the nucleus 
of a living cell. Previous studies have proposed the idea of a fractal globule hypothesis of polymers which 
leads to higher contact probability between monomers along the chain contour as compared to that obtained
from the statistics of a random coil. But recent investigations have considered the contact probability of polymer segments 
along the chain contour as a consequence of a distribution of extruded 
loops of different sizes within the polymer \cite{Alipour2012,Fudenberg2016, Polovnikov2023}.
However, we focus on if and how entropic organization of 
multiple polymers  can be achieved  by designing topological modifications within ring polymers. These topological modifications 
could result from DNA-extrusion or loops formed by linker proteins. 

To this end, we consider a system of many  ring-polymers in spherical confinement, 
where  we systematically modify the 
topology of each polymer. It is to be noted that eukaryotic chromosomes are linear polymers, 
but now it is also  established that there are multiple loops  within the linear polymer. 
Thus, one may imagine the DNA-polymer as a polymer with free ends with multiple hierarchical loops within the chain.
In this manuscript, which is an early investigation into this topic, we use a simplified model 
where we neglect the free ends and possibilities of hierarchical loops. We implement multiple specifically designed 
loops along the chain contour by introducing additional cross-links between non-neighbouring monomers on the ring-polymer 
contour. Moreover, we consider  identical polymers in confinement, each having 
the length and the same topological modifications.  We investigate whether  our designed
topological modifications  can be tuned to control  the extent of segregation and/or 
organization of polymer segments within a sphere. These segregation and organization of polymer segments is primarily 
driven by tuning entropic interactions rather than by  energetic interactions.


In our previous studies with topologically modified polymers in cylindrical confinement, 
we have established that internal loops within a ring polymer entropically repel each other, i.e., 
the polymer segments are able to take a larger number of configurations if the loops do not overlap. 
Thus,  the polymer is able to explore a larger number of microstates if the loops occupy different regions 
along the long axis. Similarly, we establish in this study,  a consequence of different-sized loops is the
emergence of an asymmetry between effective entropic interactions between the polymer 
segments. Furthermore, asymmetric interactions between multiple such polymers confined within a sphere 
lead to an emergent  organization of the loops (and thereby the polymer segments) along the radial direction, 
where smaller loops have a  preference to occupy outer regions of the sphere away from the center. 
This is in spite of flexible polymers being considered intrinsically disordered  entities. We investigate 
how the relative size of the loops within a polymer, as well as the number of smaller internal 
loops can be tuned to play a role in the emergent organisation.  
Thus, we aim unearth the key principles by which  topological  modifications of rings leads 
to the radial organization of  loops within a spherical confinement. 

We also consider cases where the topologically modified ring-polymers  can cross each other. i.e. a case where chain crossing is allowed.
We implement this by suitably choosing small diameter of the bead in our bead-spring model of polymers, such that 
excluded volume interactions between beads does not prevent chain crossing. This is relevant  because DNA-polymers 
are able to release topological constraints due to the presence of the enzyme topoisomerase-IV, which cuts ds-DNA strands 
and rejoins them after chains have crossed each other. We establish that release of topological constraints do not 
affect the radial organization of loops. Chain crossing also allows polymers to explore  concatenated configurations. On
the other hand, if one starts out with concatenations of internal loops in a packed configuration (as in a tightly folded 
chromosome configuration during mitosis), the DNA-polymers can also release concatenations as the daughter DNA-s unpack.

Hereon, we  state the plan of the paper. 
In the next section, we first describe the model that we work with. 
Thereafter, we present the Results which has two subsections,  which describe monomer organization  
where  (i)  chain crossing is not allowed and we ensure that start out with unconcatenated topologically modified 
ring polymers (ii) chain crossing is allowed for topologically modified  ring polymers.
We also plot contact maps of the different polymers, just to analyze how different segments of the polymers
are in contact with particular segments of  neighbouring polymers. Finally,  we summarize our conclusions 
and discuss the future relevance of this study in the Discussion section.

 \begin{figure*}[!hbt]
\includegraphics[width=0.95\columnwidth,angle=0]{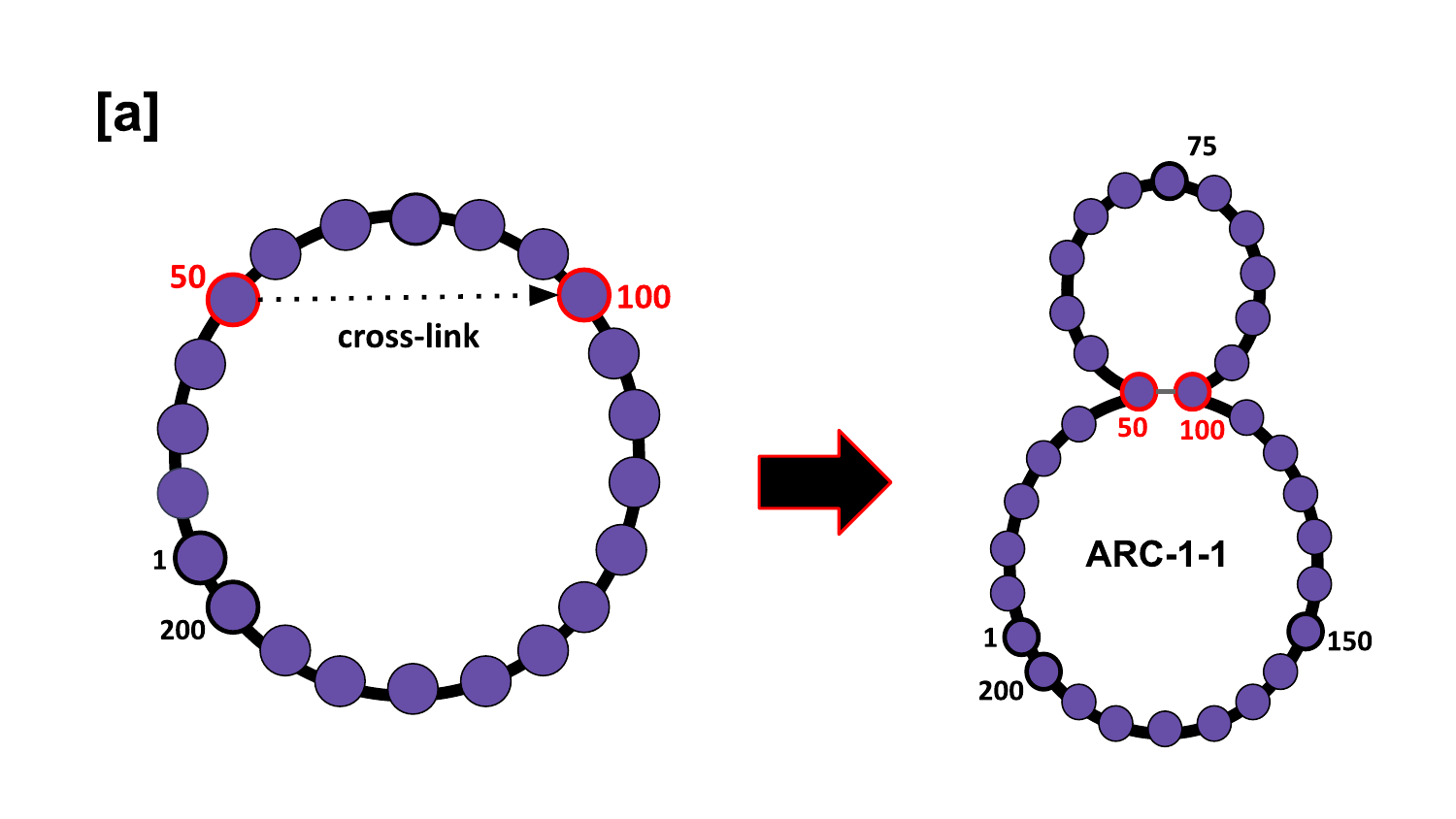}
\includegraphics[width=0.95\columnwidth,angle=0]{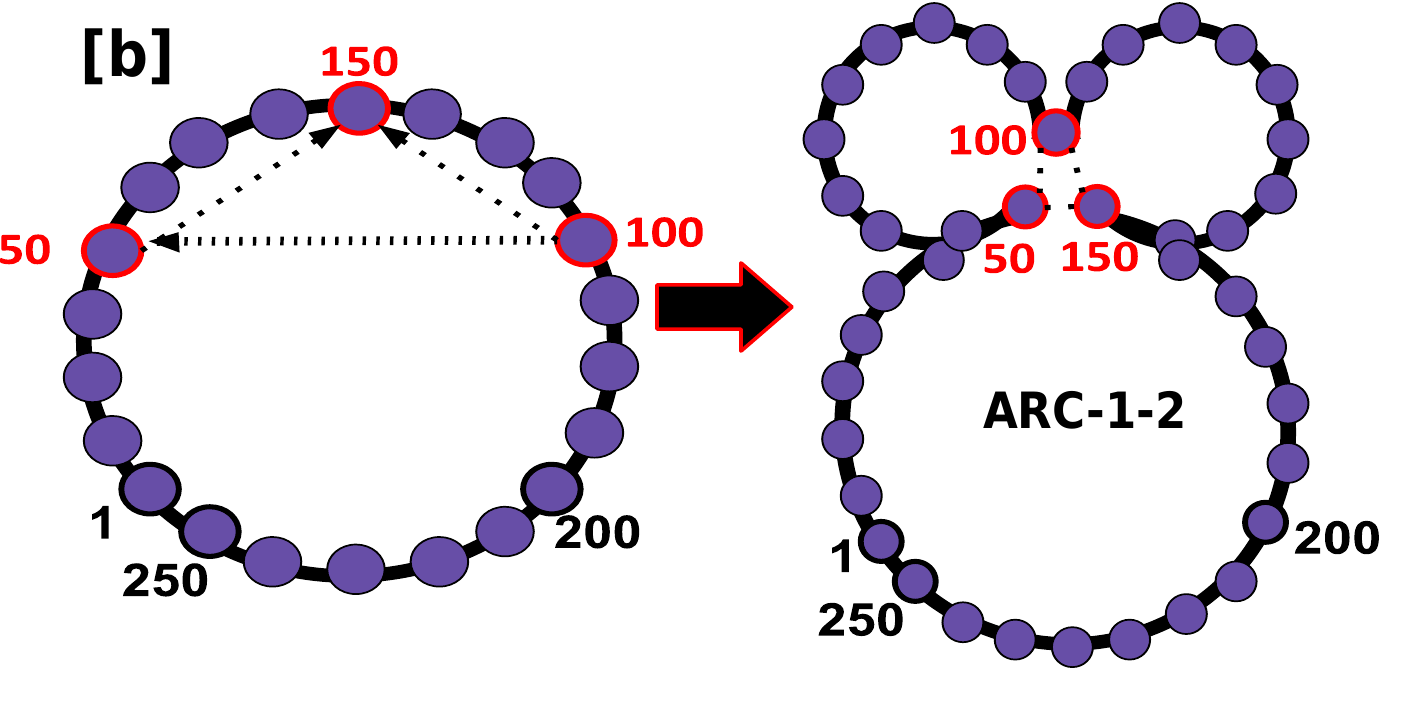}
\includegraphics[width=0.55\columnwidth,angle=0]{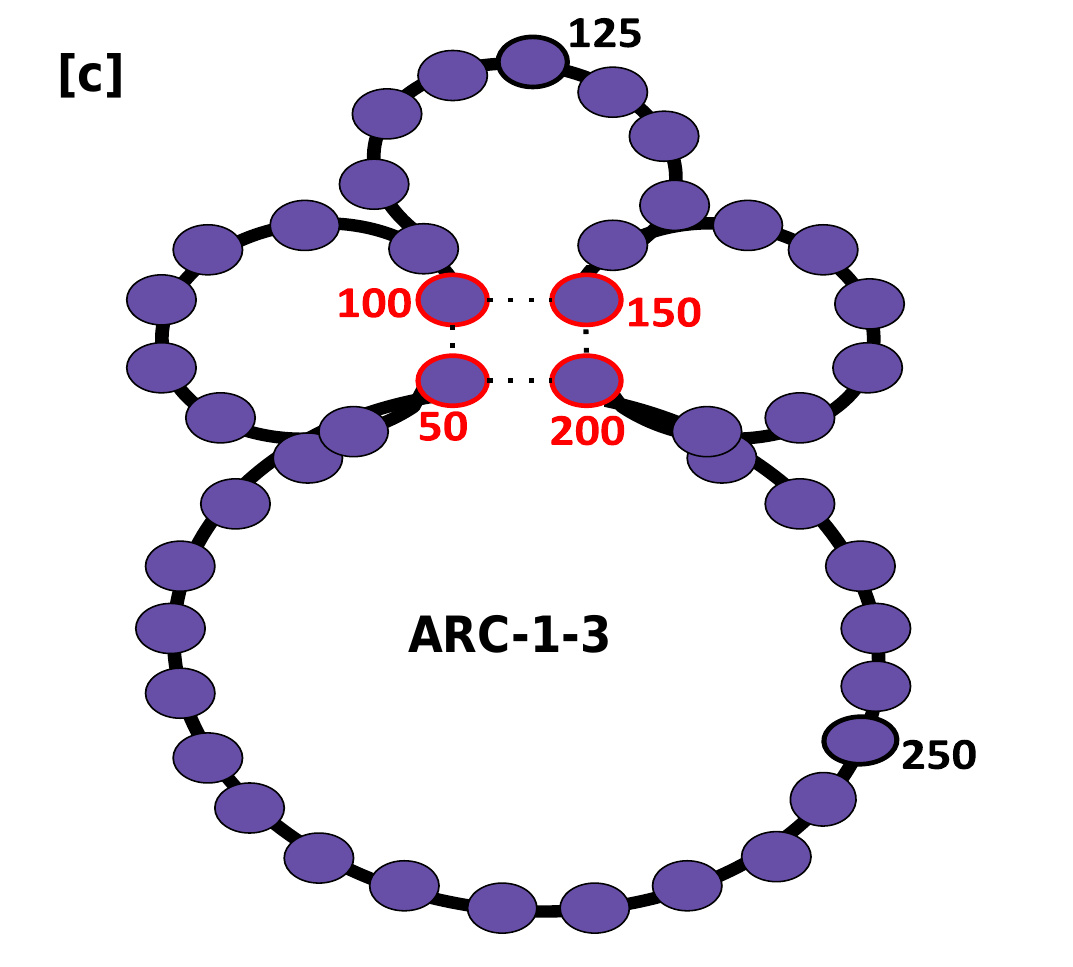}
\includegraphics[width=0.95\columnwidth,angle=0]{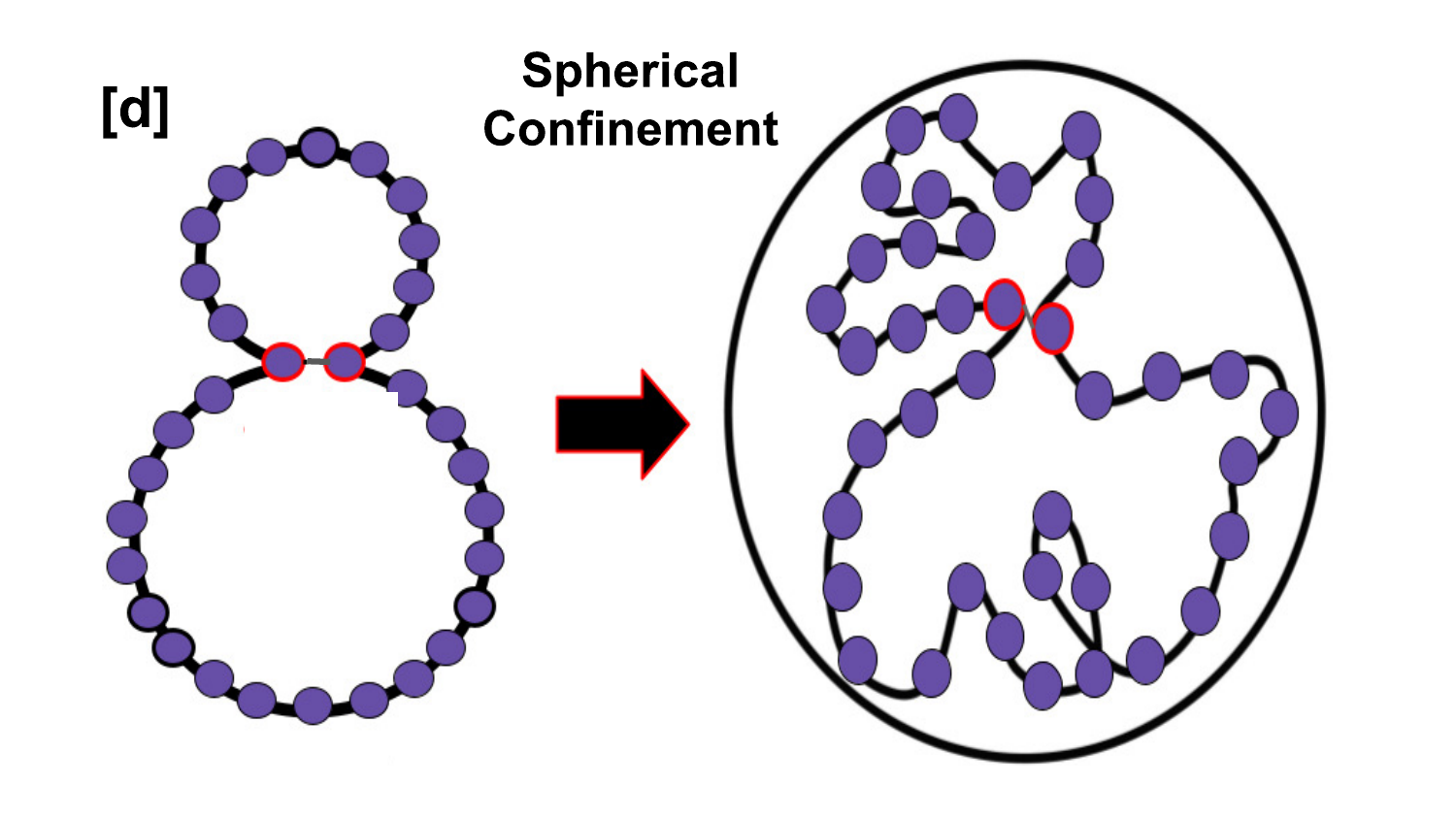}
\caption{\label{fig1}
In this figure, we show schematics of different topologically modified ring polymers. In  schematic
(a), a ring polymer with $200$ monomers is modified by cross-linking monomer $50$ and monomer $100$, 
i.e. we introduce a  harmonic spring between the  two monomers to create the topologically modified 
Arc-1-1 architecture.  The subfigures (a), (b), and (c) schematically 
show bead-spring models of the Arc-1-1, Arc-1-2, and Arc-1-3, polymers, which have one, two, and three small loops, 
respectively, created by cross-linking the appropriate pairs of monomers. Each of the small loops  have $50$ 
monomers. Each of the architectures also have a larger loop which has $150$ monomers. Hence, total number of 
monomers in Arc-1-1, Arc-1-2 and Arc-1-3 are $200, 250$ and $300$ monomers, respectively. 
The figure also highlights the specific monomers that have been cross-linked in order to create the loops. 
The light-grey lines represent the new cross-links introduced.  Subfigure (d) schematically shows an Arc-1-1 polymer 
confined within a sphere. We maintain a volume fraction of $0.2$.
}
\end{figure*}

\section{Model}
We use LAMMPS \cite{LAMMPS} to perform Langevin Dynamics simulations of flexible bead-spring model of polymers.
In the bead-spring model,  neighbouring monomers along the chain contour interact via the harmonic spring potential 
with energy: $V_{spring} = \kappa(r-a)^2$.
Here, $r$ is the distance between the adjacent monomers along the polymer contour at the particular configuration, 
$\kappa$ is the spring constant, and $a$ is the equilibrium bond length. We use $a$ as the unit of length in our system. 
Thus we express all the other lengths of other quantities, e.g. the diameter of the confining sphere in units of $a$. 
We take the value of $\kappa$ to be $100k_B T/a^2$, where $k_BT =1$ sets the unit of energy in our simulation.
The mass ($m$) of the monomers are set at $m=1$, such that the unit of time can be calculated 
as $\tau = \sqrt{ma^2/k_B T}$. Apart from the harmonic spring interactions, we also add excluded volume interactions, 
which prevent overlap between the monomers. The excluded volume interactions follow the Weeks-Chandler 
Anderson (WCA) potential \cite{allen_computer_2017}, which is given by:
\begin{equation}
V_{WCA}=4\epsilon[(\sigma/r)^{12} - (\sigma/r)^6] +   f_c r +  \epsilon_0,  \\  \forall r< r_c.
\end{equation} 
The cut-off distance $r_c$ of the WCA potential is given by $2^{1/6}\sigma$, and  $V(r)=0$ for $r>r_c$.  
The addition of $f_c r$ and $\epsilon_0$ to the equation with suitable values of $f_c$ and $\epsilon_0$, 
ensures that both the potential  and the force goes smoothly goes to zero at the cut-off distance $r_c$. 
 The diameter of the monomers is $\sigma$, which we take to be $0.8a$. This choice of $\sigma$ also ensures 
 that chains do not cross during the simulation. However, the choice of $\sigma=0.4a$, used for {\em some} 
 of our calculations implies that chains can easily cross each other, as a monomer from a different segment 
of the polymer can be fit in   between two adjacent monomers along the chain. Note that the (phantom) 
spring potential just maintains the mean distance between adjacent monomers along the contour but we do 
not have any real material spring  or chemical  bonds which maintain an average distance of $a$ between 
adjacent monomers along the chain contour.  In our model, chain crossing is prevented as  a consequence  
of excluded volume interactions and can be  controlled by a suitable  choice of $\sigma$. We integrate 
the equations of  motion  using a  time step  $\delta t = 0.01 \tau$. 

\begin{table}
\begin{tabular}{ | c | c | c | c | c |} 
\hline
Architecture & $N_m = N_{sm} + N_{bg}$ & $N_p$  &  $R_s$  &  $R_{actual}$  \\ 
\hline
\multirow{6}{5em}{Arc-1-1} & \multirow{6}{5em}{$200 = 50+150$} & 1 &  $4a$ & $4.4a$  \\ 
 & & 2 &  $5a$ & $5.4a$  \\ 
 & & 4 &  $6.25a$ & $6.65a$  \\ 
 & & 6 &  $7.25a$ & $7.65a$  \\
 & & 8 &  $8a$ & $8.4a$  \\
 & & 10 &  $8.5a$ & $8.9a$  \\ 
 \hline
\multirow{6}{5em}{Arc-1-2} & \multirow{6}{5em}{$250 = 2 \times 50+150$} & 1 &  $4.25a$ & $4.65a$  \\
 & & 2 &  $5.5a$ & $5.9a$  \\ 
 & & 4 &  $6.75a$ & $7.15a$  \\ 
 & & 6 &  $7.75a$ & $8.15a$  \\
 & & 8 &  $8.5a$ & $8.9a$  \\
 & & 10 &  $9.25a$ & $9.65a$  \\
  \hline
\multirow{6}{5em}{Arc-1-3} & \multirow{6}{5em}{$300 = 3 \times 50+150$} & 1 &  $4.5a$ & $4.9a$  \\
 & & 2 &  $5.75a$ & $6.15a$  \\ 
 & & 4 &  $7.25a$ & $7.65a$  \\ 
 & & 6 &  $8.25a$ & $8.65a$  \\
 & & 8 &  $9a$ & $9.4a$  \\
 & & 10 &  $9.75a$ & $10.15a$  \\
 \hline
\end{tabular}
\caption{\label{Tab:Tcr}
The table lists (a) the different architectures, (b) total number of monomers $N_m$ in each polymer (c) the number of monomers 
in the bigger (smaller) loops  $N_{bg}$ ($N_{sm}$), (d) the number of total polymer chains in the sphere $N_p$, (e) the effective 
($R_s$) and  actual radii ($R_{actual}$) of the confining sphere so that volume fraction is equal to $0.2$ with $\sigma=0.8a$.
At the end of this manuscript, we also use $\sigma=0.4a$ to allow chain crossing, keeping the above numbers unchanged.
}
\end{table}
{\em Polymer nomenclature:} We primarily  simulate ring polymers with $N_m$ monomers (beads) in a ring. For a ring polymer, 
the last monomer along the contour is connected to the first monomer by a harmonic spring interaction, with the same 
spring constant as mentioned above. We start with $ N_m=200$ monomers per chain,
and subsequently consider chains with larger number of monomers as we introduce topological modifications. To introduce 
topological modifications, we cross link specific  monomers of the ring-polymer using harmonic springs of equilibrium 
length $a$. This effectively creates distinct internal loops within the ring polymer. 
The size of the loops, i.e. the number of monomers in the  internal loops are  dictated by the choice of 
monomers which we cross-linked. As an example, for a ring polymer with $200$ monomers,  we create the Arc-1-1 
architecture by adding a cross-link between the $50$-th and $100$-th monomers, refer Fig.\ref{fig1}a for a schematic
of the architecture (topology). This effectively creates {\bf one} big loop of $150$ monomers and {\bf one} small loop having $50$ monomers. 
Hence, the name: Arc-1-1. Similarly, Arc-1-2  (and Arc-1-3)  each have one big loop of $150$ monomers and two (and three) 
small loops, respectively.  Each of the small  loops have $50$ monomers, unless specified otherwise. Refer Fig.\ref{fig1}(b,c)
for schematics of the corresponding architecture.  We choose the large loop in each polymer to have  $150$ monomers. 
Thus, when we need to be more specific about the number of monomers in each loop, we label the
the polymers as Arc-1-1[150-50],  Arc-1-2[150-50] and Arc-1-3[150-50] such that 
we can calculate the total number of monomers in the polymers as $N_m = 200, 250, 300$, respectively. 
Later we also consider Arc-1-3 polymers with $150$ monomers in big loops and $25$ monomers in the small loop. 
Thus an Arc-1-3[150-25] will have $225$ monomers in the polymer.

{\em Radius of confining sphere:} We equilibrate $N_p$ number of polymers in a confining sphere 
to study the radial organization of loops and positional distribution of monomers which are part of different 
internal loops of the ring polymer. The choice of the number of polymers  with different topological modifications 
and the corresponding radius of the confining spheres  has been listed in Table \ref{Tab:Tcr}.

The effective radius of the confining sphere($R_s$) is calculated in order to keep the 
monomer volume fraction to be equal to $0.2$ with $\sigma=0.8a$ with $N_p$ polymers, refer Fig.\ref{fig1}d. 
Chromosomes in living cells have been reported to have volume fractions close to this value, 
and this  inspires our choice of volume fraction. In our modelling, we ensure that the centres of 
the monomers to be able to access distances of $R_s$ from the center of the sphere,  
i.e. the monomer surface farthest from sphere center can be at $ R_{actual} = R_s + \sigma/2$ from 
the center sphere. This is implemented by  the WCA potential between each of the monomers and the wall 
of the confining sphere. We then use this value as the actual radius of the sphere($R_{actual}$) in the simulations.
The effective and actual sphere radii of the confining sphere used in all of our different cases have also summarised 
in Table \ref{Tab:Tcr}. When we use $\sigma=0.4a$, we use the same values 
of $R_{actual}$ as listed in Table \ref{Tab:Tcr} for $N_p$ polymers, such that we have identical 
number density of monomers.

{\em Initialization and Equilibration:} 
We aim to investigate ring-polymers without concatenations between loops. We have
taken special care to ensure that we initialize polymers within the sphere without concatenations. Thereafter,
excluded volume interactions keep them unconcatenated during the course of the simulations.
 We initialize  by placing the monomers of each chain along compact circular ring configuration,
 without the consideration of topological modifications due to  cross-links or strong excluded volume 
 interactions due to overlaps between the monomers in the initial configuration. 
 The diameter of the ring is chosen to be less than that of the confining sphere. 
 If there are multiple polymers in the sphere, then the monomers of each polymer  are stacked in a ring configuration 
 one over each other. The interaction energies are initially extremely high in the system because of excluded volume interactions
 between overlapping monomers, as well as stretched/compacted springs as we force the monomer to occupy adjacent positions
 along a circles. To relax this system, we first do off-lattice Metropolis Monte Carlo (MC) simulations from 
 the initial configuration  for $10^5$ MC steps. The interaction potentials for the springs and the LJ interactions 
 are kept the same as those to  be used for Langevin simulations using LAMMPS. Simulations with $10^5$ MC steps are
 enough to relax the system and bring the springs to their equilibrium lengths. We explicitly check 
 by simulations in a much larger box that the polymers are free to diffuse away from each other and remain 
 unconcatenated  at the end of the Monte Carlo relaxation  run. Thereafter, we  use this configuration
 to  create ten different initial conditions for independent  runs, which are then used for the Langevin simulation 
 production runs using LAMPPS. Each  Langevin simulation production run is for 
 $2 \times 10^8$ iterations and  we collect data  only after $10^8$ steps, 
 as we allow equilibration  in the presence of other polymers in the sphere for the first $10^8$ iterations. We collect 
 the position  of each monomer in the system every $10^4$ iterations (i.e. every $100 \tau$) to  ensure 
 statistical independence of polymer configurations (microstates), and use this data to 
 calculate averaged statistical quantities to decipher the conclusions presented in the next section.

\begin{figure*}[!hbt]
\includegraphics[width=0.45\columnwidth,angle=0]{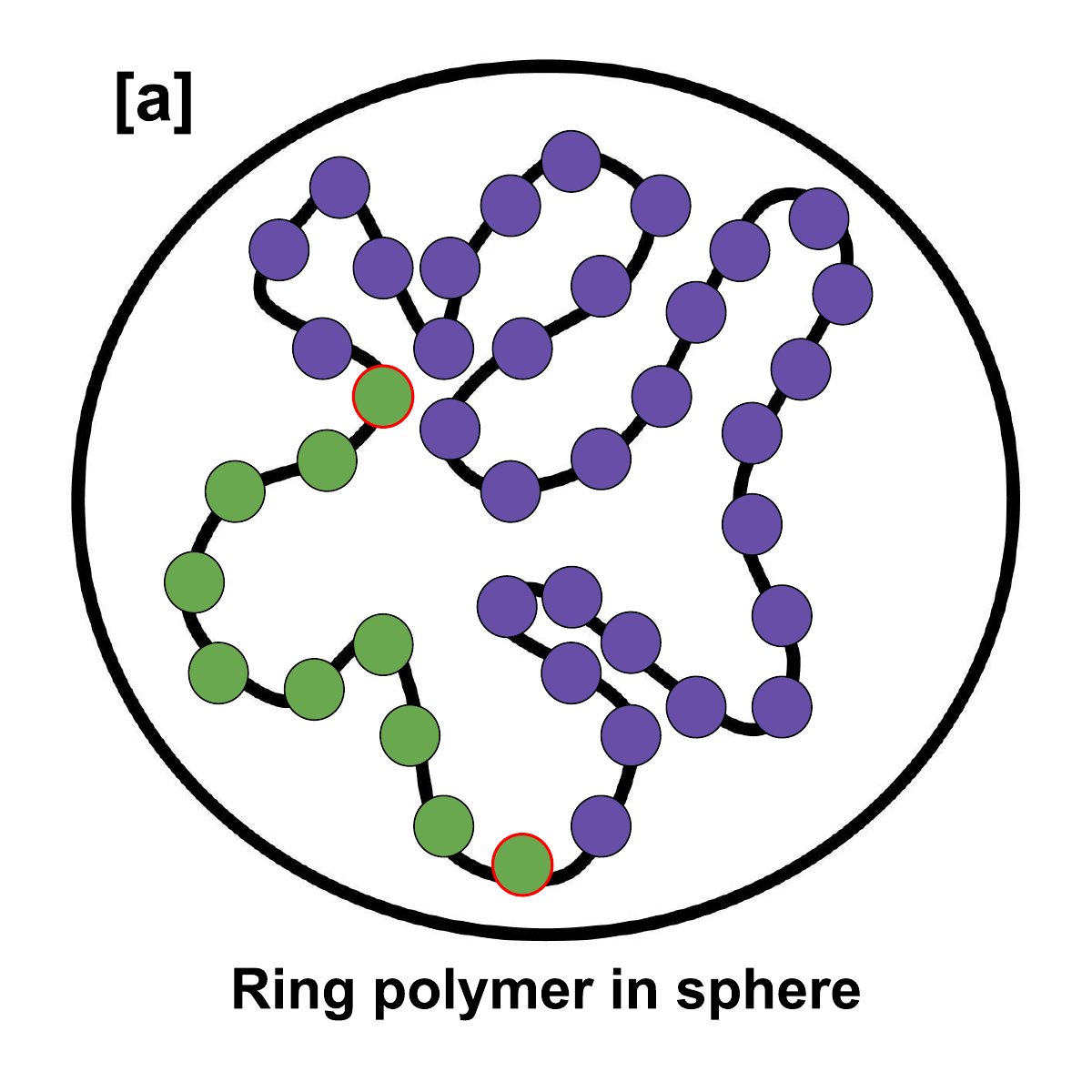}
\includegraphics[width=0.47\columnwidth,angle=0]{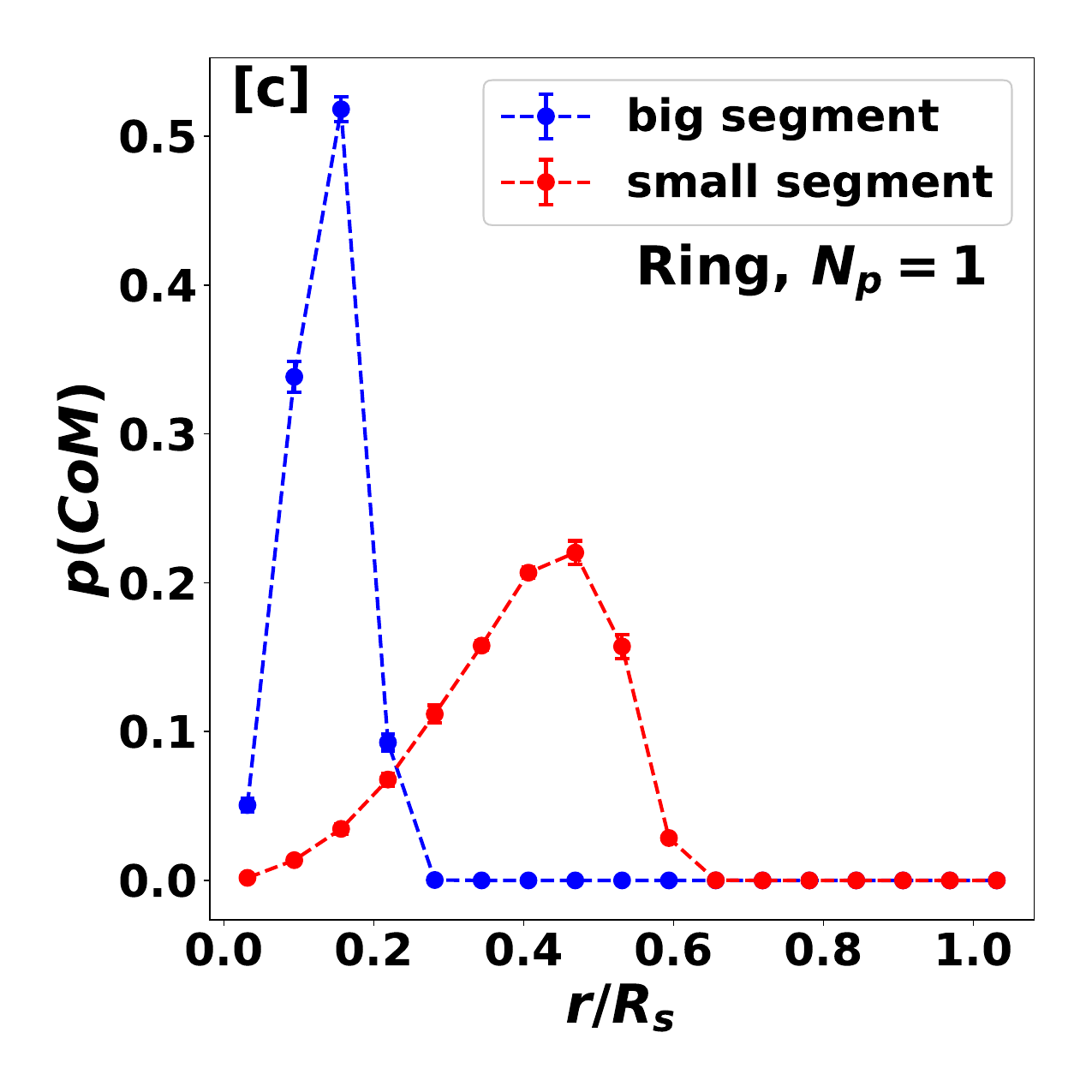} \hskip0.3cm
\includegraphics[width = 0.47\columnwidth]{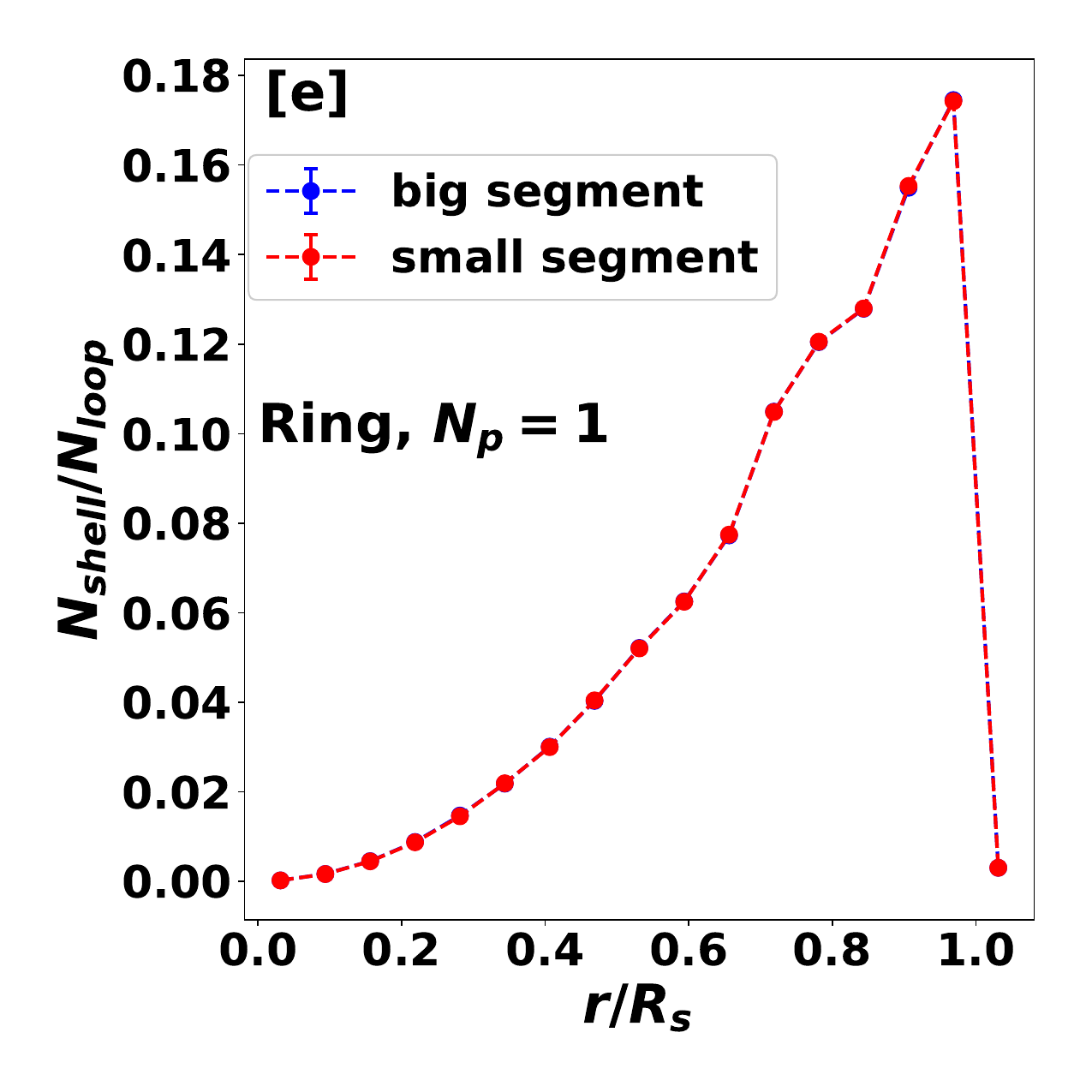} \hskip0.3cm
\includegraphics[width=0.47\columnwidth,angle=0]{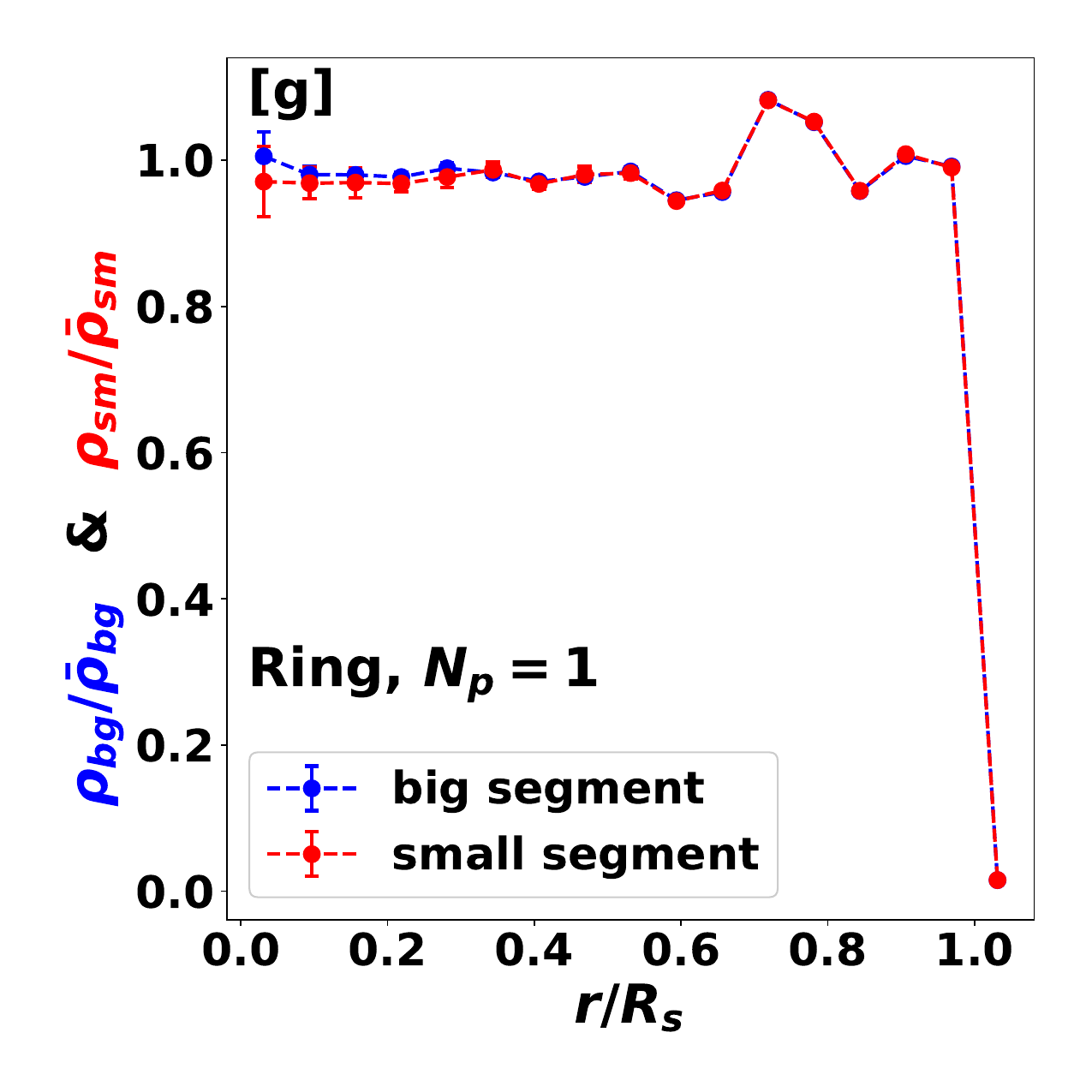} 
\includegraphics[width=0.45\columnwidth,angle=0]{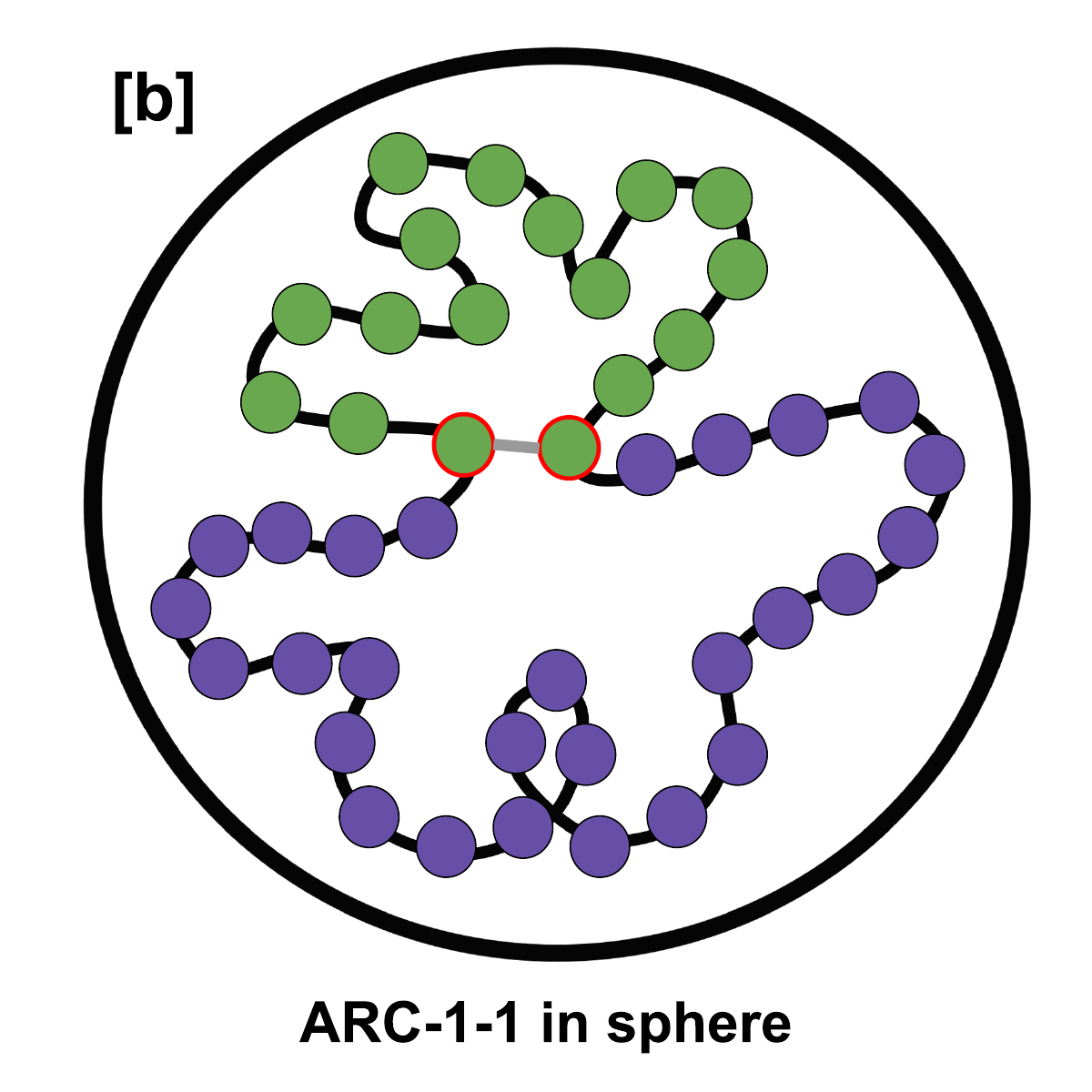}
\includegraphics[width=0.47\columnwidth,angle=0]{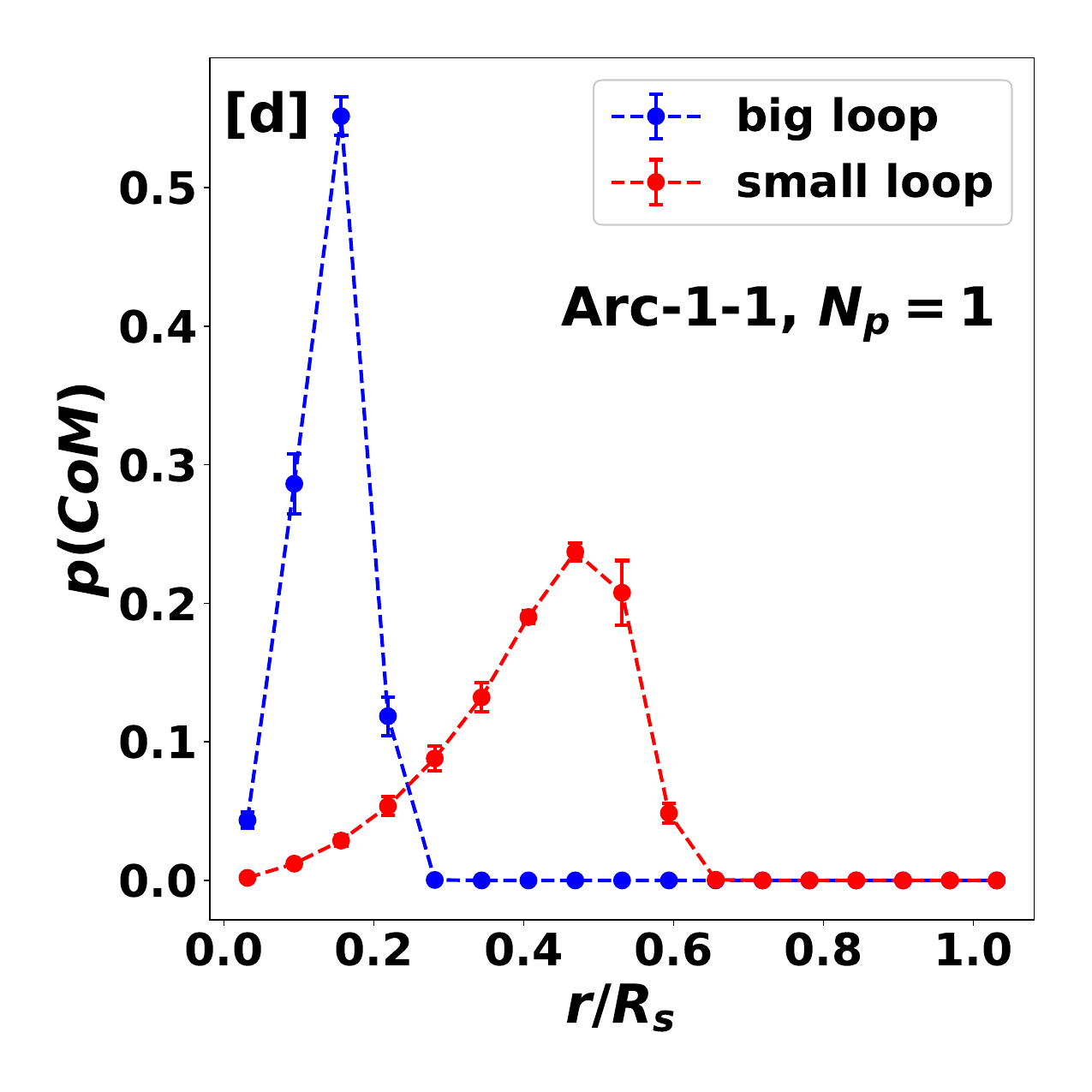} \hskip0.3cm
\includegraphics[width=0.47\columnwidth,angle=0]{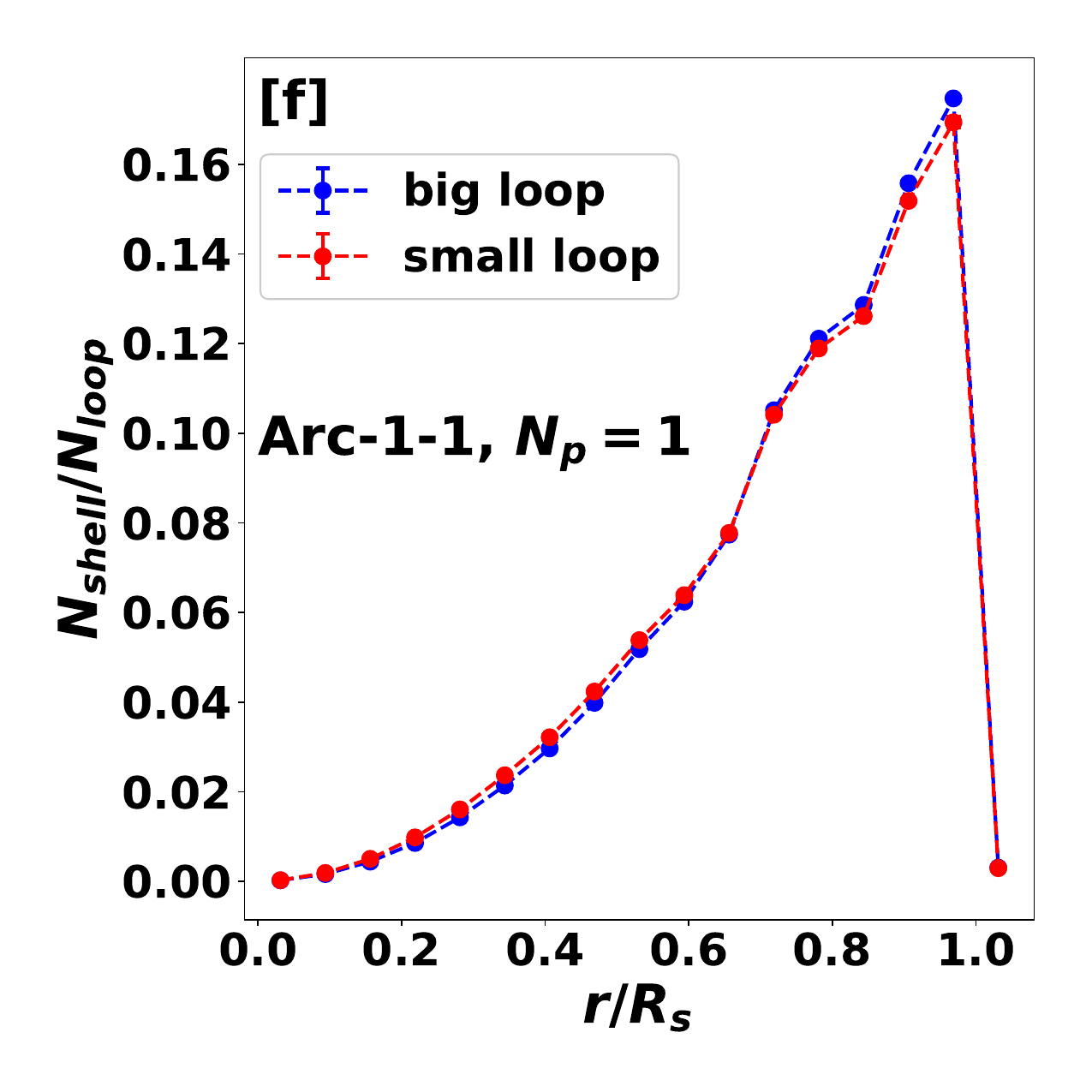} \hskip0.3cm
\includegraphics[width = 0.47\columnwidth]{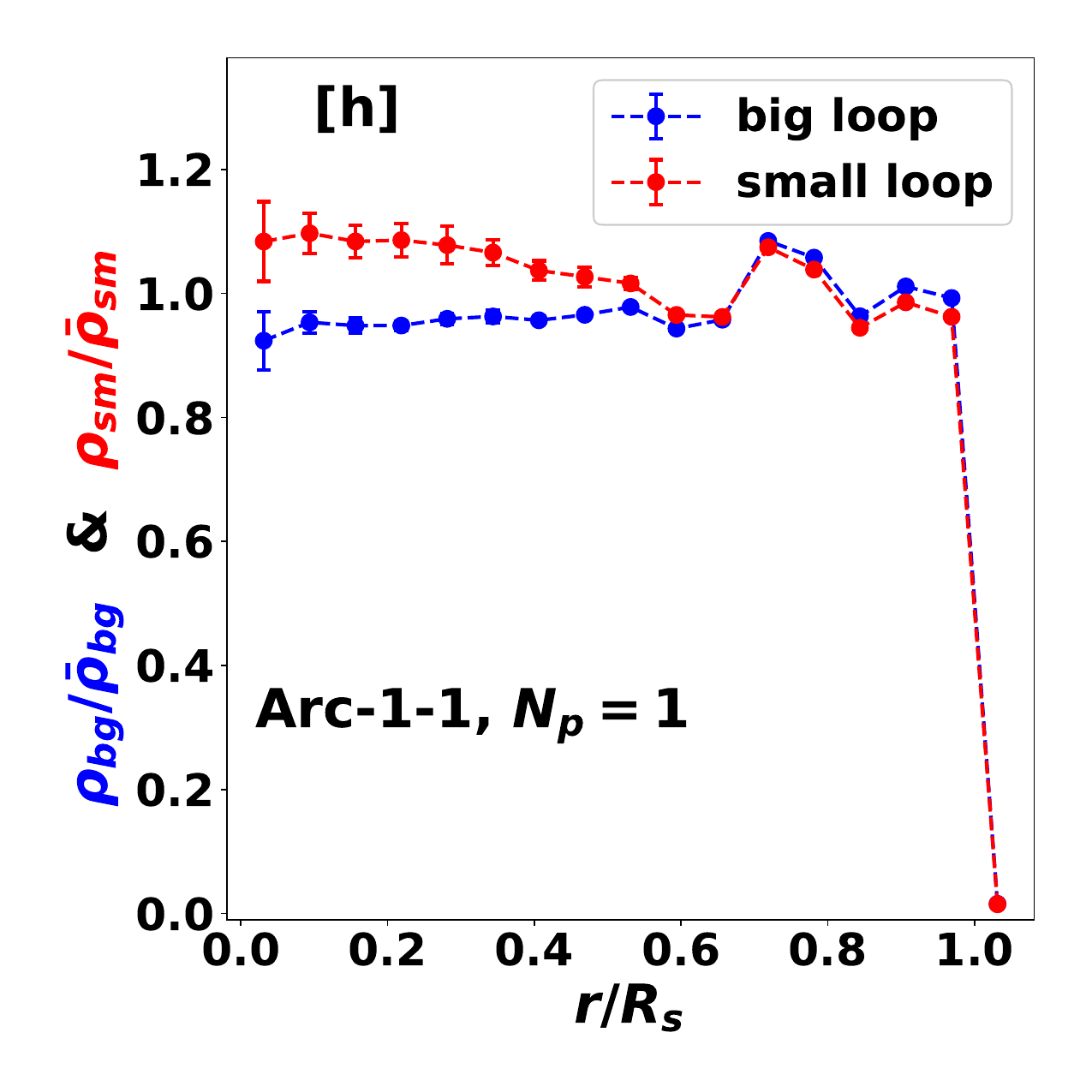}
\caption{\label{fig2}
This set of subfigures shows the difference in organization between the two different segments of a Ring-polymer 
and the corresponding  two segments (loops) of the Arc-1-1 polymer confined within in a sphere. 
Schematics (a) and (b) show a ring polymer and Arc-1-1 polymer within the sphere. Monomers between $50$ to $100$
correspond to the {\em small segment} (green circles) and is of length $50a$ in the ring polymer.
The other {\em big segment} (purple segments) of the  ring polymer has $150$ monomers. 
In subfigure (b), the  same  two monomers, i.e. the $50$-th and the $100$-th monomer  
(marked in red outline in both schematics) have been cross-linked to form two loops with $50$ 
monomers ({\em small loop}) and $150$ monomers ({\em big loop}) of the Arc-1-1 architecture.  
Each polymer is confined in a sphere of radius $R=4a$, as mentioned 
in Table \ref{Tab:Tcr}. Subfigures (c) (and [d]) show the radial probability distribution of the 
Center of Mass (CoM) of the small and big segments (and correspondingly, small and big 
loops)  of the ring  polymer (of the Arc1-1 polymer). The sphere is divided into concentric spherical shells of width 
$\delta r =0.25a$, and we plot the probability of finding the CoM within each shell.  Subfigures (e) and (f) 
show  the average fraction of monomers $N_{shell}/N_{loop}$
of a particular  segment (or loop) which found in each shell, at a distance $r$ from the center of sphere. 
Finally, subfigures (g) and (h) show the statistically averaged  normalized monomer number densities 
$\rho_{bg}/\bar{\rho}_{bg}$ and $\rho_{sm}/\bar{\rho}_{sm}$ in each shell for the ring and the Arc-1-1 polymer,
respectively.  Refer text for calculation details.
The error bars show the standard deviation calculated from the $10$ independent runs.  In each subfigure, 
we also  mention the polymer architecture (topology) and the number of polymers $N_p$ confined within the sphere, 
for the reader's quick reference.}
\end{figure*}
\section{Results}
\subsection{Organization of polymer segments in a sphere.}
 We now look at how the internal loops of topologically modified polymers
are distributed radially in the sphere, and compare this with the distributions of segments of simple ring polymers 
having the same contour length and under similar confining conditions. To this end, we first compare a 
single Arc-1-1 architecture  and a ring polymer of the same length,i.e, $N_m=200$ monomers. 
The single polymers are each confined within  a sphere $R_s=4a$. The choice of $R_s$ maintains 
the volume fraction of monomers at $\rho_s =0.2$ for $\sigma=0.8a$.
The radius of gyration $R_g$ of $200$-monomer ring polymer in a dilute solution with good 
solvent without confinement is  $\approx 6.7a$ \cite{Ravi2022} (also refer Appendix)
The reader may refer to Table \ref{Tab:Tcr}
for the values of $R_s$  for different cases henceforth. Figures \ref{fig2}(a) and 
(b) schematically show the ring polymer and an Arc-1-1(150-50) polymer that we compare
at the  start of our investigations. 
\begin{figure}[!hbt]
\includegraphics[width=0.98\columnwidth,angle=0]{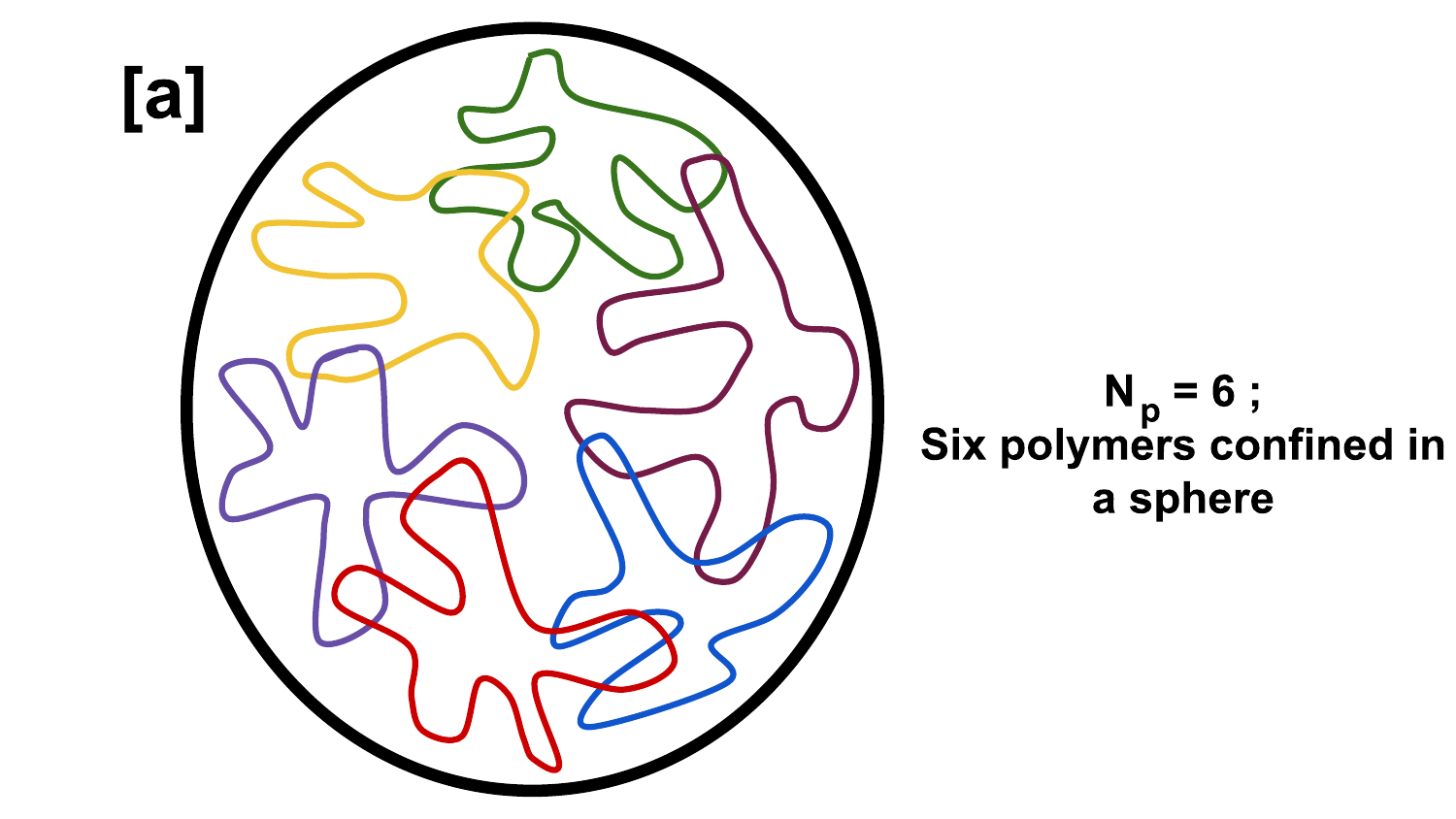}
\includegraphics[width=0.49\columnwidth,angle=0]{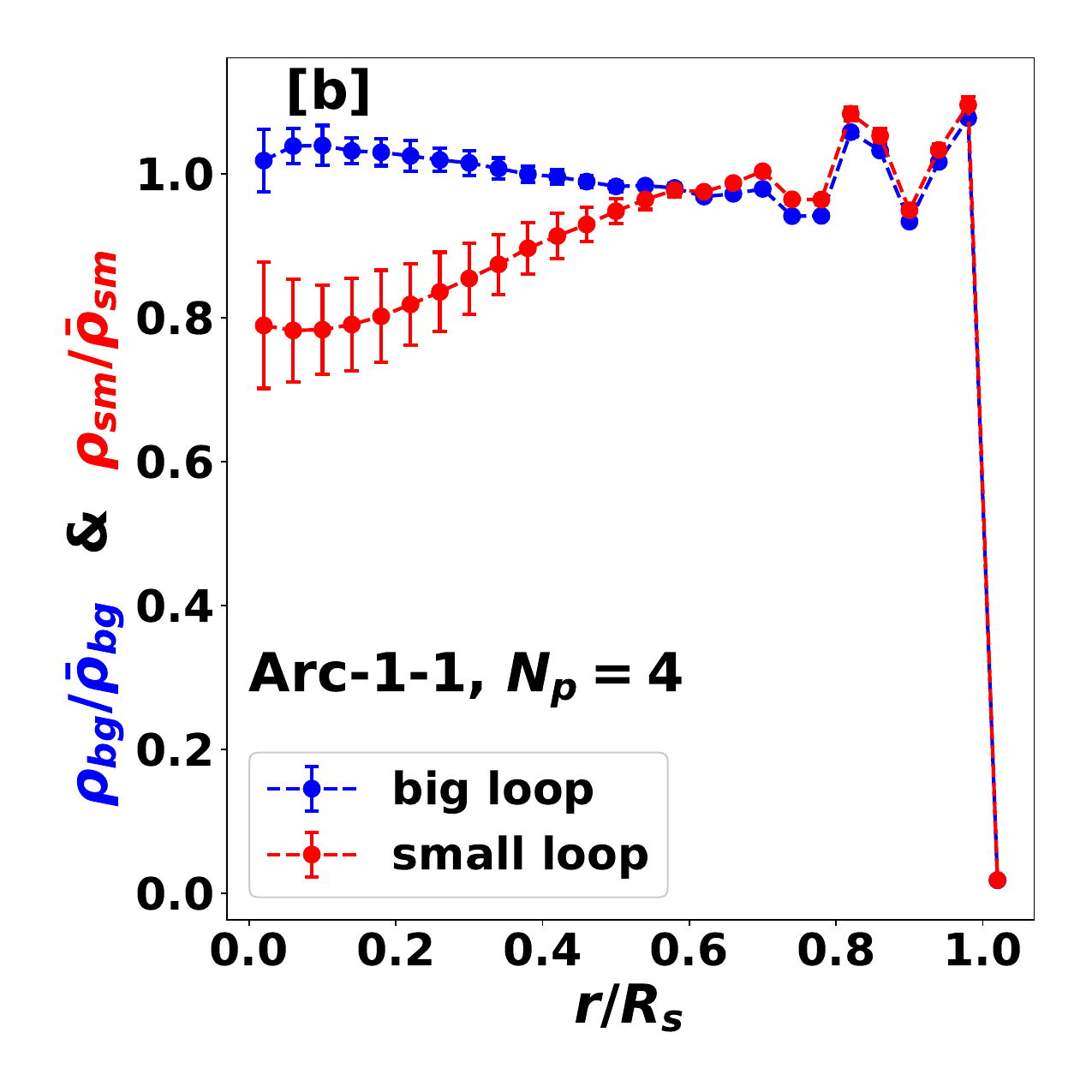}
\includegraphics[width=0.49\columnwidth,angle=0]{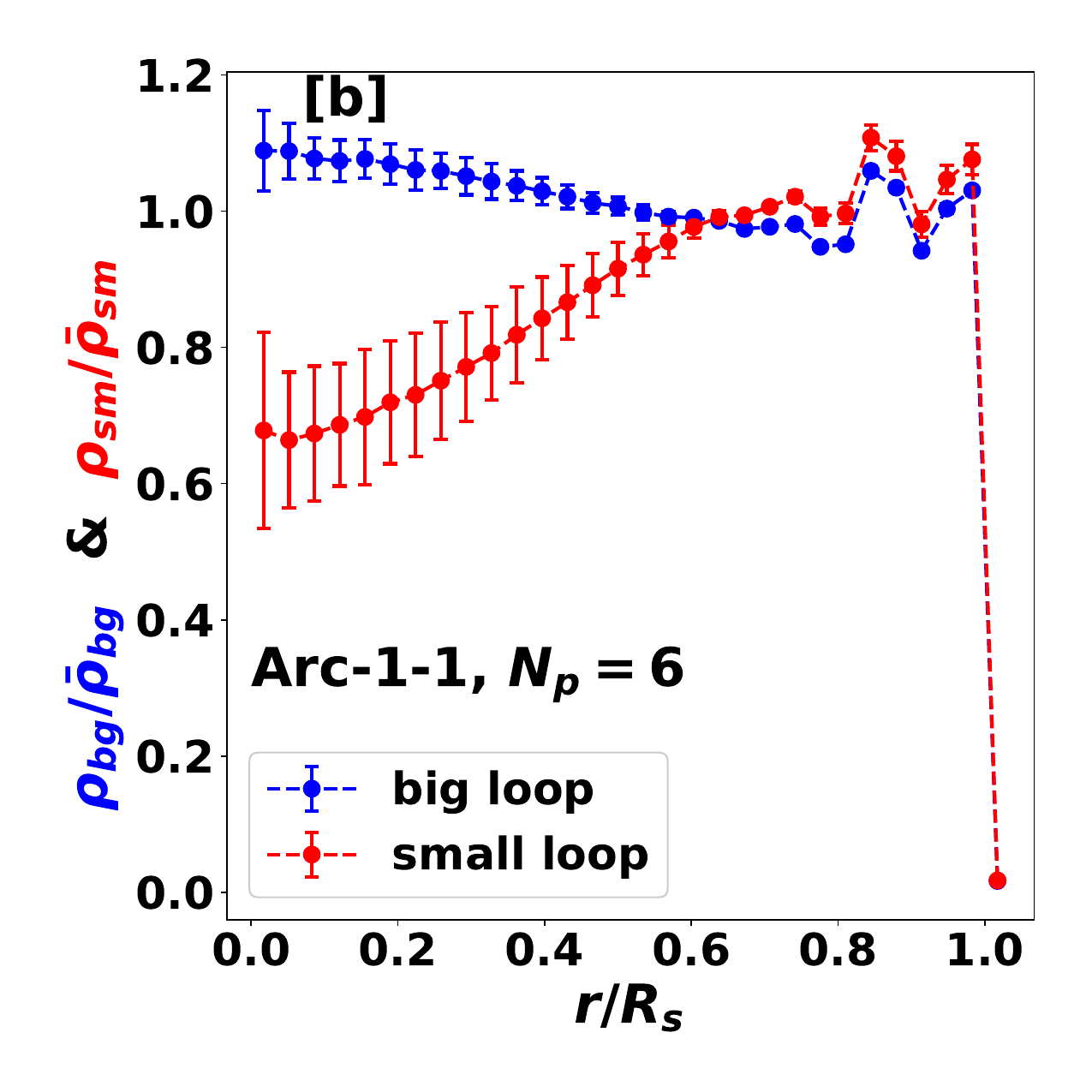}
\includegraphics[width=0.49\columnwidth,angle=0]{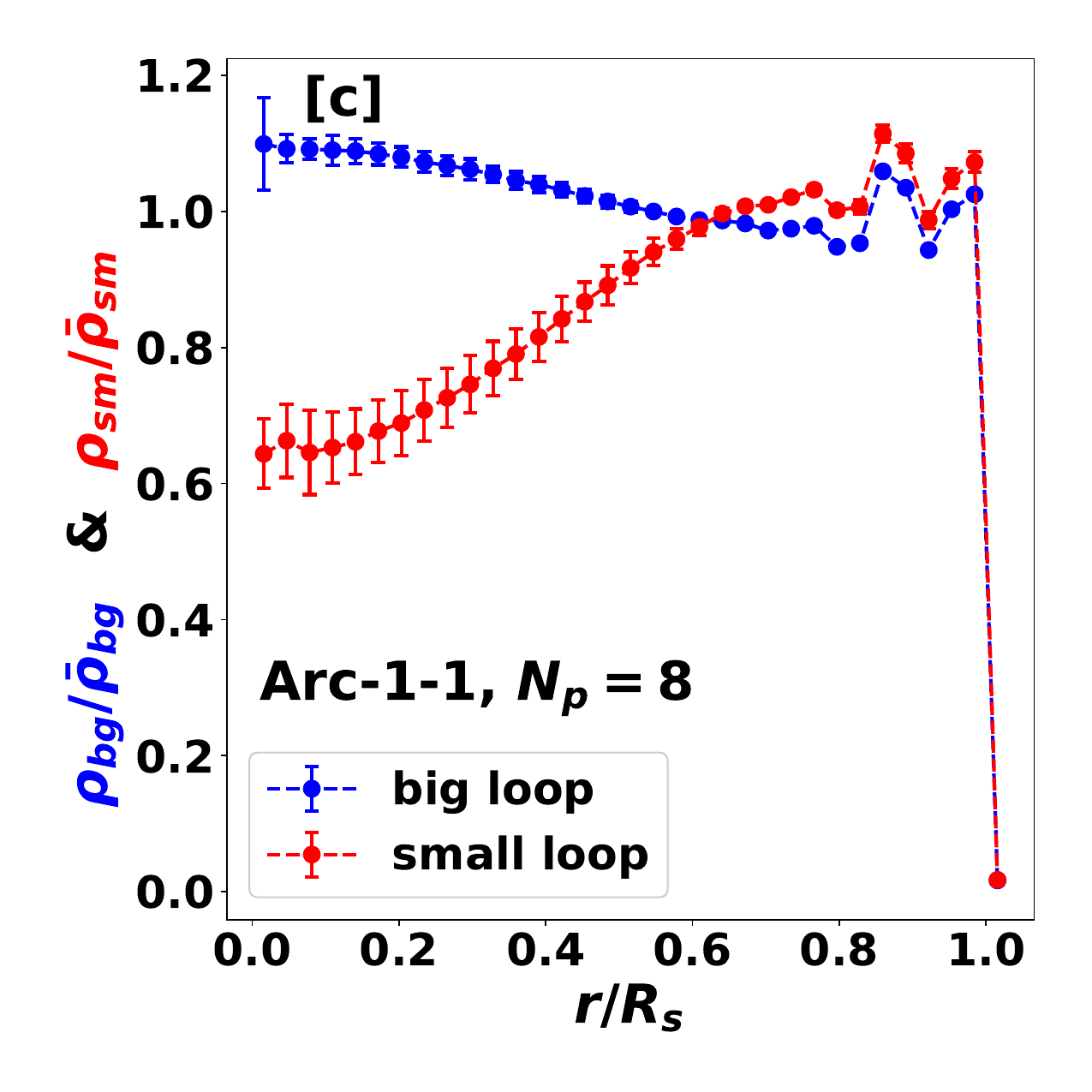}
\includegraphics[width=0.49\columnwidth,angle=0]{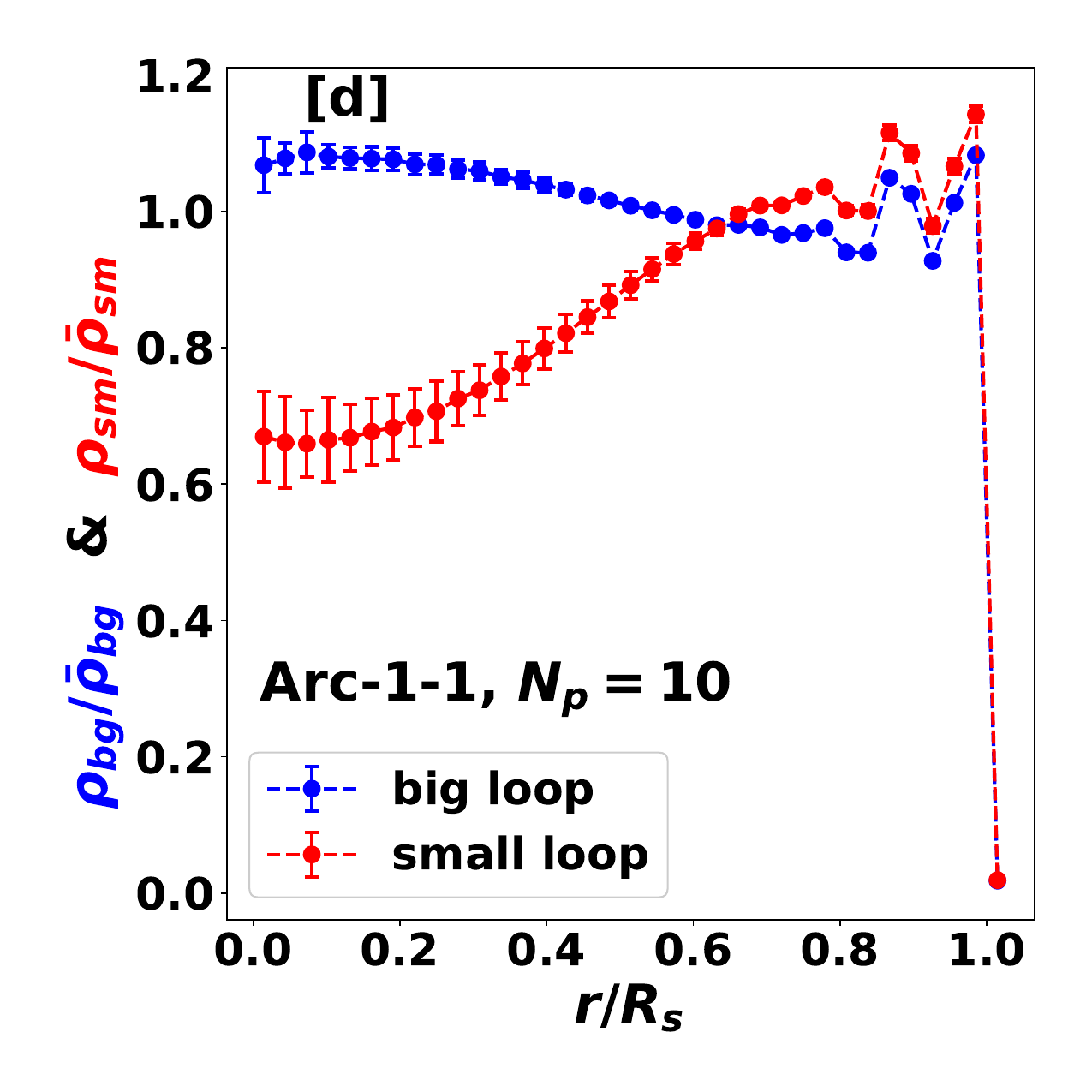}
\caption{\label{fig3} 
This figure shows the organization of the small and big loops when there are multiple Arc-1-1 polymers in the chain. 
The schematic in subfigure(a) shows a particular case where there are six polymers are confined in a sphere. All the polymers confined in the sphere have the same topology as we just vary the numbers of polymer 
chains in the sphere for different cases. Subfigures (b), (c), (d), and (e) show the normalized monomer densities,
belonging to the small or big loops of the $N_p$ Arc-1-1 polymers, as a function of the (normalized) radial distance 
$r$ from the center of sphere. The number of polymers $N_p$ within the sphere 
is equal to $4$, $6$, $8$, and $10$ respectively. The values of normalized monomer density of the small(or big) loop represented in the plots is the average of the values found for the small(or big loop) of each individual polymer confined in the sphere. Each subfigure has a different 
value of $N_p$, and we vary the $R_s$ accordingly to keep volume  fraction fixed. Since we are keeping the 
width of the shells as $0.25a$ in every case, spheres with larger radii can contain a larger number of shells, 
and consequently have a larger number of data points are plotted. We use relevant monomers from all $N_p$ polymers  
to calculate  $\rho_{sm}$ or $\rho_{bg}$.
}
\end{figure}

To ensure equivalent comparisons such that we identify changes (if any) to the statistical 
distributions are solely due to the introduction of an additional CL between monomers $50$ and $100$ in 
Arc-1-1, we do the following. We choose to notionally divide the $200$ monomer ring polymer 
into two segments with $N_{sm}=50$ monomers (small segment)   and $N_{bg}=150$ monomers (big segment), 
respectively.
This is equal to the number of monomers in the small and big loops of Arc1-1(150-50), respectively.
We compare  radial distributions of the center of Mass (CoM) of the two segments for the ring polymer 
with that of the Arc1-1 loops in  Figs.\ref{fig2}(c) and (d). We see no significant difference between the 
CoM distributions for the Arc1-1 and the ring polymer. Furthermore,  for both cases, the distribution of the CoM 
of the bigger loop (segment) is closer to the center of the sphere than the CoM distribution of
the smaller loop (segment). 

 However, the astute reader will soon realize that the CoM distribution does not reveal any information about 
 the positional distribution of the monomers in the loops.  The observation of the CoM of the big loop 
 being close to the sphere center may be a consequence of two different scenarios. It may 
 mean that the monomers of the big loop are localized close to the center. However, monomers distributed 
 evenly close to the sphere periphery in an isotropic manner may also lead to the position of CoM being 
 observed to  be close to the center. Hence, to understand where the monomers of each
segment are primarily located inside the sphere, we divide the sphere into concentric shells of different 
radii. We then calculate the mean of the fraction of monomers from each  segment (loop) $N_{shell}/N_{loop}$ 
present in each  of the shells for the ring (Arc-1-1)  polymer. The quantity $N_{shell}$ is the number of monomers 
of each segment (loop) of the ring polymer (Arc-1-1 polymer) which are in a particular shell, 
and $N_{loop}$ is the total number  of monomers in the corresponding  segment (loop). 
These plots are shown in Fig.\ref{fig2}(e) and (f). 
We observe that for the ring polymer, monomers of the small and big segments
are equally distributed throughout the sphere. However, by modifying the ring topology to Arc1-1, a slight 
difference can be seen between the distributions of the monomers of the two different loops.

As we go radially outwards, each of the spherical shells have increasingly larger volumes, and can thus accommodate 
more monomers. Thereby, the relevant quantity to  plot is the mean number density of 
monomers from each segment (loop) as we move radially outwards, and moreover, normalize the mean number density 
of monomers in each shell by the mean density of monomers of each segment (loop) in the entire sphere. 
The mean number density $\bar{\rho}_{bg}$ ($\bar{\rho}_{sm}$) of the big (or small)
loop in the sphere is given by $N_{bg}/V_s$ (or $N_{sm}/V_s$), where $V_s$ is the volume 
of the sphere with radius $R_s$. The subscripts $bg$ and $sm$ denote the big and small segments (loops), respectively,
and $N_{bg}$ and $N_{sm}$ are the number of monomers in big and small segments (loops) for ring (Arc-1-1) polymers. 
This quantity $\rho_{sm}/\bar{\rho}_{sm}$ or $\rho_{bg}/\bar{\rho}_{bg}$  is the normalized monomer density  for 
the small and big segments (loops), and is plotted in Fig.\ref{fig2}(g) and (h). 
This quantity measures the degree of deviation of the monomer densities of each
loop/segment from the mean monomer density at different radial locations within the sphere. 
We first discuss the reason for observing peaks in the normalized monomer density near the wall, 
starting from a distance of $\approx 0.75 r/R_s$.  These peaks arise because of the presence of the wall, 
which creates a layer of monomers first at a distance $\sigma/2$ from the wall surface, and further layers at distances that are
multiples of $\sigma$ from there (i.e., $3\sigma/2$, $5\sigma/2$, and so on) \cite{Jun2007,calleja_spherical_pore}. At very large volume fractions(very high confinements) we would be able to observe multiple peaks. 
At the level of confinement that we have chosen($\phi = 0.2$), we can only distinctly observe two of the peaks. 
These peaks disappear when we consider 
lower volume fraction of monomers within the sphere, as is the case when we release topological modifications of the system in a future section.

For the  ring polymer, the normalised monomer densities 
from the big and small  segments overlap as  one moves out radially. This implies  that the 
two different segments of the ring 
polymer have no preference to  be positioned either near the center or the periphery of the sphere. 
However, the big and small loops of Arc-1-1 show  a  clear difference between the normalized monomer 
densities in interior  (near the center of sphere) and peripheral  (near the wall of sphere) regions 
of the sphere. Monomers of the  big loop are more probable to be found 
near the periphery.  This is because the shells near the periphery have more volume,
and the monomers can explore more configurations, i.e. higher number of  microstates. 
Monomers of the smaller loop are relatively pushed   closer to the centre of the sphere. 

These plots further show that the centre of mass distribution does not give the complete picture 
of how monomers are localised in the sphere. The CoM distribution of the ring polymer segments are nearly identical 
to the distribution of CoMs   of Arc1-1 loops. Since the smaller-loop and smaller-segment have 
equal number of  monomers, the radial distribution of CoM  is solely  dependent on the number of monomers
we consider in each segment. We further establish this observation in the following sections. 
Figures \ref{fig2}(g) and (h) firmly establish that modifying the topology by adding CLs indeed re-organises
the  radial distribution of probabilities of finding monomers from different segments within the sphere. 
Monomers of smaller loop are found preferentially near the center of sphere.
\begin{figure}[!hbt]
\includegraphics[width=0.8\columnwidth,angle=0]{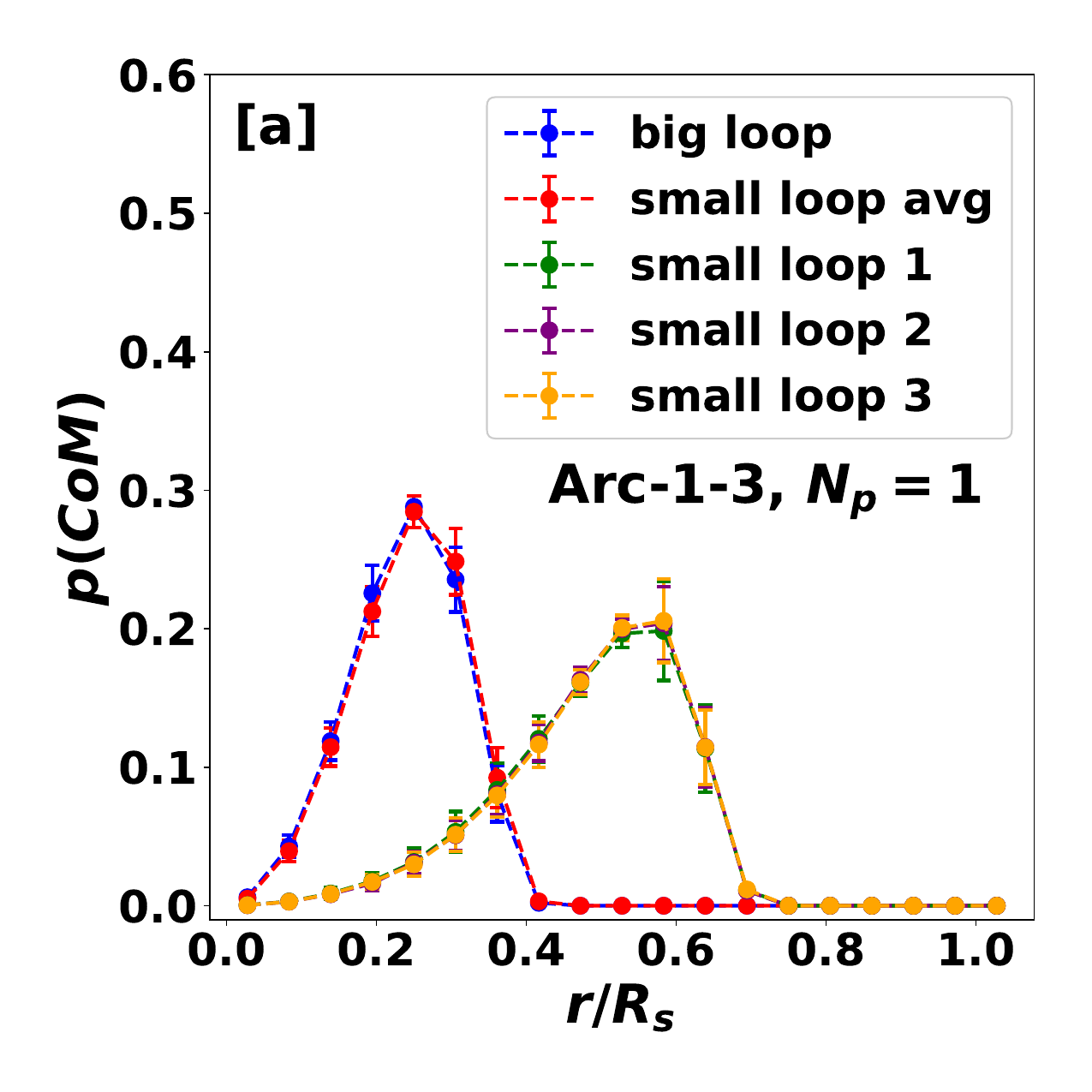}
\includegraphics[width=0.8\columnwidth,angle=0]{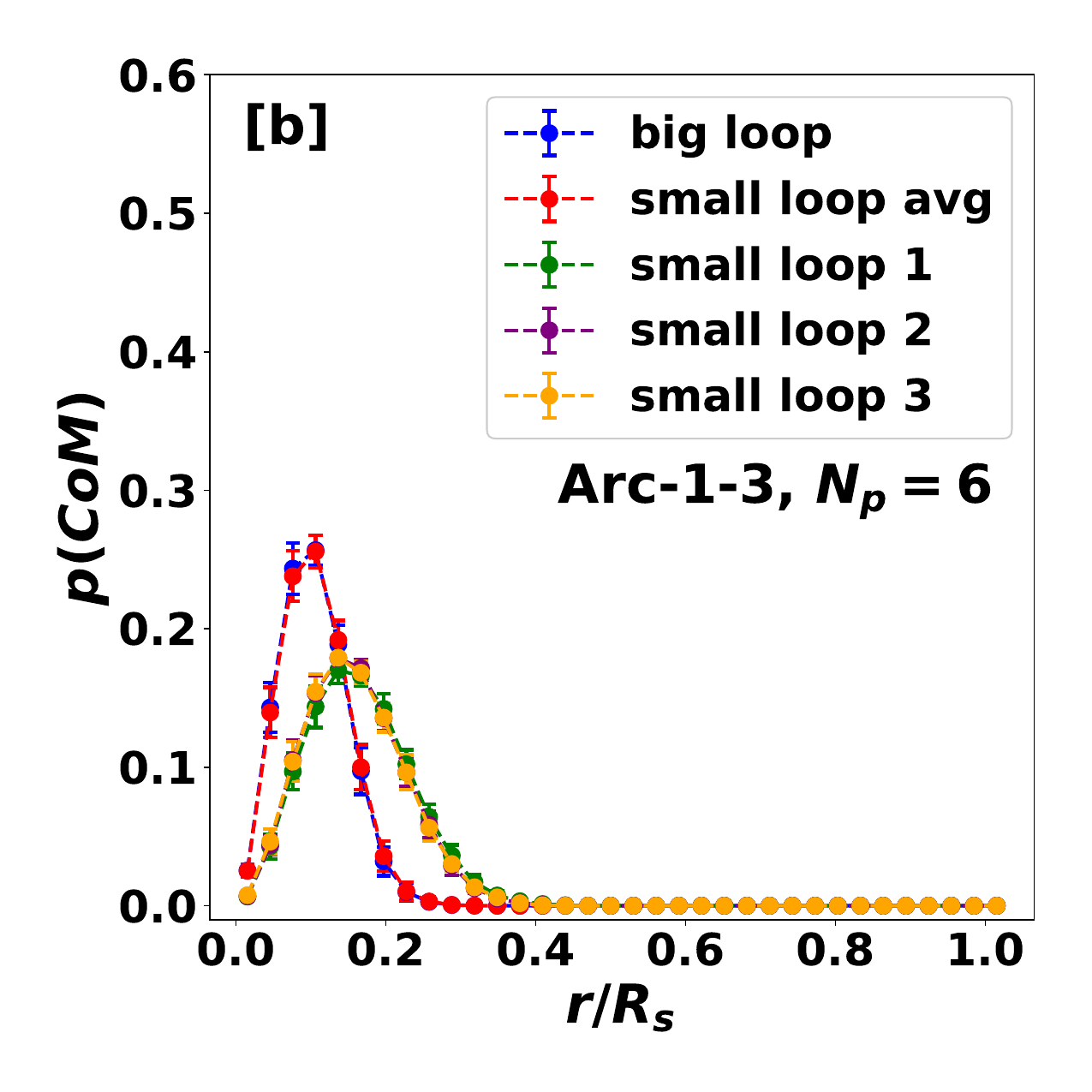}
\caption{\label{fig4}
This figure shows the distribution of the CoM's of the different loops for  Arc-1-3 polymer(s) confined in 
a sphere. Subfigure(a) shows the CoM distribution for Arc-1-3 polymer when $N_p=1$,
whereas subfigure(b)  shows the distribution when there are  $N_p=6$  Arc-1-3 polymers in the sphere. 
We have plotted the distribution of the CoM of the big loop, as well as the 
three separate CoMs of each of the three small loops of Arc-1-3. In addition, we
have also averaged the CoM coordinates of the three small loops, essentially giving us the the combined CoM 
of the small loops. The distribution of this average CoM of the small loops is also shown above and labelled 
`small loop average'. For the $N_p=6$ case, the plotted distributions represent the CoM distribution 
of big and  small loops  averaged over all the six different Arc-1-3 polymers.
}
\end{figure}

\begin{figure*}[!hbt]
\includegraphics[width=0.64\columnwidth,angle=0]{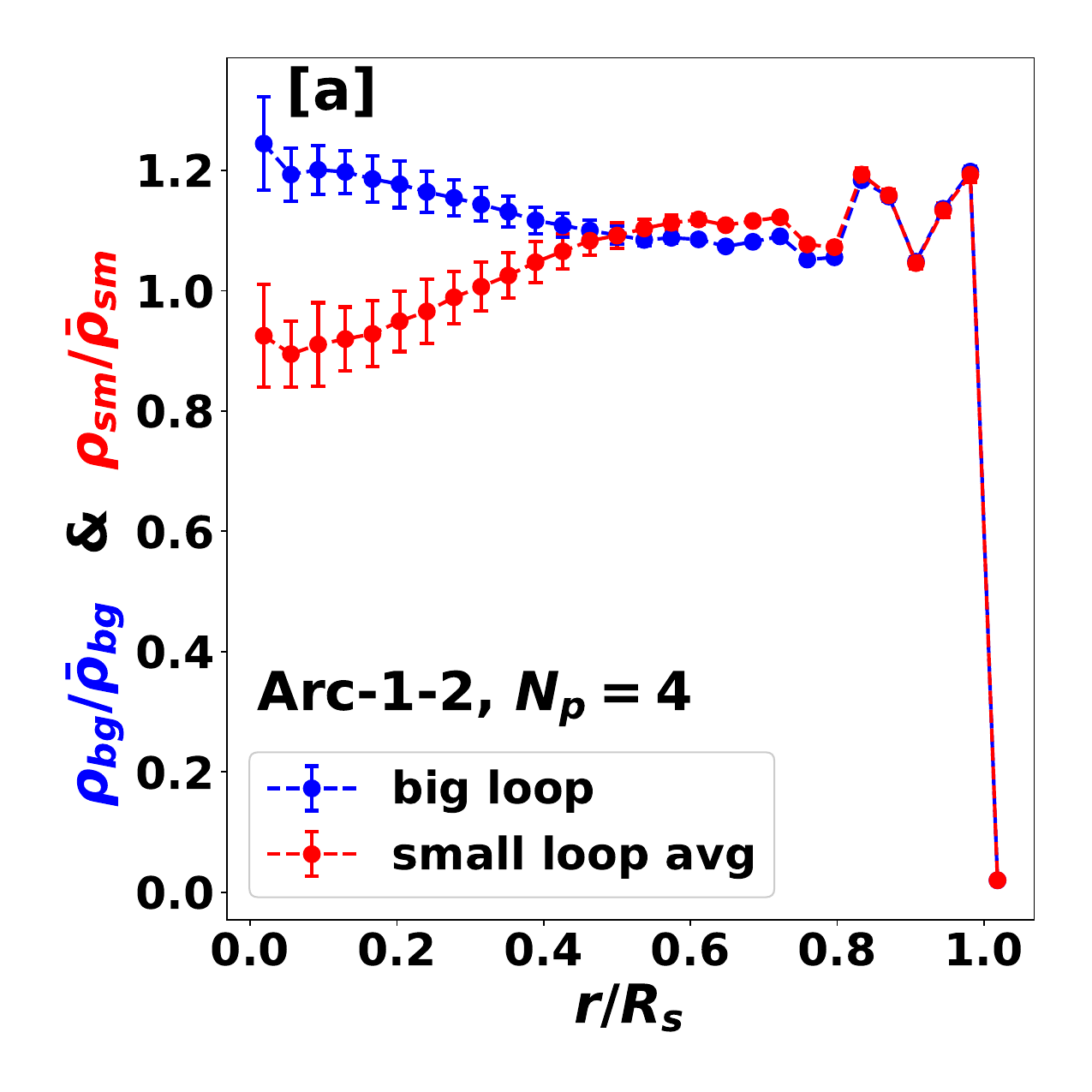}
\includegraphics[width=0.64\columnwidth,angle=0]{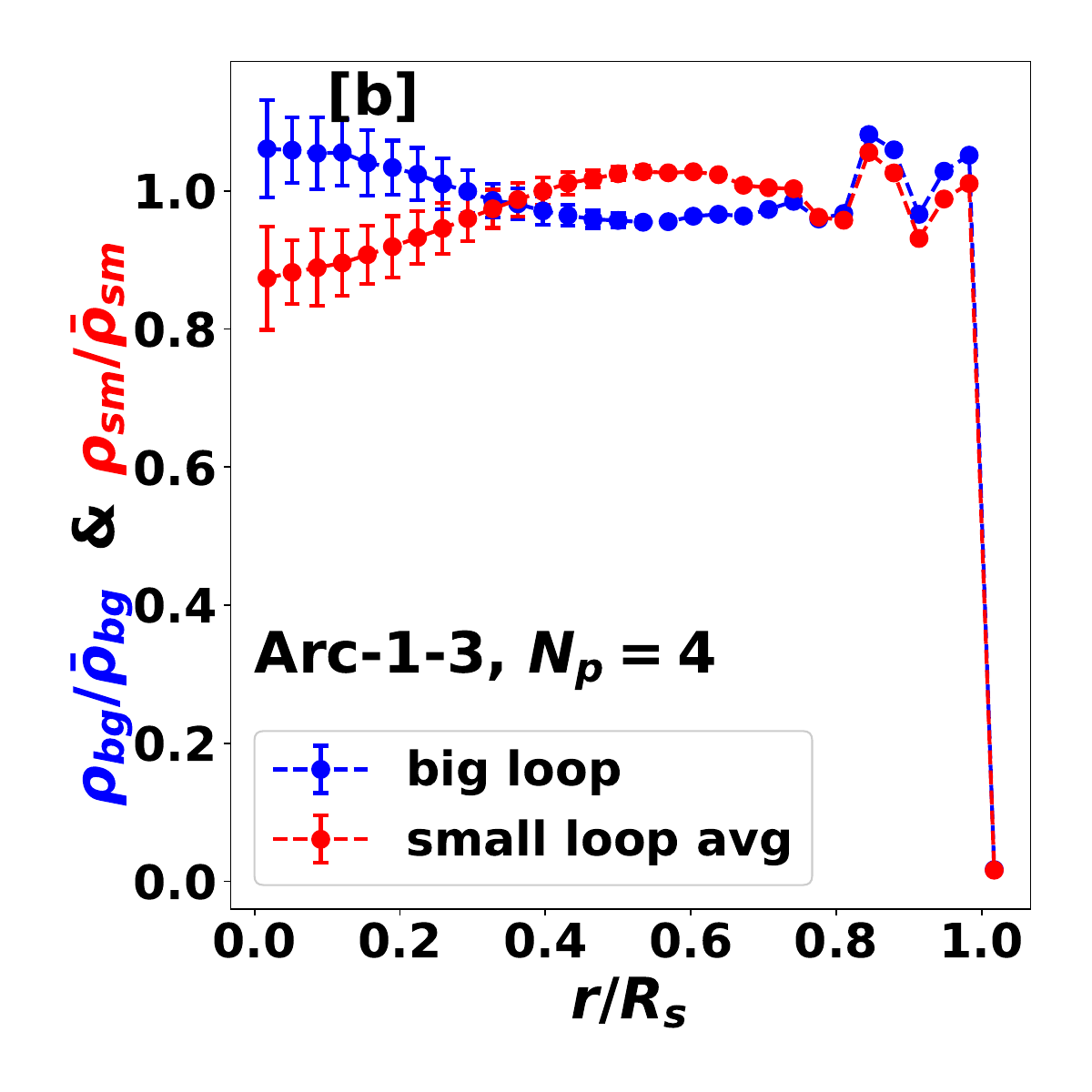}
\includegraphics[width=0.64\columnwidth,angle=0]{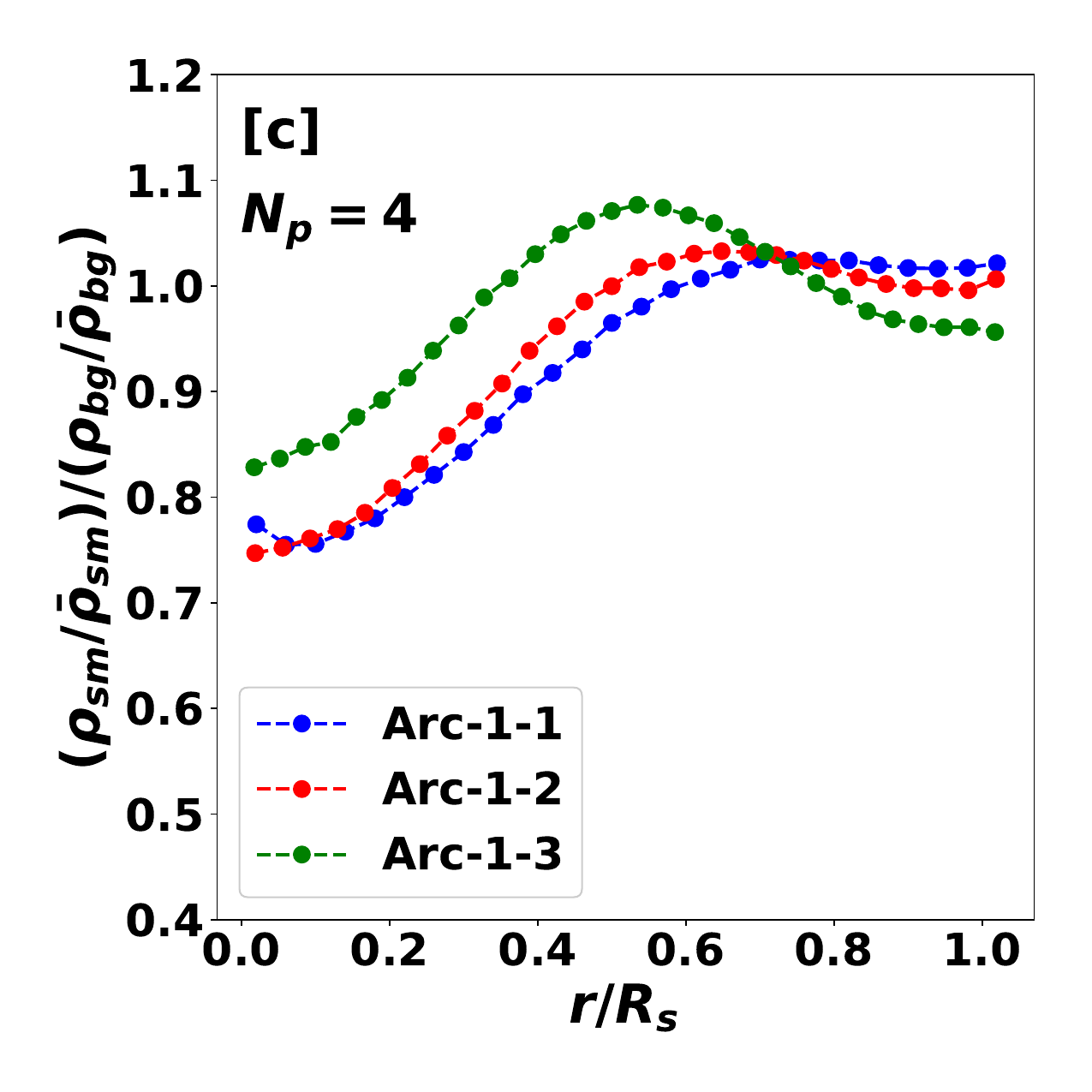}

\includegraphics[width=0.64\columnwidth,angle=0]{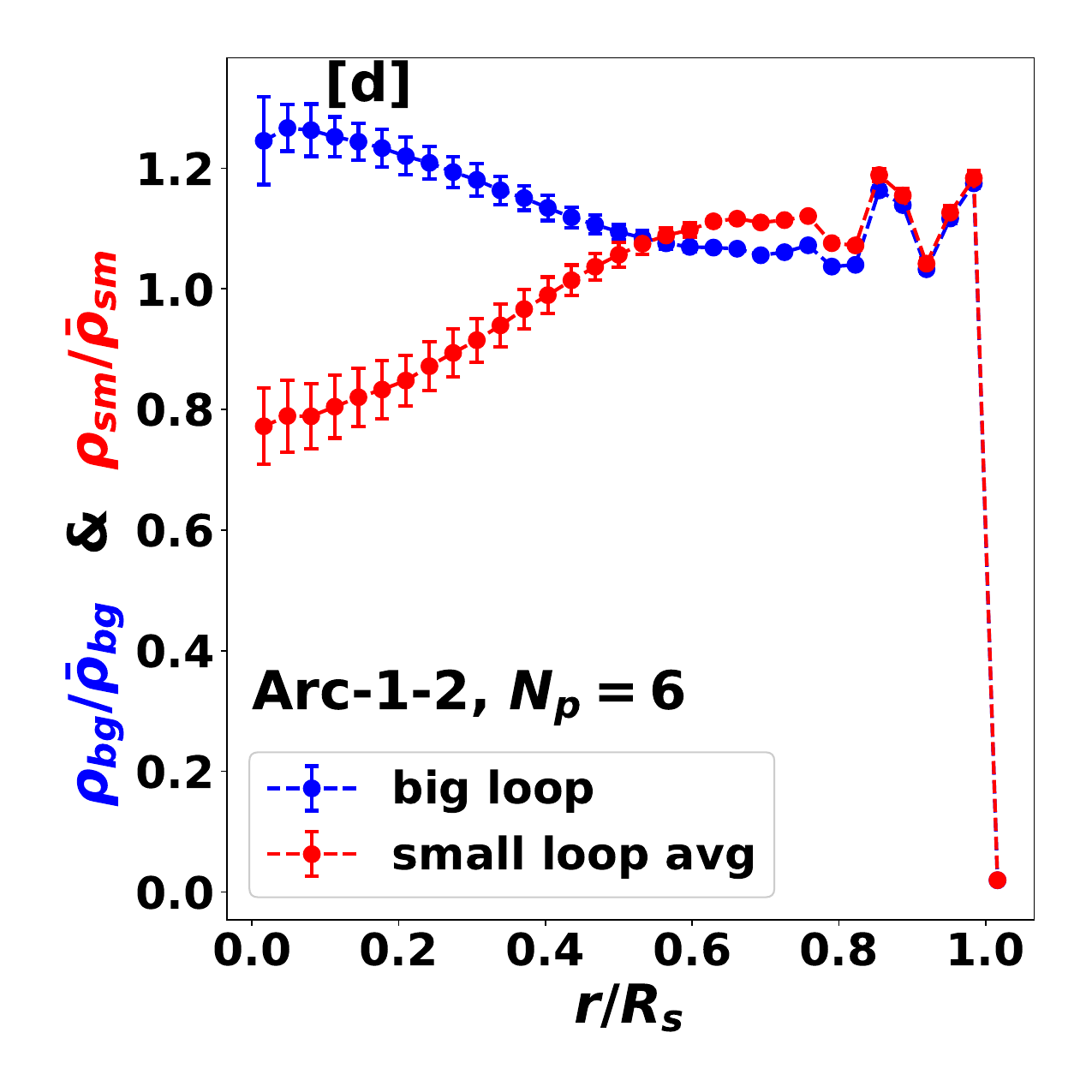}
\includegraphics[width=0.64\columnwidth,angle=0]{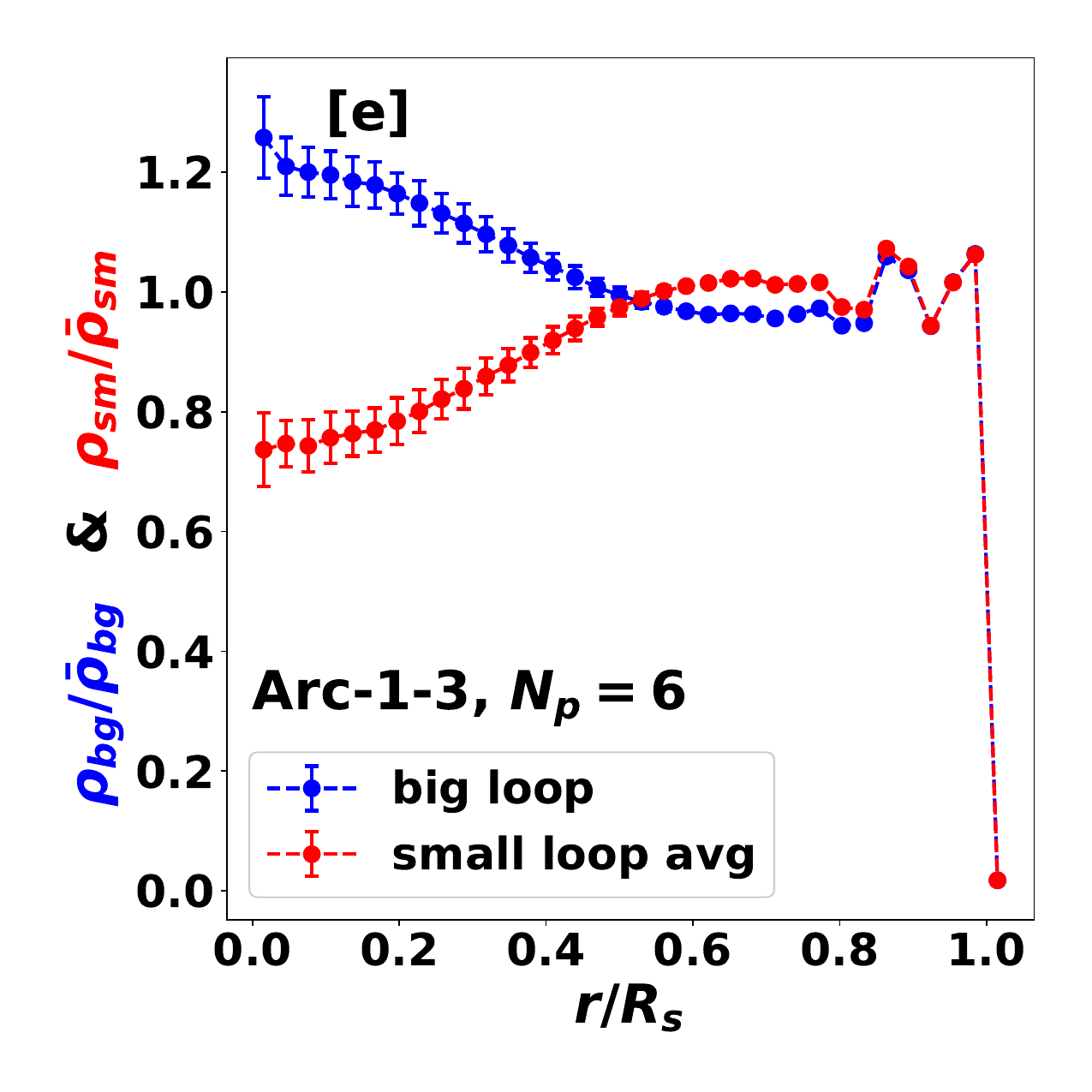}
\includegraphics[width=0.64\columnwidth,angle=0]{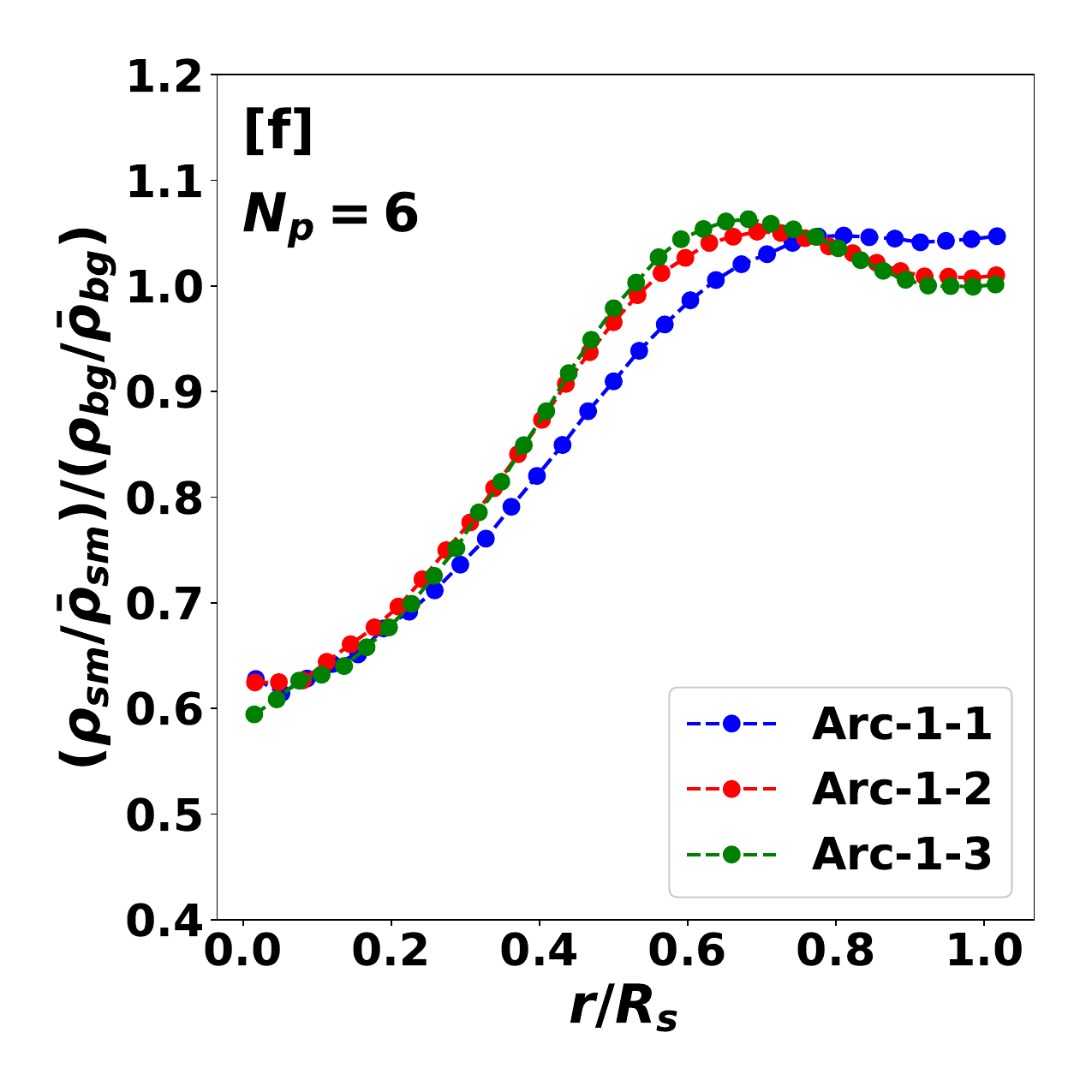}
\caption{\label{fig5}
This figure shows the normalized monomer densities of small and big loops of Arc-1-2 and Arc-1-3 polymers for 
different cases of $N_p$. We also show the ratio of normalized monomer densities  for different architectures, so that 
we can compare localization of smaller loops for different architectures. Subfigures (a) and (b) show the normalized monomer densities of the loops of Arc-1-2 and Arc-1-3, respectively when $N_p = 4$. Both Arc-1-2 and Arc-1-3 have more than one small loop, and therefore the small loop normalized monomer density plotted in the graph is the average of over monomer densities of different loops from different polymers. Subfigure (c) shows the ratio of the normalized monomer density of the small loop($\rho_{sm}/\bar{\rho}_{sm}$) to that of the big loop($\rho_{bg}/\bar{\rho}_{bg}$), for $N_p=4$, in the three different architectures that we consider. Subfigures (d), (e), and (f) show the normalized densities of Arc-1-2, Arc-1-3 and the normalized density ratios for the three architectures, respectively, for the case of $N_p=6$.
}
\end{figure*}

Thereafter we investigate how packing a larger number of polymer chains within the sphere changes the radial
distribution of monomers. Figure \ref{fig3} shows the normalised monomer density for $N_p=$ $4$, $6$, $8$, 
and $10$ Arc1-1 polymers in a sphere of the appropriate radius. 
Any quantity showing the behaviour of the small (or big) loop in these cases is the average 
over loops from different polymers.
Increasing the number of polymers shows a complete reversal in the 
nature of organisation that we found for a single Arc-1-1 in the sphere. In these cases, 
the normalized monomer density of the small loops is higher at the
peripheral  regions of the sphere, and  the big loops are statistically found closer to the 
sphere centre. This means that the bigger loops remain  localised closer to the centre.
For  multiple polymers in the sphere, the sphere diameter is
proportionally larger. In this situation, the region around the sphere center has enough volume to accommodate
the monomers of the bigger loops, allowing them to take up multiple configurations even as they overlap. 
Past studies 
show that rings in contact behave like soft spheres of radius approximately the size 
of $R_g$ of the ring, and energy of overlap is $\approx 5 k_BT$. This is has been shown 
in both dilute as well as semidilute conditions \cite{Narros2010,Bernabei2013,blobtheory,Narros2013}. They 
have found the effective interaction potentials between two ring polymers of differing sizes. 
It has been shown that the amplitudes of the effective interaction potential decreases as size 
of each ring polymer increases. Thus, the effective repulsion between each small loop is expected 
to be more between two smaller more compact rings than between two larger ring polymers. 
Thus, with multiple polymers present in the sphere, it is favourable for the big loops 
to overlap near the center. The small loops tend to avoid each other more and consequently, 
get pushed towards the periphery. The mutual avoidance of smaller loops 
can also be verified later when we provide the contact maps at the end of the manuscript.
The difference in the normalized monomer distribution of big  loops and small loops is maximum 
for $N_p=6$.

Next, we aim to investigate the consequences of increasing the number of smaller loops in the ring polymer 
architecture. We expect  that this will increase the asymmetry  within a single polymer chain. 
To that end, we add two or three smaller loops, having $50$ monomers in each, to a big loop of fixed length of $150$ monomers. That is, we look at the properties of Arc-1-2(150-50) and
Arc-1-3(150-50).  In Fig.\ref{fig4}, we begin by analyzing the CoM distribution of the loops of Arc-1-3 for 
two distinct cases:  (i) when there is  one polymer in the sphere, refer Fig.\ref{fig4}(a), and (ii) when there 
are six polymers in the sphere,  refer Fig.\ref{fig4}(b). For both the cases, 
we see again that the  CoM distributions of the individual small loops (each having $50$ monomers)
lie relatively away from the center of the sphere, as compared to the distribution of CoM of 
the big loop. In contrast, when we consider the  distribution of CoM calculated using 
the monomers of all the three small loops, the distribution almost overlaps with that of the 
distribution of CoM of the big loop.

Thus, we definitively establish our previous understanding  that the relative positions of the CoM
of any section of the polymer is independent of the architecture and solely depends on the size of the polymer 
segments/loops that we consider. Segments  having equal number of monomers have identical CoM distributions.
If we compare the CoM of two segments (or loops) of unequal lengths,  the centre of mass of the segment 
with fewer monomers will be away from the center in the statistical sense as compared to the other. 
This observation is independent  of whether we confine  $N_p=1$ polymer or $N_p>1$ polymers within the sphere.

This also explains why the CoM distributions of the ring polymer 
segments and the Arc-1-1 loops were identical:  the segments of ring and the loops of Arc-1-1
had an equal number of monomers.  This effect occurs because the CoM of the entire polymer 
(or a large segment of the polymer) will always fluctuate
very close to the sphere's centre.  Thereby, if we take separate segments of the polymer, 
the CoM of the larger segment will be closer to the centre of the sphere.

\begin{figure}[!hbt]
\includegraphics[width=0.49\columnwidth,angle=0]{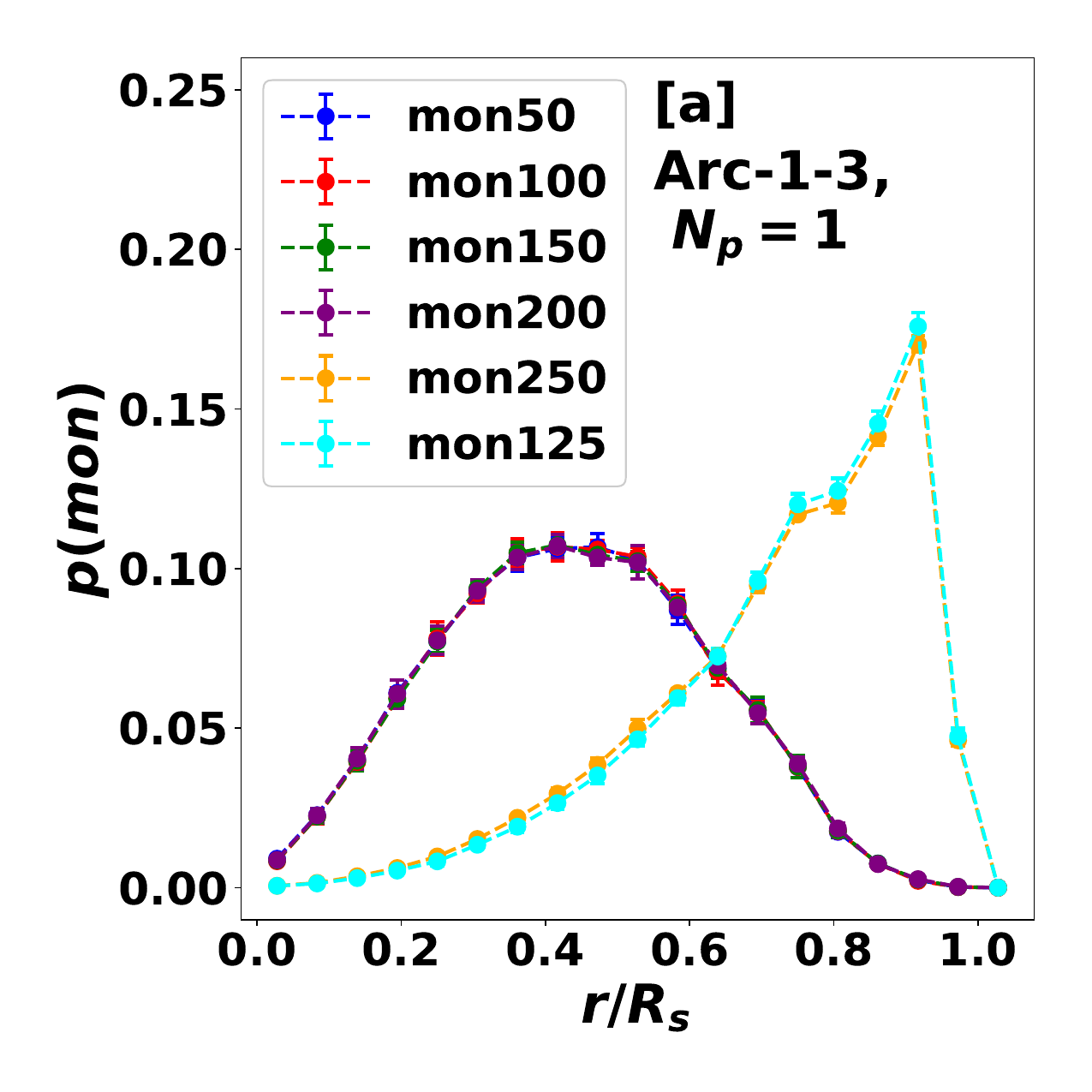}
\includegraphics[width=0.49\columnwidth,angle=0]{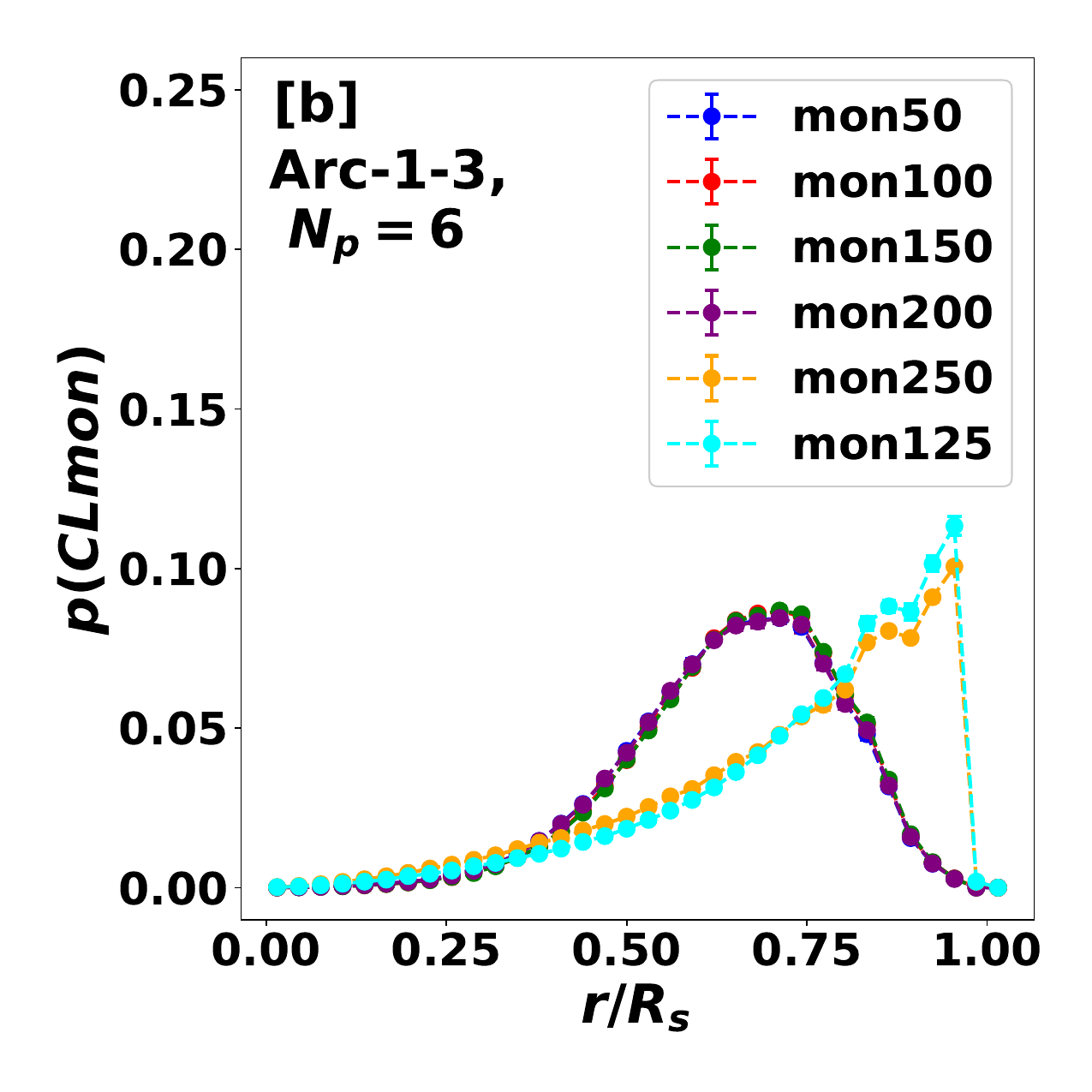} \\
\includegraphics[width=0.49\columnwidth,angle=0]{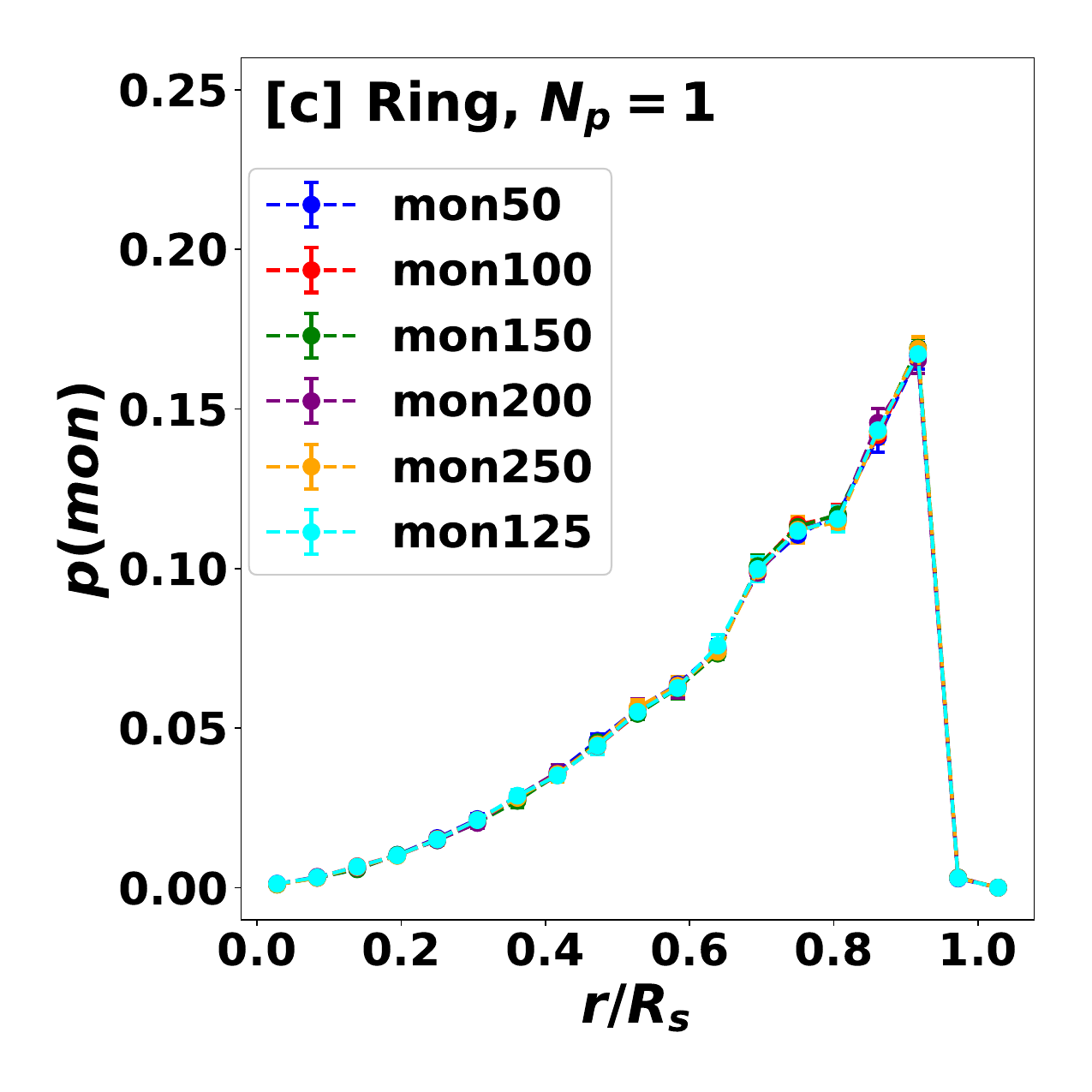}
\includegraphics[width=0.49\columnwidth,angle=0]{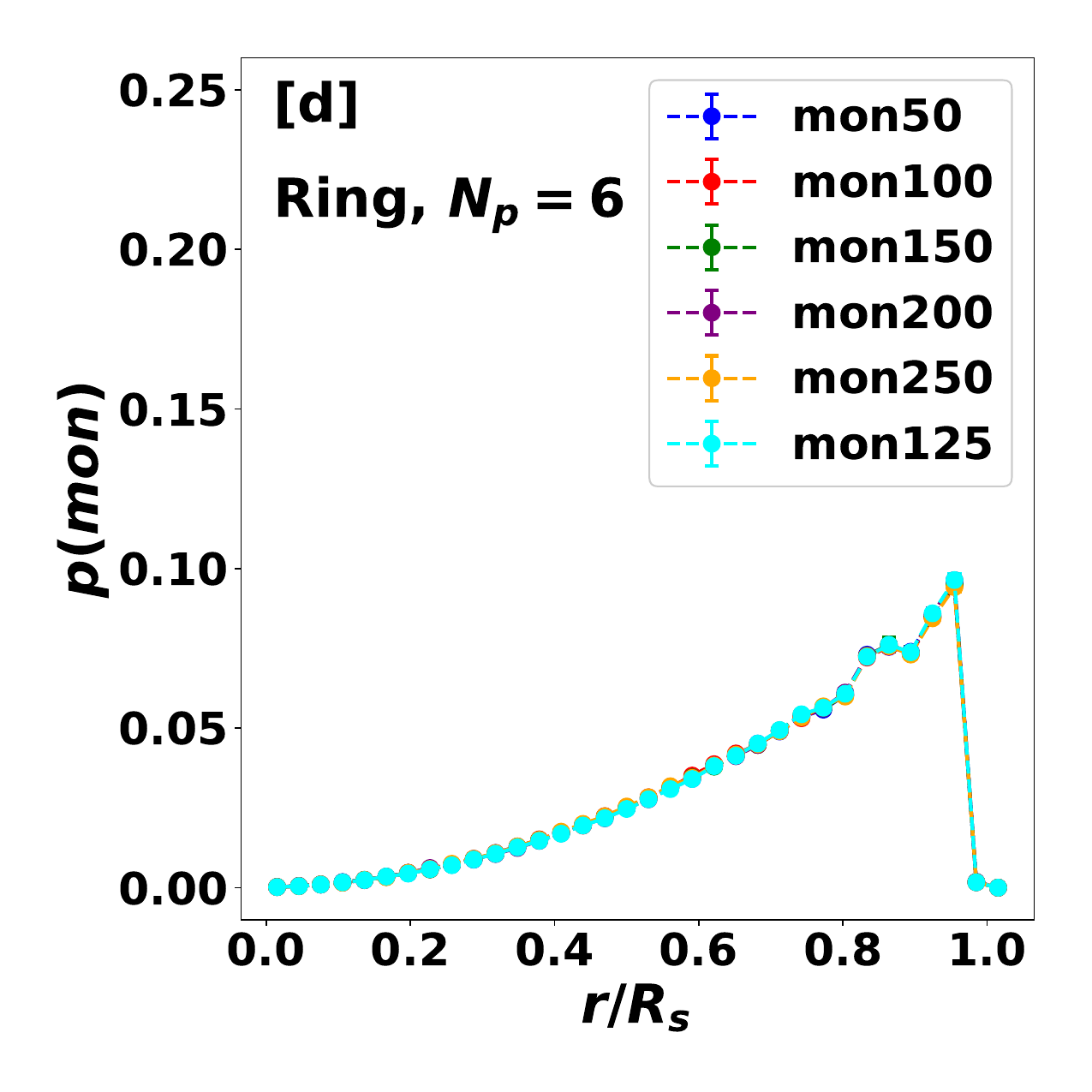}
\caption{\label{fig6}
This figure shows the probability distributions of the position of  particular monomers along the contour of the 
Arc-1-3(150-50) polymers. We also give data for the same monomers when we have just a ring 
polymer(s).  Monomers with IDs $50, 100, 150$, and $200$ are the monomers present at the 
cross-links. Monomers $125$ and $250$ are in the middle of one of the small loops and the big loop along the contour, 
respectively. Subfigures (a) and (b) show the location of the cross-links in contrast to  monomers in the middle of the 
loops, when $N_p=1$ and $N_p =6$, respectively. Subfigures (c) and (d) show the distribution of the same tagged
monomers for ring polymers of equal length. Errors bars calculated from $10$ independent runs 
have been plotted, but can be seen for only  few of the data points.
}
\end{figure}

In the rest of manuscript, we further look only at the normalized monomer densities since they show the 
pattern of organization of the monomers in the sphere.  Figures \ref{fig5} (a),(b),(d),(e) show the individual
normalized monomer density plots of Arc-1-2 and Arc-1-3 for $N_p=4$ and $N_p=6$. We see that, for both 
Arc-1-2  and Arc1-3 polymers, the wall effects dominate at distances very close to the wall of the sphere, 
again leading to peaks in the normalized monomer densities.  Due to these effects, the normalized monomer 
densities  of both loops have nearly identical values near the wall ($r/R_s >0.8$). 
Although we observe that big loops have higher normalized monomer densities near the 
center($r/R_s =$ $0$ to $0.2$) as compared to that of small loops, we should interpret this difference with care. 
This is because the  inner shells have small radii and hence accommodate relatively
fewer monomers. So small changes in the number of monomers would lead to
higher fluctuations of statistical quantities at low $r$. Thus, we focus more on the values of 
normalized monomer densities $\rho_{sm}$ and $\rho_{bg}$ in the range: $r/R_S = 0.2$  to
$r/R_S= 0.8$. The small-loop monomers are  found with higher probabilities at $r/R_s > 0.5$  than 
the monomers of the larger loops. In contrast, the central inner regions of sphere are
primarily occupied by the  monomers of the larger loops. The ratio of the normalized monomer densities 
from the small and big  loops can go as low as $\approx 0.6$ near $r/R_s=0.2$, as can be seen 
in Fig.\ref{fig5}(c) and (f). In these two subfigures,  we compare data from three different  
architectures as it helps us evaluate the relative  difference brought about in the organization 
by addition of smaller loops.

\begin{figure}[!hbt]
\includegraphics[width=0.65\columnwidth,angle=0]{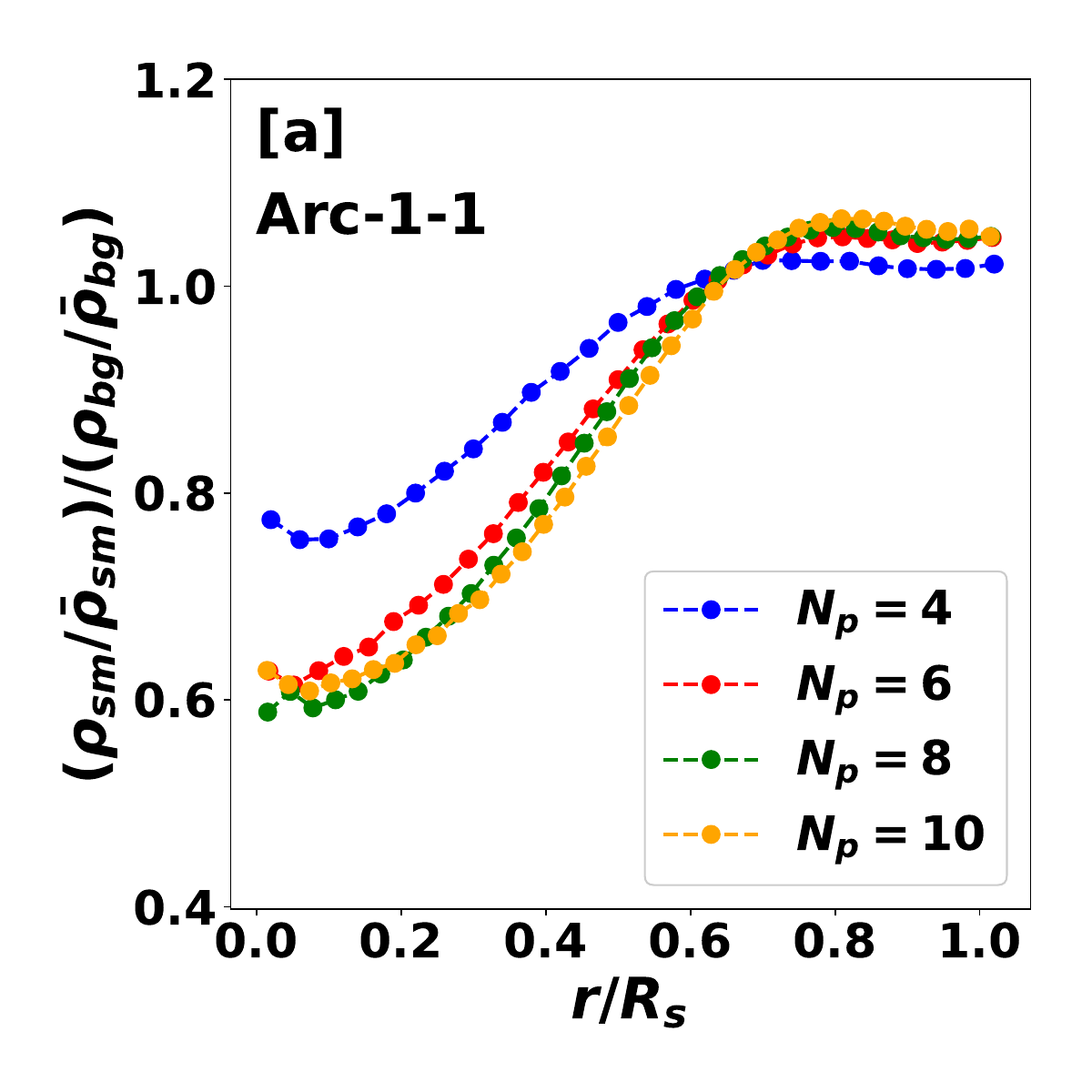}
\includegraphics[width=0.66\columnwidth,angle=0]{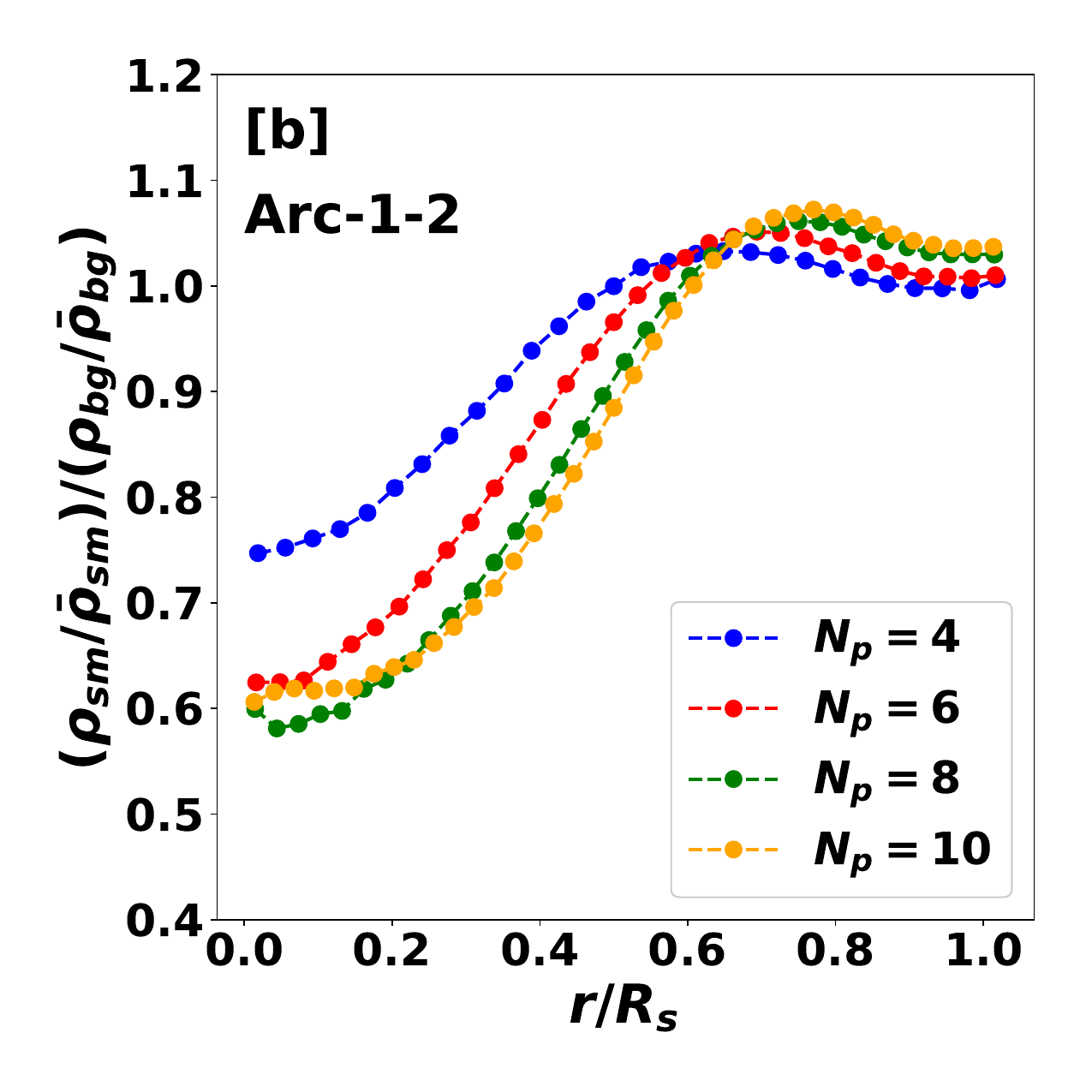}
\includegraphics[width=0.65\columnwidth,angle=0]{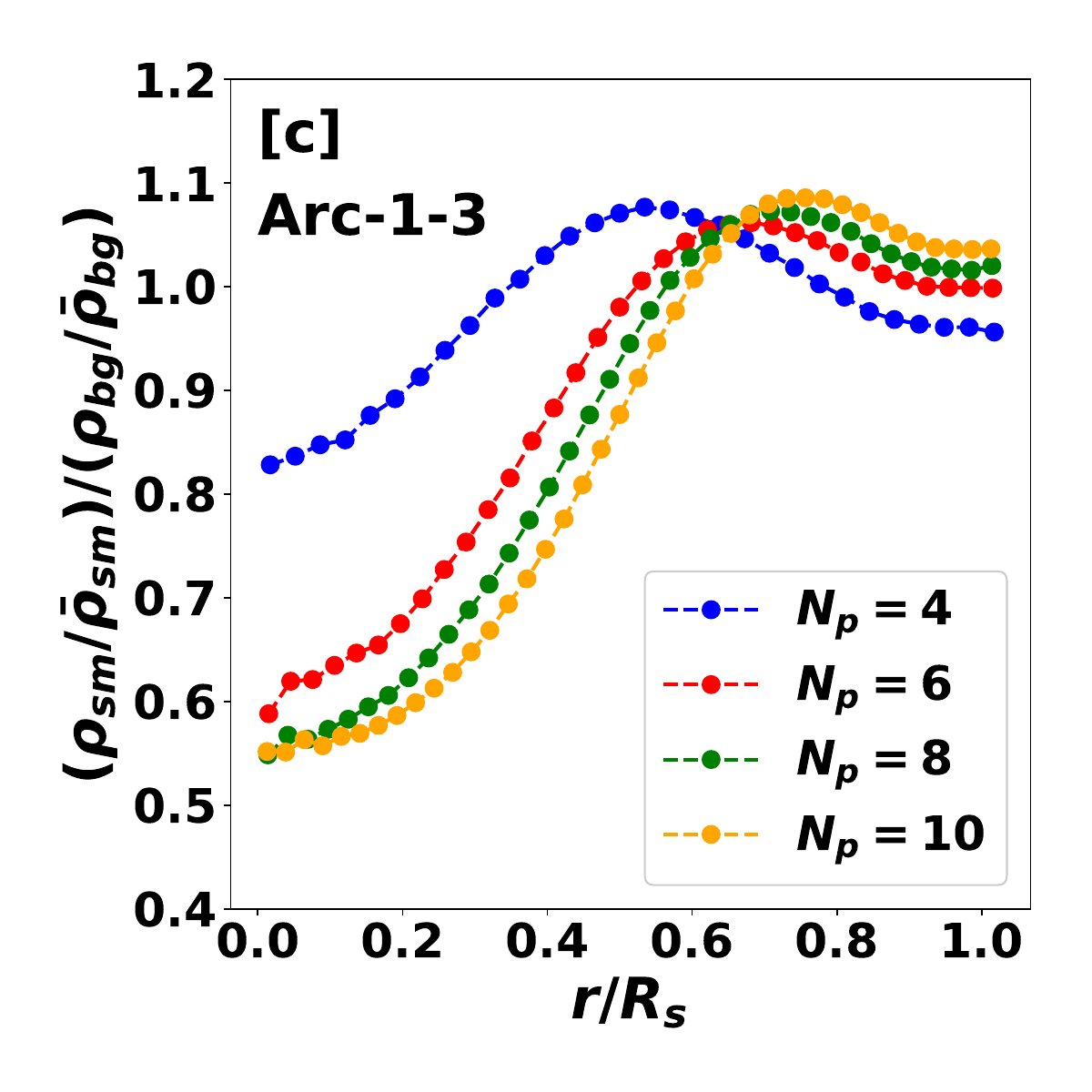}
\caption{\label{fig7}
This figure shows the variation caused in the preferential localization of different loops in the sphere on changing the number of polymers in the sphere. The subfigures show the ratio of the normalized density of the small loop to the normalized density of the big loop for different cases of $N_p$. We confine different numbers of the same kind of polymer and (a), (b), and (c) shows the case when $N_p$ number of Arc-1-1, Arc-1-2, and Arc-1-3 polymers, respectively, are confined within a sphere of appropriate radius.
}
\end{figure}
\begin{figure*}[!hbt]
\includegraphics[width=0.64\columnwidth,angle=0]{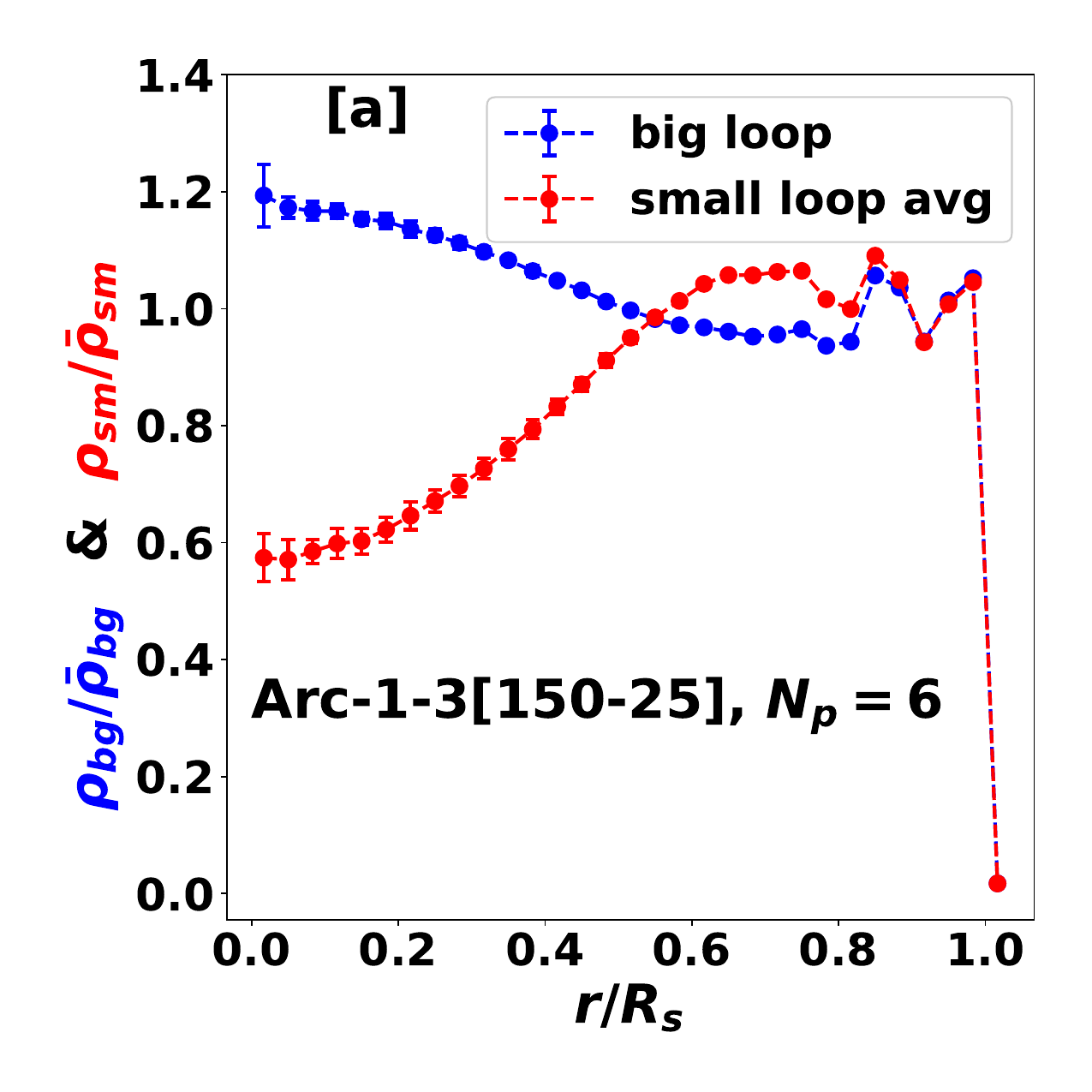}
\includegraphics[width=0.64\columnwidth,angle=0]{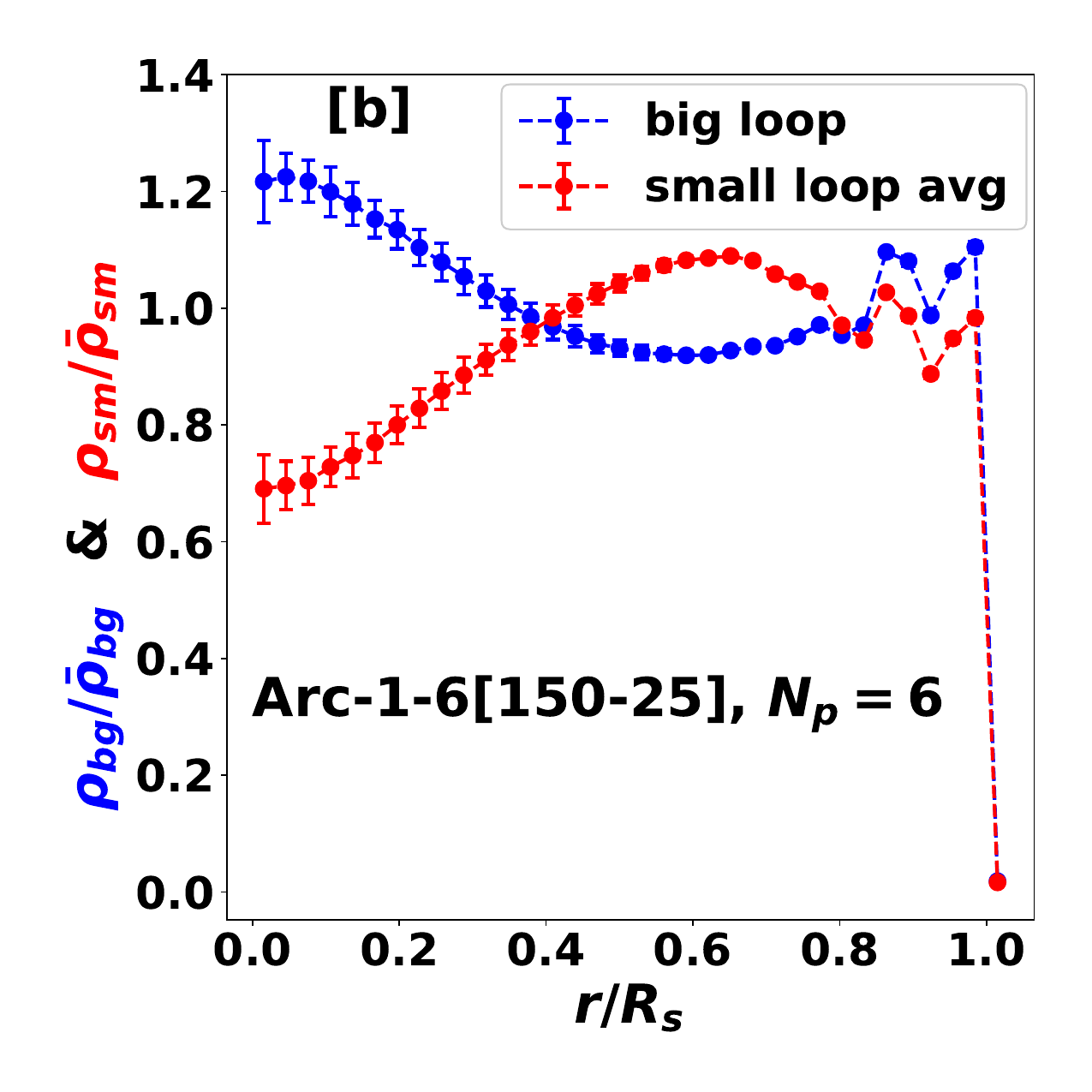}
\includegraphics[width=0.64\columnwidth,angle=0]{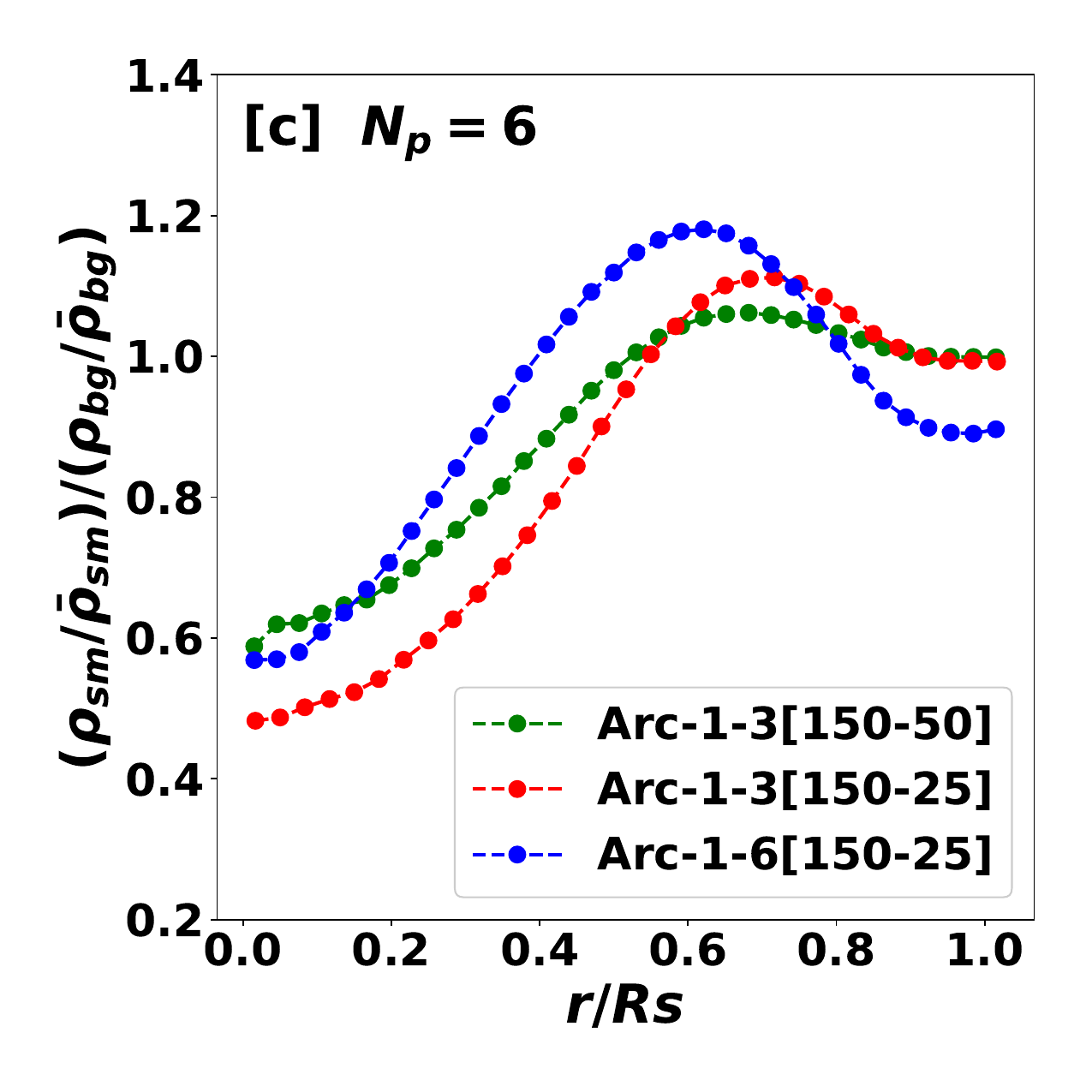}
\caption{\label{fig8}
Subfigure(a)  shows the normalized monomer densities of loops for $N_p=6$ polymers of the  Arc-1-3[150-25] 
architectures with $3$ small loops of $25$ monomers each. This data is compared with that for Arc-1-6[150-25] 
in subfigure (b) to check if the number of the small loops have any effect in the localization of small loops. 
We also compare data of Arc-1-3[150-50] polymers with Arc-1-3[150-25] in subfigure (c) which show the 
ratio of normalized monomer densities for small and big loops.
In subfigure (c), data for Arc-1-6[150-25] is also shown to compare with the other two topologies.
}
\end{figure*}

We can conclude from these plots that within the range: $r/R_S = 0.5$ to $r/R_S= 0.8$, 
the ratio of the monomer densities of the small loop to the monomer densities
of the big loop is $>1$ for all three cases. 
Moreover, the ratio is lowest for Arc-1-1 and takes higher values for Arc-1-2 and Arc-1-3, indicating that adding more smaller loops to the architecture increases the level of organization. Looking further radially outwards near the walls, i.e. between $r/R_s =0.8$ and $ r/R_s=1$ the ratio of small loop to big loop monomer density is 
less than $1$. This ratio is lowest for the 
Arc-1-3 topology. We understand this as follows: the small loops entropically repel each other 
and behave as soft spheres which exclude each other, as well as the wall. Moreover,
the chain segments of larger loops entropically occupy the space between 
the soft-balls of smaller loops and the wall. For Arc-1-3, there is more space between 
the rosettes of smaller loops from different polymers, which is occupied by the monomers 
of the larger loops. The center of the sphere is also preferentially occupied 
by the monomers of larger loops for $N_p \ge  4$  (refer Appendix  for data with $N_p=2$ and $N_p=3$ 
polymers).

 We also plot the distribution of radial 
position of monomers which constitute the cross-links(CLs), and compare it with the position of monomers 
which are away from the CLs. In particular, in Fig.\ref{fig6} we plot the radial position distribution of 
the specific monomers indexed $50$, $100$, $150$ and $200$ monomers, as  these constitute the CLs 
in an Arc-1-3 architecture.  We then  compare the position distrubution of these to the position
distribution  of monomers indexed as $125$ and  $250$, which are monomers  positioned away from the CLs
belong to  the small loop and big loop, respectively.  
From the radial distributions plotted in Fig.\ref{fig6}(a,b), we can conclude that the monomers which constitute 
the CLs stay away from the walls as compared to the other monomers under consideration. 
We present data for $N_p=1$ and $N_p=6$ cases. When $N_p=1$, the CLs are closer to the sphere center.
But when $N_p=6$, the peak of the distribution for  CLs lie in between $r/R_s= 0.6$ and $r/R_s = 0.8$.
In any case, cross links cannot be located too close to the periphery as it is entropically unfavourable 
for the three small loops which are joined at the cross-links. This is because the three 
smaller loops have soft sphere-like behaviour  and shall surround the cross-linked monomers.
Thus, monomers $125$ and $250$, which are away from the cross-links along the chain contour, have higher 
probabilities of occupying the peripheral shells. 
However, note that at $r/R_s \approx 0.9$,  the value of the probability distribution 
for monomer $250$ is less than the value of probability distribution
of monomer $125$ for both $N_P=1$ and $N_p=6$ cases. This appears counter-intuitive apriori 
for $N_p=1$ as monomers of the bigger loops are expected to be at the periphery.
However, the cross-linked monomers are close to the center and the other 
monomers of the smaller loops radially distribute themselves outwards from this central point. The monomers
farthest away from the cross-linked monomers along the contour of the small loops thus are to be found away from the 
center near the peripheral regions. The data for Arc-1-3 can be compared to the control 
system of ring polymers  with $N_p=1$ and $N_p=6$ polymers in a sphere. Here we see that 
all the tagged monomers have identical probability distributions for their positions, 
refer Fig.\ref{fig6}(c,d).

\begin{figure*}[!hbt]
\includegraphics[width=0.5\columnwidth,angle=0]{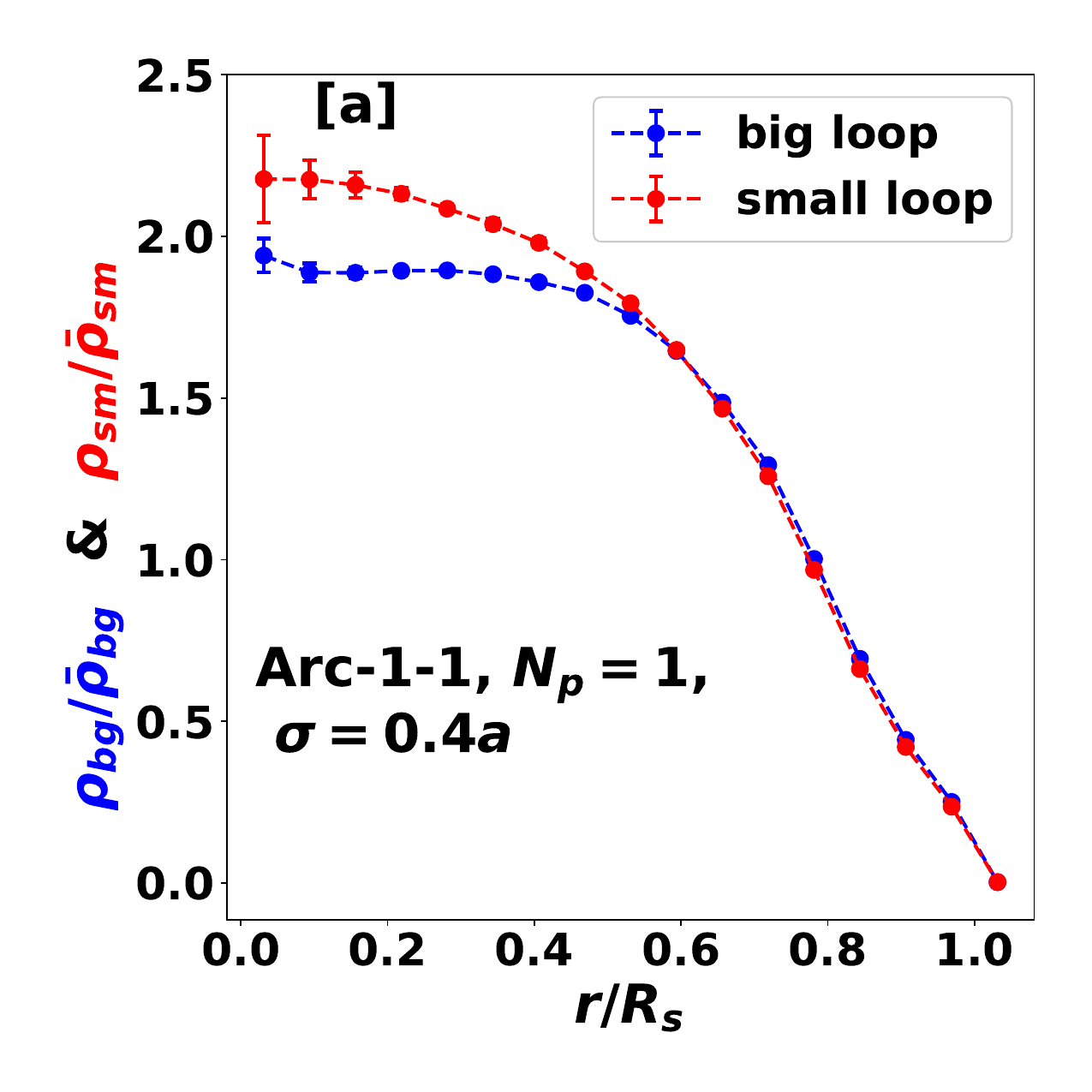}
\includegraphics[width=0.5\columnwidth,angle=0]{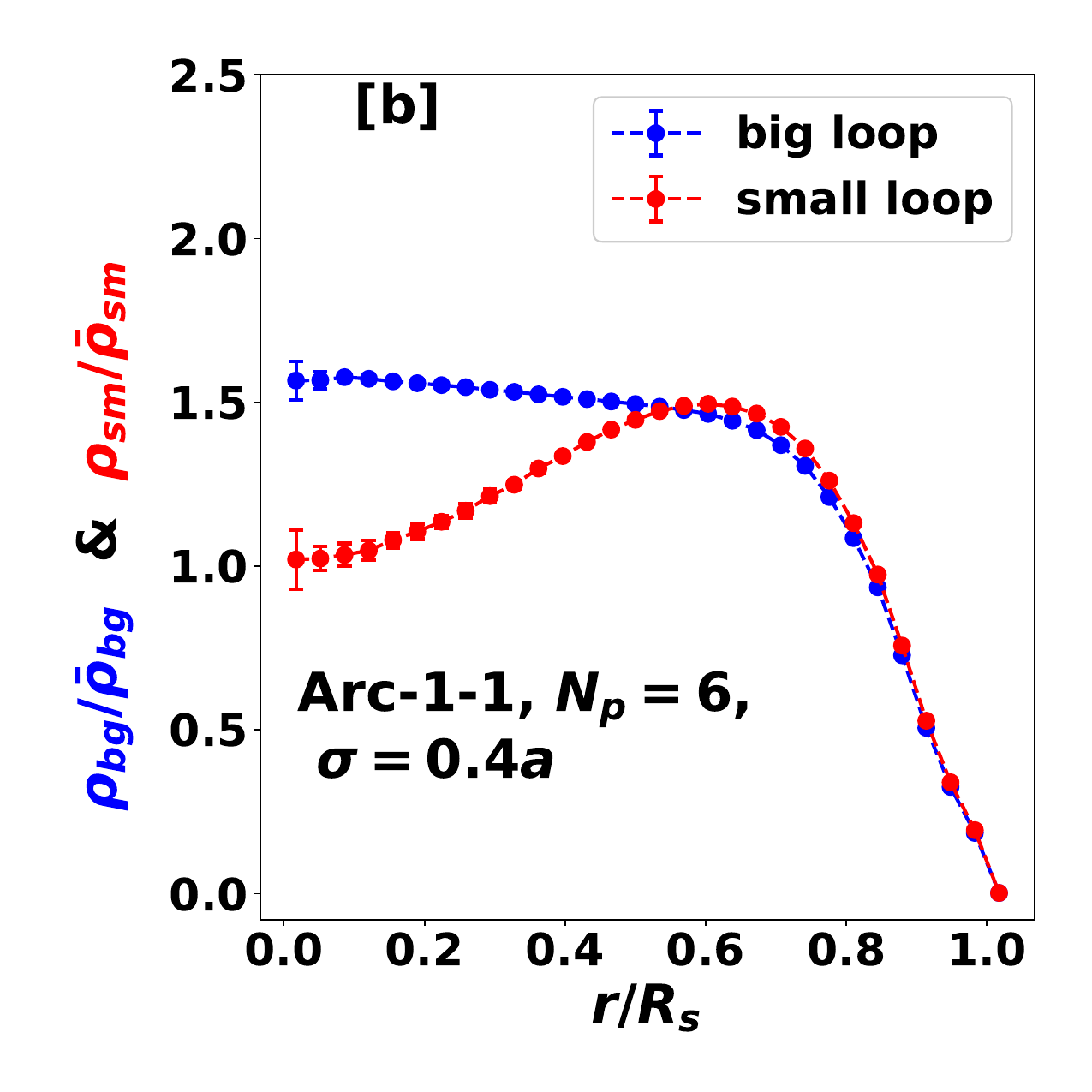}
\includegraphics[width=0.5\columnwidth,angle=0]{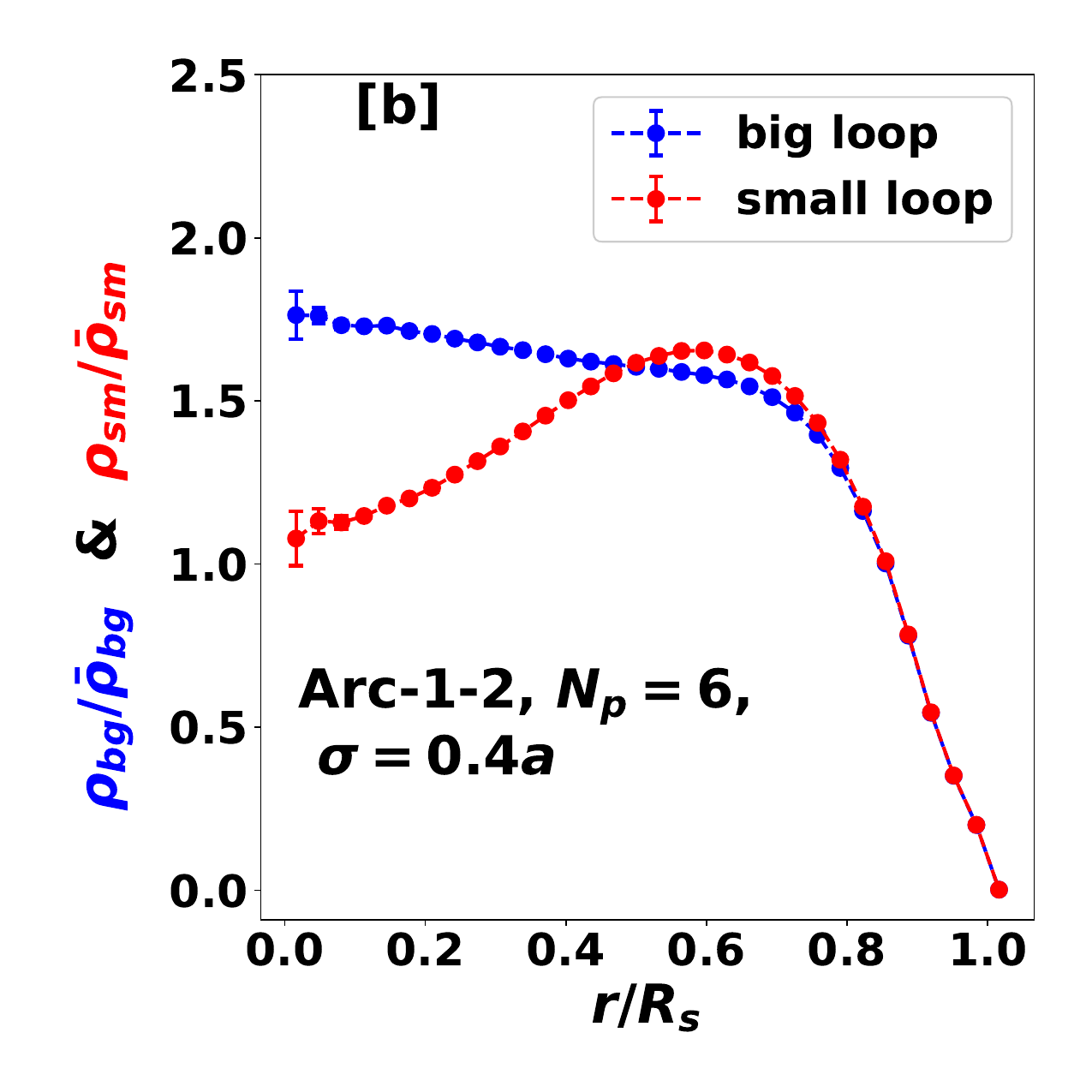}
\includegraphics[width=0.5\columnwidth,angle=0]{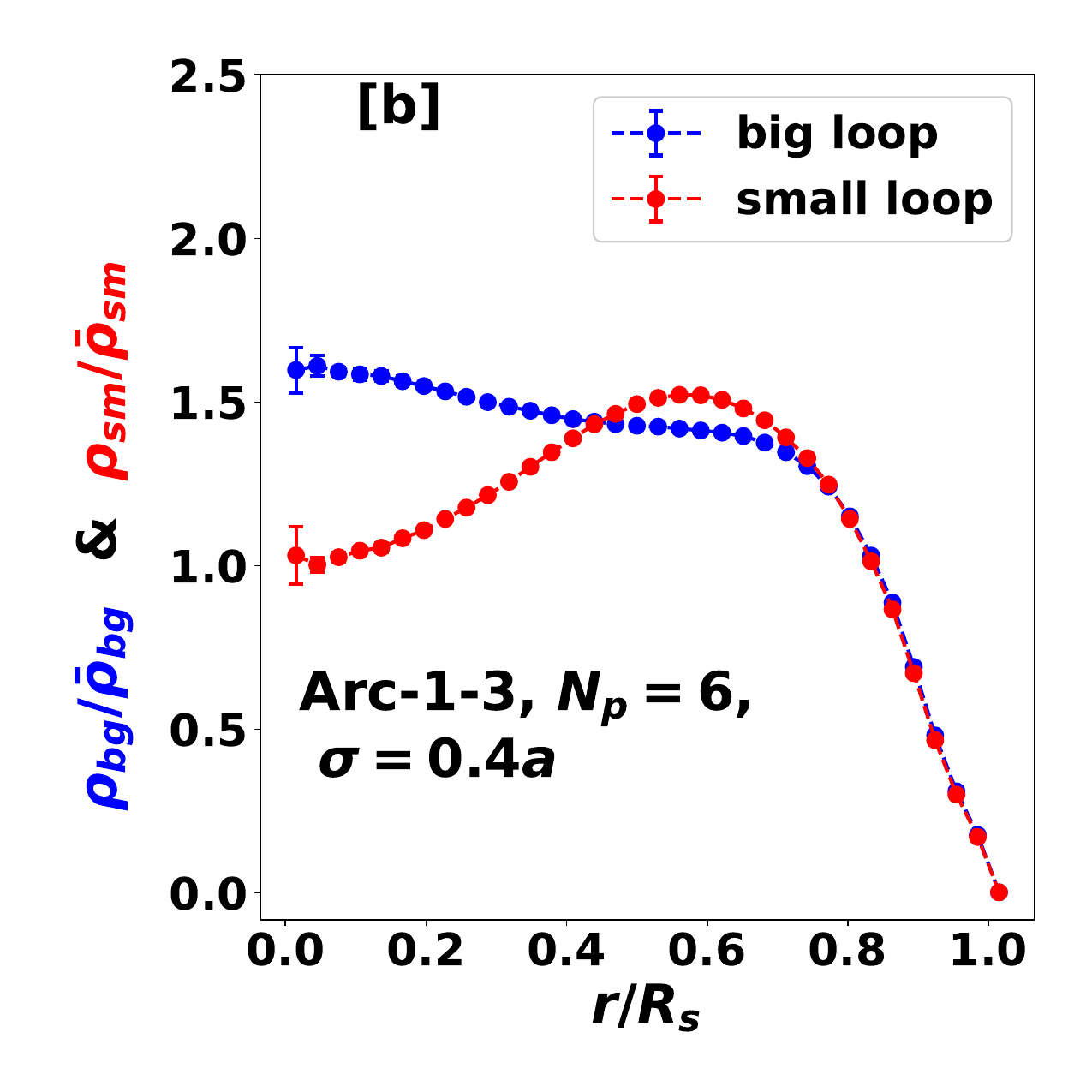}
\caption{\label{fig9}
This figure shows the normalized monomer densities of the small and big loops of different architectures 
in runs where topological constraints have been released by allowing chain-crossing. Subfigure (a) shows
the data for of one Arc-1-1 with $N_p=200$ monomers,  when crossing of chains is permitted by choosing
$\sigma=0.4a$. The data is to be compared with Fig.\ref{fig2}h, which considers the case where chain
crossing is disallowed. Similarly, subfigures (b),(c) and (d) of this figure consider six polymers of 
Arc-1-1, Arc-1-2, Arc-1-3, respectively, and plot the normalized monomer densities of small and big loops 
with chain crossing allowed. These plots are to be compared with  Figs.\ref{fig3}c, \ref{fig5}d and 
\ref{fig5}e, respectively, which plot the same quantities with chain crossing disallowed.
The volumes of the spheres in different cases have been kept the same as given in Table- \ref{Tab:Tcr} even when 
$\sigma$ of the monomers is chosen to be $0.4$. This means that the volume fraction in the cases where
chain crossing is allowed is less than $0.2$.
}
\end{figure*}
 In  Figs.\ref{fig7} (a),(b) and (c), we establish that the localization of smaller loops
 to the peripheral regions of the sphere increases if we increase 
 the number of polymers $N_p$ of Arc-1-1 and Arc-1-2 within the sphere, but not significantly. This is shown by 
 the comparison  of the ratio of  normalized monomer densities  of small loops and  big 
 loops for different values of $N_p$ for  Arc-1-1, Arc-1-2, and Arc-1-3 polymers respectively.
 For Arc-1-3 architecture,  the outlier is the $N_p=4$ case as the ratio
 becomes $> 1$  at $r/R_s \approx 0.4$, but in other cases the ratio crosses $1$  at $r/R_s \approx 0.55$.
The reason for smaller loops being closer to the center is that we use a smaller 
confining sphere for $N_p=4$ as compared to systems with higher $N_p$.
As a consequence the $R_s \approx 7a$ is  comparable to twice the
radius of gyration $R_g \approx 3a$ of the smaller loops. Thereby the data 
shows that the monomer density of smaller loops is closer to the center for $N_p=4$.

\begin{figure}[!hbt]
\centering
\includegraphics[width=0.64\columnwidth,angle=0]{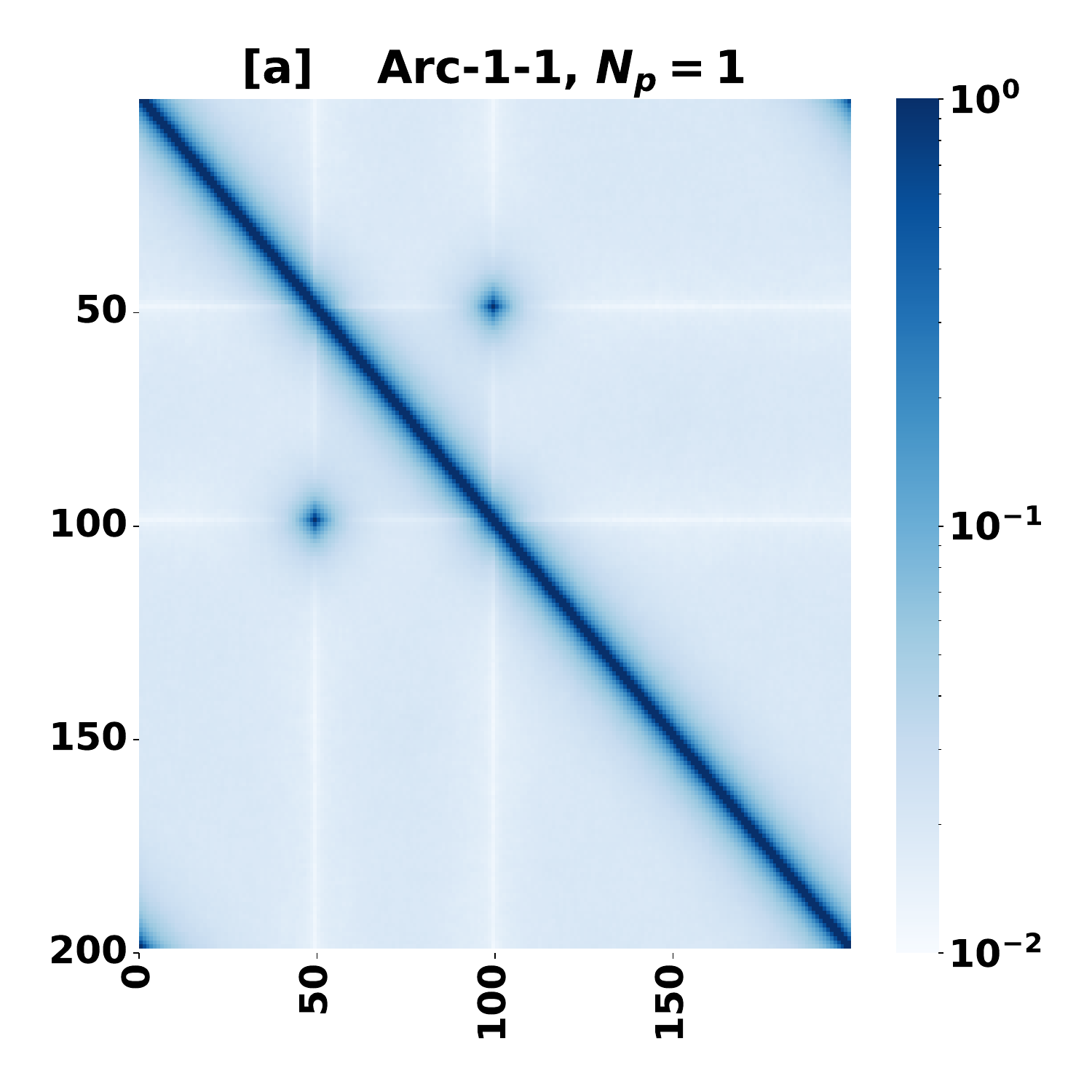}
\includegraphics[width=0.64\columnwidth,angle=0]{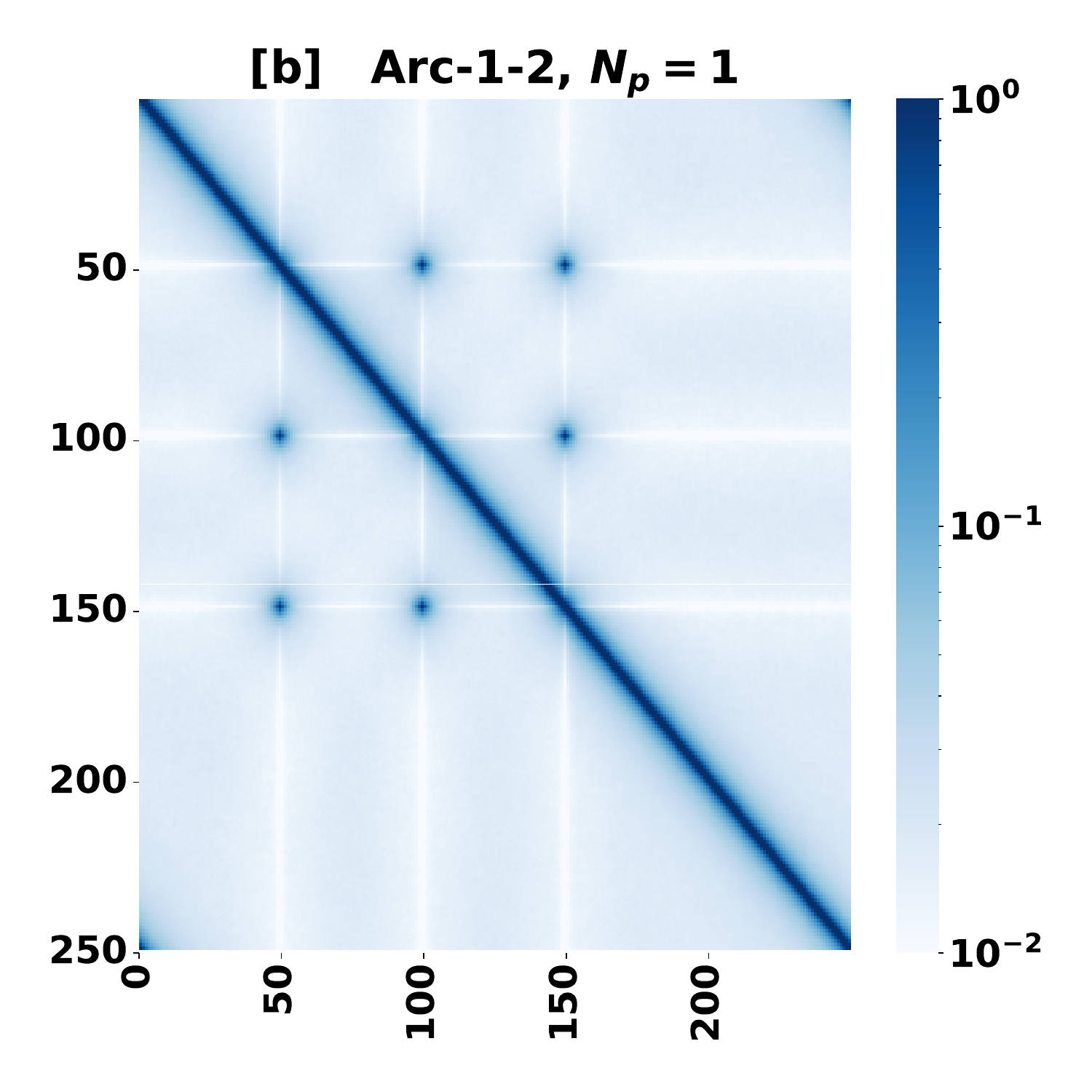}
\includegraphics[width=0.64\columnwidth,angle=0]{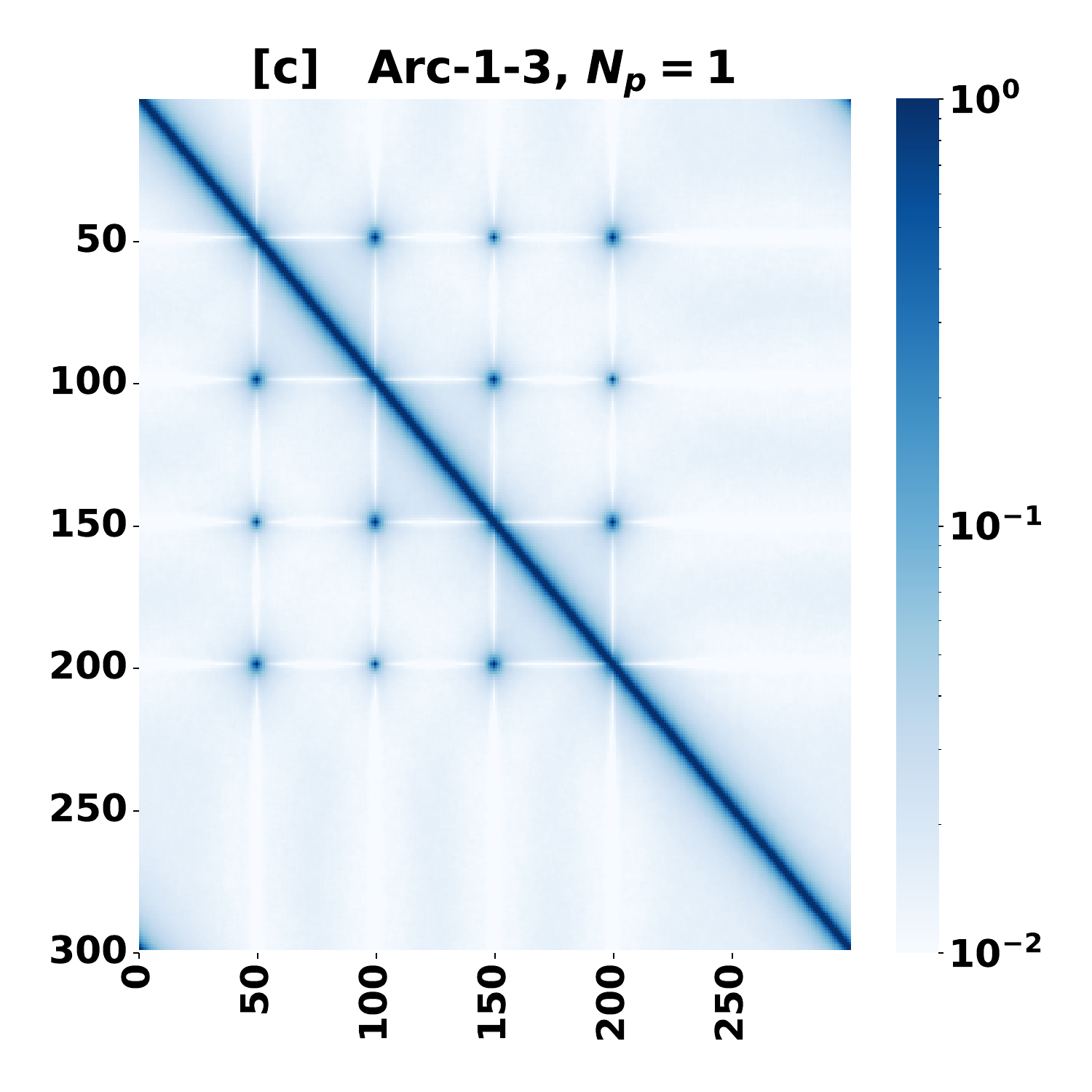}
\caption{\label{fig10}
The above figure shows the contact maps of (a) Arc-1-1  (b) Arc-1-2, and (c) Arc-1-3 polymers,
when only one polymer is confined within the sphere of radius $R_{actual}$, refer Table \ref{Tab:Tcr}.
The color bar is in log-scale and is limited to the range $10^{-3}$ to $10^{-1}$. 
Note that the first pair of CLs link the $51$-st monomer to the $100$-th monomer.
}
\end{figure}

To increase the localization of the smaller loops at the peripheral regions more significantly, we 
must increase the  asymmetry within the polymer further. We do this 
by creating a new architecture of polymer with $6$ small loops of $25$ monomers each. The normalized monomer 
densities of Arc-1-3[150-25]  and Arc-1-6[150-25] are plotted in Figs.\ref{fig8} (a) and (b). The system 
with $6$ loops shows more significant segregation of  small and large loops 
along the radial direction. In Fig.\ref{fig8} (c), we plot the ratios of the normalized monomer
to compare the values of Arc-1-3[150-25] with that of our previously considered Arc-1-3[150-50].
Detailed monomer density plots of Arc-1-3[150-50] have been previously shown in Fig.\ref{fig5}(e). 
We see that when number of small loops is kept the same, the radial organization is affected by the 
size of the small loops. By having smaller loops of $25$ monomers, the level of organization
increases. The data for Arc-1-6[150-25] has also been added  such that we can also compare with
Arc-1-3[150-50]. These two topologies have the the same number of monomer per polymer and are 
thereby confined in the same sized sphere.
\subsection{Polymers with topological constraint release.}
Within the nucleus of a cell, there exists a machinery to release any topological 
constraints that may arise between entangled segments of the chromosomes. The enzyme Topoisomerase releases 
entangled configurations arising in the DNA by cutting single strands, allowing chains to cross, and then joining 
the strands back together. Thus, it is pertinent to cross-check and establish that such release of
topological constraints do not affect the organization of
internal loops that we have found within a sphere.  To that end, we also carry out 
polymer simulations where the mean distance between neighbouring
beads in a polymer chain is maintained at $a$, but the diameter  
of the beads is chosen to be $\sigma =0.4a$. This reduced bead size allows  chains to
cross each other as the excluded volume of each monomer is significantly reduced.

 Figure \ref{fig9} once again shows the radial distribution of suitably normalized monomer densities
 for different polymer topologies in systems where crossing of chains is allowed. In Fig.\ref{fig9}(a), 
 we show data when there is just one chain(Arc-1-1) within   the sphere, before presenting data for
 multiple polymers within a sphere (Figs. \ref{fig9} (b)-(d)). The qualitative features
 of the positioning of loops remain unchanged, though there are quantitative differences between this 
 data and the corresponding data  where chains cannot cross. For example, there is no peak in the 
 normalized monomer density near the periphery of the sphere, in contrast to studies with $\sigma=0.8a$.
 Furthermore, the monomer densities gradually decay to zero near the sphere surface.
 This is due to the reduced excluded volume of the beads, which leads to the volume fraction in this 
 case to be much less than $0.2$. We can infer that if chain crossing is allowed only intermittently 
  such that it occurs with low probability than considered in this  section, we expect loops will be positioned 
  in the same manner as observed for the cases where chain crossing 
 is not allowed.

\section{Contact Maps}
To find out more information about the contact probabilities between polymer segments as we modify 
the polymer topology inside the sphere, we plot a contact map.  If there are $N$ total monomers in the system, a contact map of the system will
be a $N \times N$ matrix mapping the probability of close contacts of each monomer with every other monomer. 
The contact map strategy is inspired by Hi-C Map experiments that reveal close 
contact information in chromosomes \cite{Chahar2023}.

To calculate the contact map, for each micro-state we calculate the distance from of a  monomer to all the 
other monomers in the system. If the distance between the $i$-th monomer and the $j$-th monomer is  
less than a selected threshold distance $r_T$, then we consider those two monomers to be in close contact.
Thus $i-j$ th index show the probability that two monomers are in contact with each other. 
The probabilities of contact is then shown as a heat-map for all pairs of 
monomers. Our choice of threshold distance is $r_T=2\sigma = 1.6a$. Thus, we only highlight the frequencies
of very close contacts, and hence there there can be pairs which rarely come in contact with each 
other and can have probabilities $ \ll 1$. Thereby, the scale of the color bar is in log-scale
for Fig.\ref{fig10};  these figures  shows the contact  maps of a single Arc-1-1, 
Arc-1-2, Arc-1-3 polymers confined in a sphere. 

In  the figures, the main diagonal  is dark in color as they show  contacts of 
neighboring monomers along the chain contour. The off-diagonal dark spots correspond to the monomers which are 
cross-linked and  are at a (mean) distance of $a$. Because of the proximity of cross-linked 
monomers with each other, there arises steric hindrance between the pair and other monomers of the polymer. 
As a consequence, we also see significantly less contact between 
these pair of  monomers and the other monomers: these show up as horizontal and vertical 
white lines  in the contact map. As previous studies with unconfined polymers  
have also shown,  the monomers of a loop have a higher contact probability 
with each other as compared probability of  contact with other monomers. These intra-loop 
high contacts show up as darker squares along the diagonal. Subfigures Fig.\ref{fig10}a,b,c
have one, two, three dark squares, respectively corresponding to the number of smaller loops. 
\begin{figure*}[!hbt]
\centering
\includegraphics[width=0.48\columnwidth,angle=0]{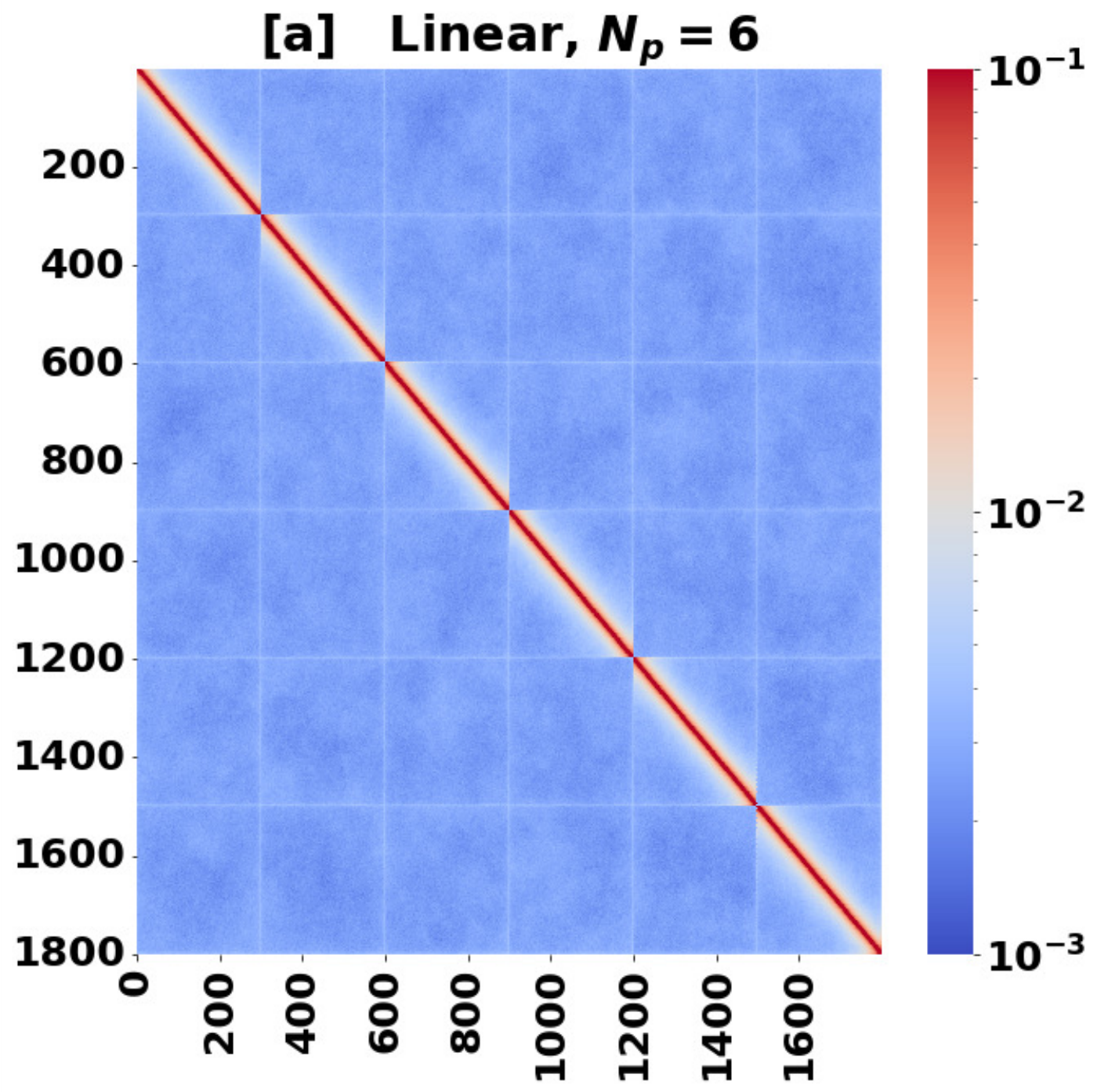}
\includegraphics[width=0.48\columnwidth,angle=0]{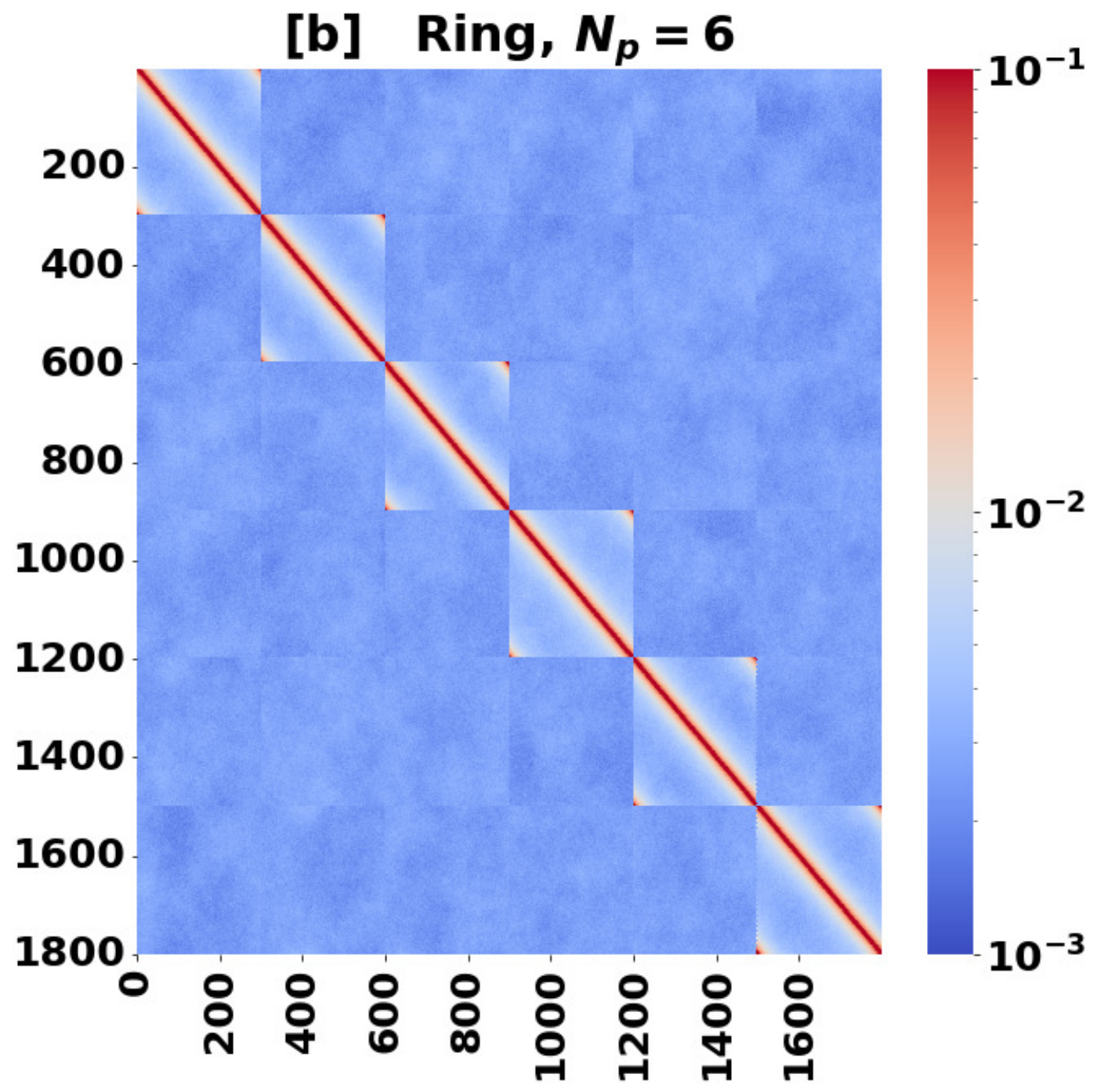}
\includegraphics[width=0.48\columnwidth,angle=0]{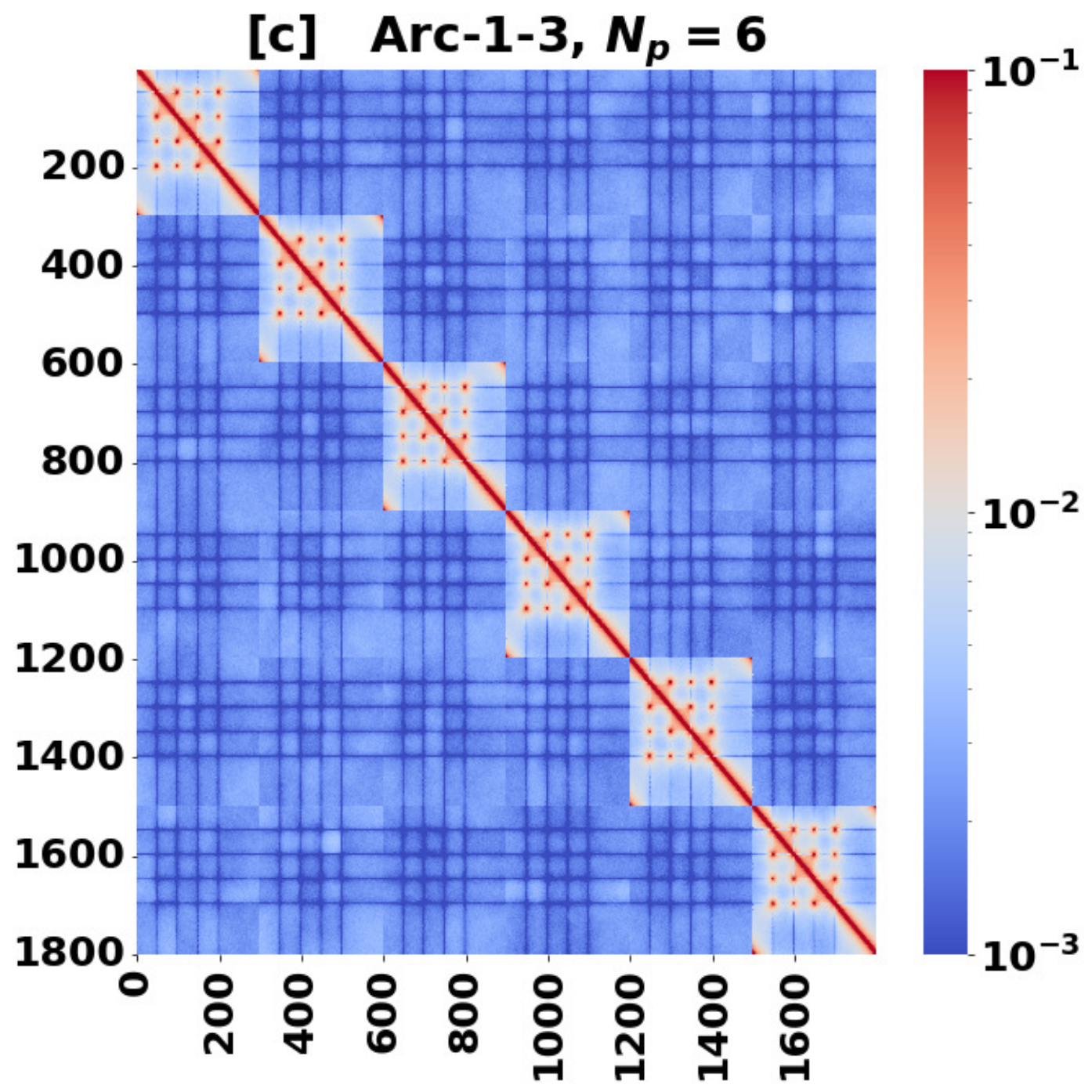}
\includegraphics[width=0.48\columnwidth,angle=0]{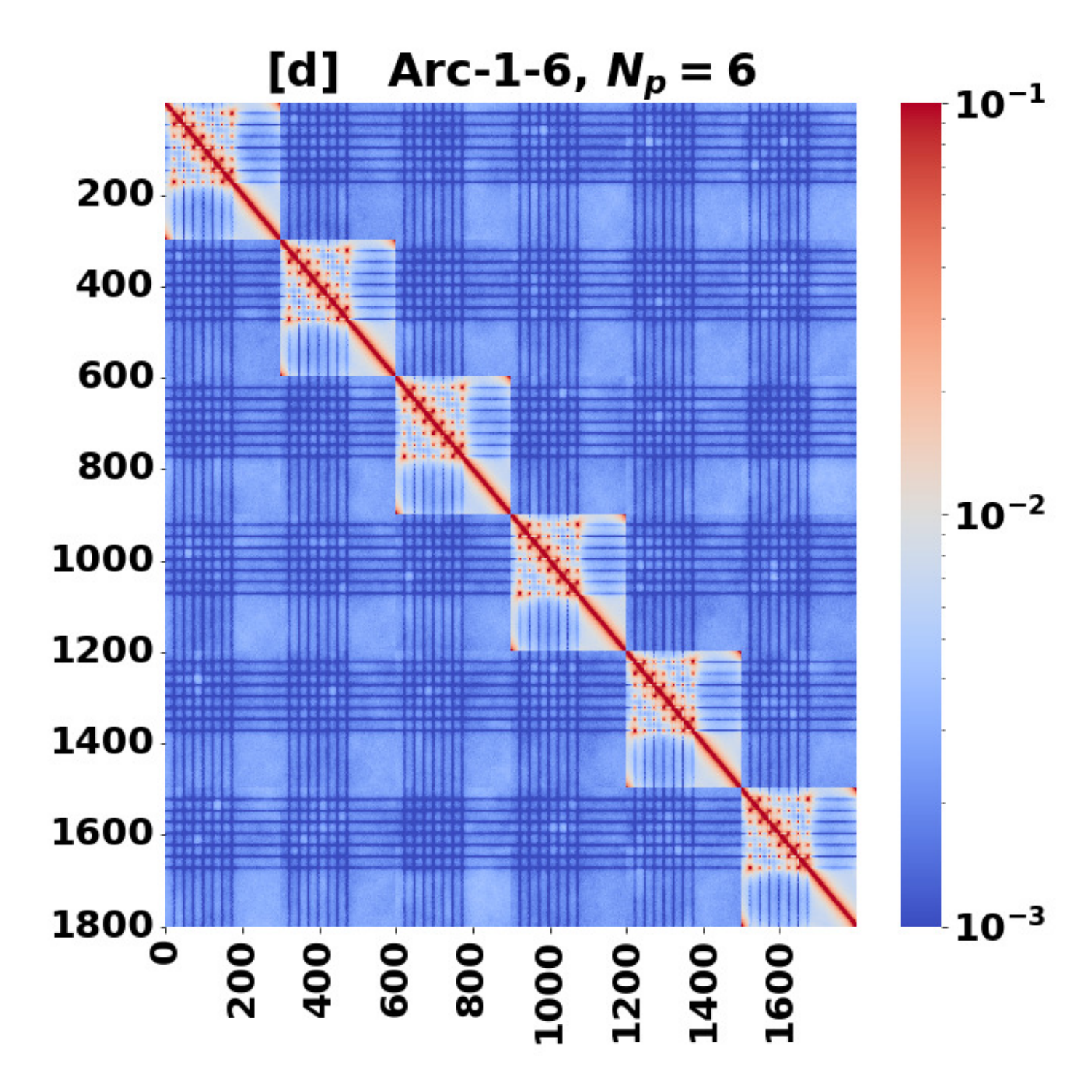}
\caption{\label{fig11}
In this figure, the subfigures (a), (b), (c), and (d) respectively show the contact maps of linear polymers, ring polymers, Arc-1-3[150-50], and Arc-1-6[150-25] polymers confined in the sphere. Each case has $N_p=6$ polymers in the sphere. The Arc-1-3 and Arc-1-6 polymers show distinct polymer demixing, whereas the linear and ring polymers are more mixed.
The color bar is in log-scale, and has been set to $10^{-3}$ to $10^{-5}$.
The color scale has been chosen such that red represents high contact(warm), blue represents low contact(cool), and white represents intermediate contact.
}
\end{figure*}

Next, we would like to investigate how inter-polymer contacts get modified as a consequence of 
topological modifications when there are multiple polymers in the sphere. Figure \ref{fig11} 
shows the contact maps for different topologies of polymers, ensemble averaged over $10$ 
independent runs, where each run is $10^6 \tau$. The same protocol was followed to obtain data shown in 
Fig.\ref{fig10}. We consider six polymers of the same 
topology and length,  confined within a sphere of the same radius.  
We consider four different topologies: linear polymers, ring polymers, Arc-1-3, and Arc-1-6,
and for every case there are $300$ monomers in a  single polymer. 

As expected, the diagonal is in a dark red color corresponding to very high contact probability
in  all the subfigures of Fig.\ref{fig11}. In figure \ref{fig11}a, the relatively uniform color in the off-diagonal
squares of the contact maps of the linear  polymers shows that they are mixed. 
Differences if any, cannot be identified especially since 
the color bar is on logarithmic scale. The squares on the diagonal are slightly lighter-blue in colour. 
This means that monomers from the same polymer
have a higher probability to be in contact with each other as compared to that of monomers 
from other polymers. The free ends
of the linear polymers shown up as horizontal and vertical white lines in the contact map.
In ring polymers, shown in Fig.\ref{fig11}(b), diagonal squares are lighter, and show more self-contact
within ring polymers than in linear polymers. The off-diagonal squares are almost of the same shade.
In Fig.\ref{fig11}(c), showing data for Arc-1-3, squares 
showing inter-polymer contact are deeper blue than squares showing intra-polymer contact. However, there is some
non-uniformity that appears in the off-diagonal squares as well indicating that not all polymers(or loops of each polymer) are equally overlapped.
In Arc-1-6(Fig.\ref{fig11}(d)), the difference in color in the off-diagonal squares becomes more distinguishable. Moreover, we can distinguish that big loop-big loop contacts are more probable than small loop-small loop contact, further
supporting our assumption that topology radially organizes big loops at the centre while small loops are pushed towards the periphery.
Thus, on adding loops to the architectures, 
the polymers confined in the sphere are going from a mixed state to a relatively more segregated, 
or demixed state. Further systematic studies of the demixing of polymers with loops are 
currently underway and will be reported in a future publication.



\section{Discussions}
In this work, we have shown that the segments of a ring polymer can be entropically arranged 
(in a statistical sense) within a sphere by suitably modifying the internal topology. 
Topological modifications by introduction of internal loops of different sizes can be used to  
position the differently sized loops at different distances from the center of the confining sphere. The primary 
underlying mechanism of this phenomenon is that the loops of different sizes behave like soft repulsive spheres.
The smaller loops are pushed to the outer peripheral shell of the sphere, and the inner volume is shared by the monomers
of the big loops,  where they can share space and explore multiple configurations to maximize entropy, or equivalently 
minimize free energy. Internal energy contributions due to WCA interactions contribute minimally to the free
energy; refer Appendix. However, when we have one or two polymers in the sphere, the free energy minimization is 
achieved by monomers of the big loops being preferentially positioned towards the periphery. 

\begin{figure*}[!ht]
\centering
\includegraphics[width=0.5\columnwidth,angle=0]{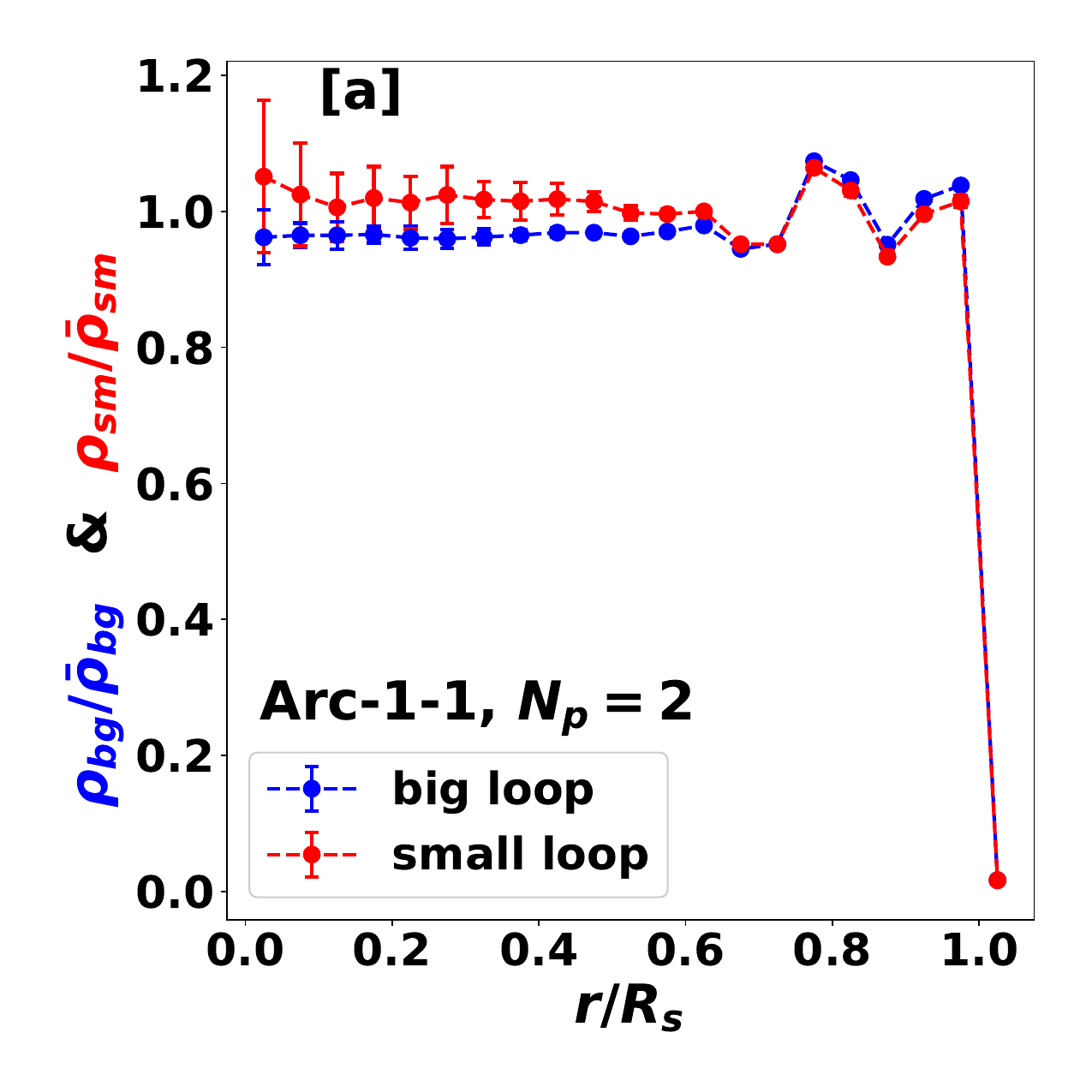}
\includegraphics[width=0.5\columnwidth,angle=0]{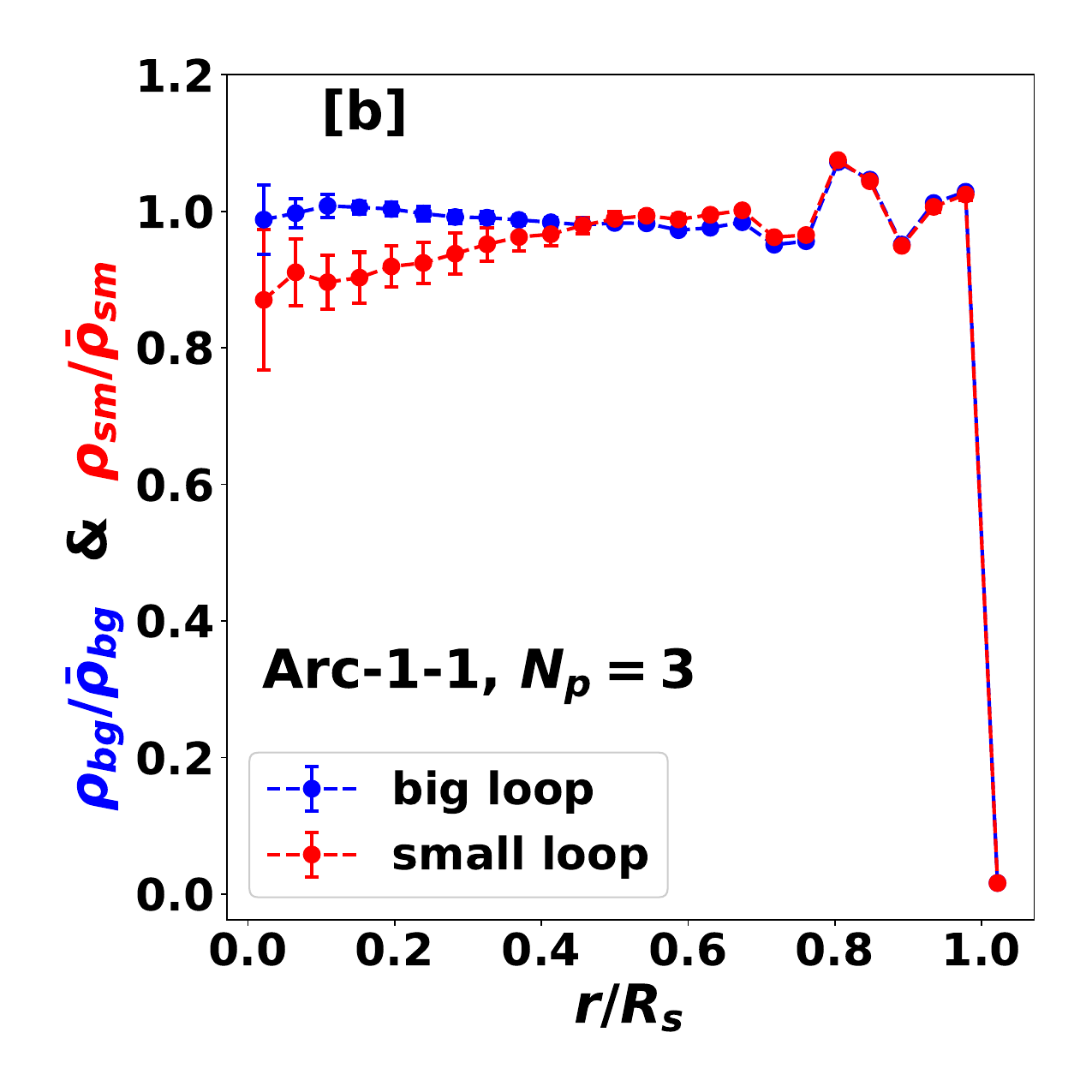}
\includegraphics[width=0.5\columnwidth,angle=0]{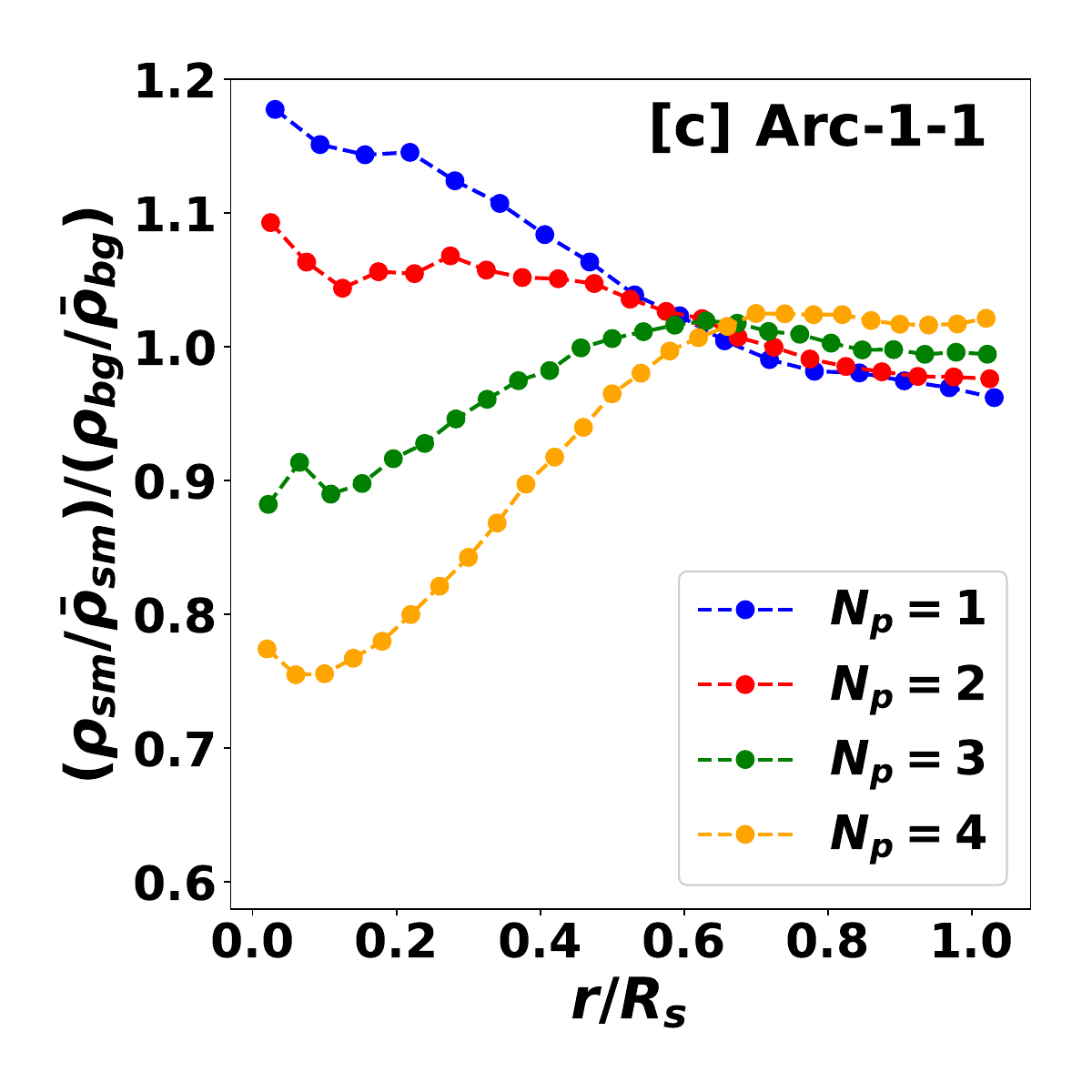}\\
\includegraphics[width=0.5\columnwidth,angle=0]{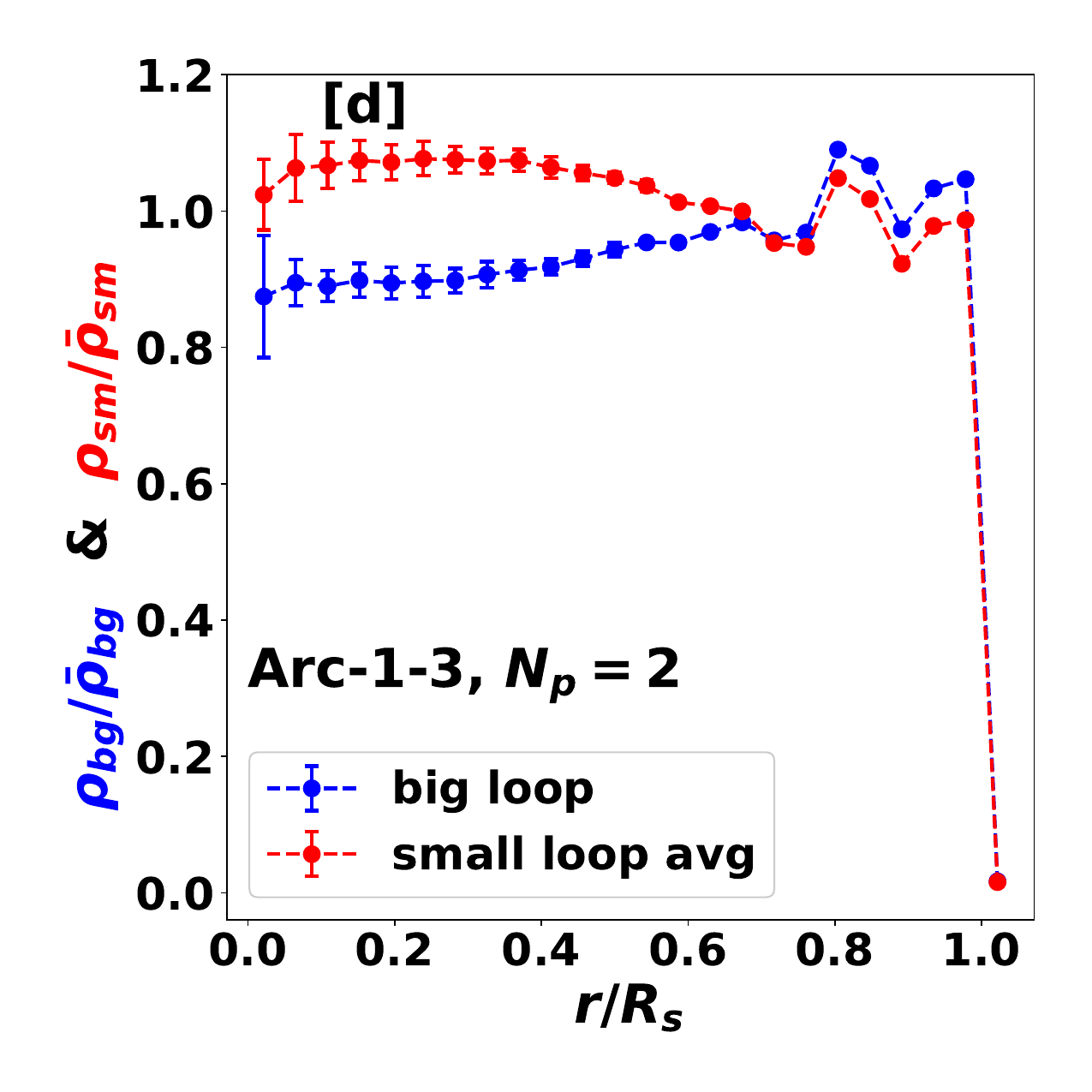}
\includegraphics[width=0.5\columnwidth,angle=0]{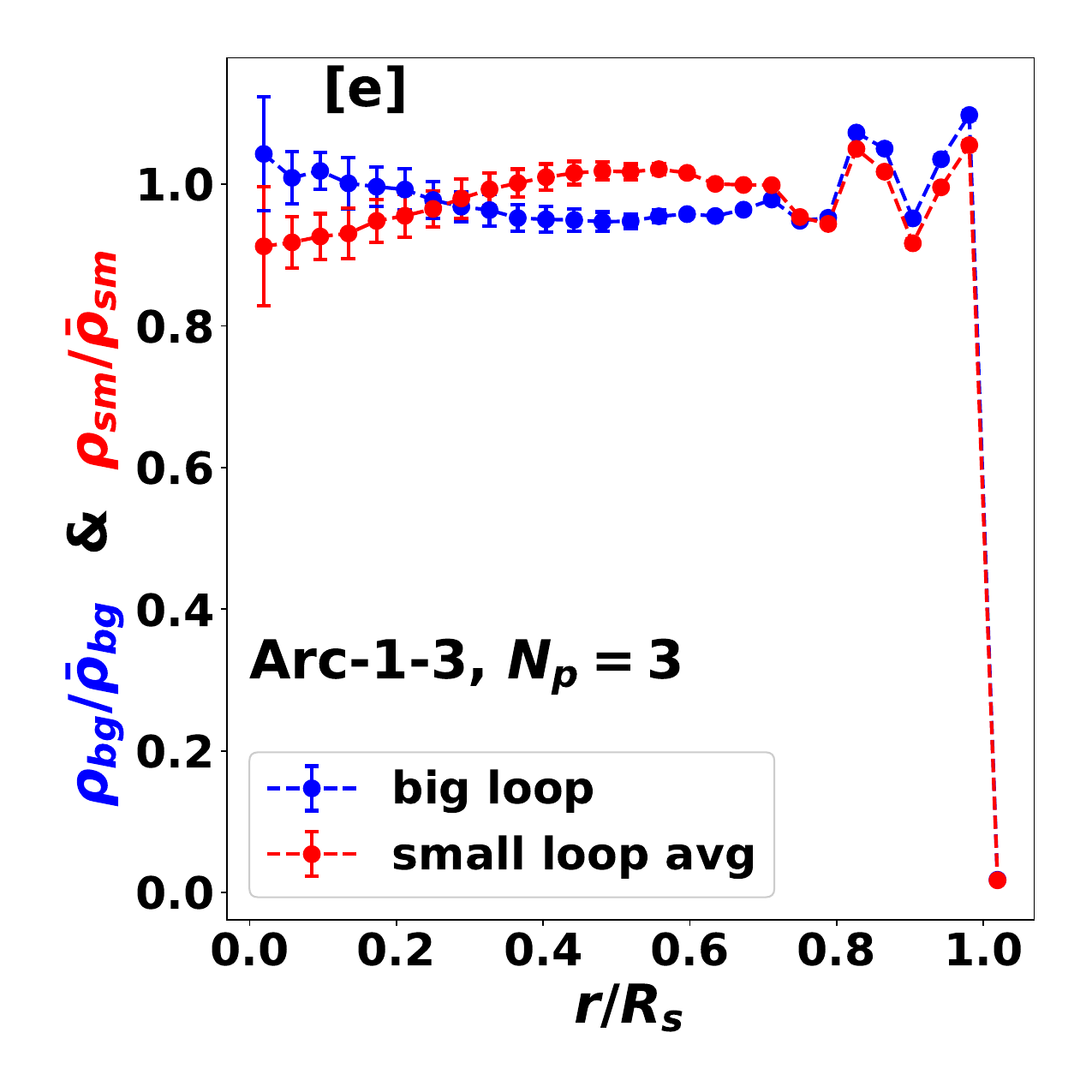}
\includegraphics[width=0.5\columnwidth,angle=0]{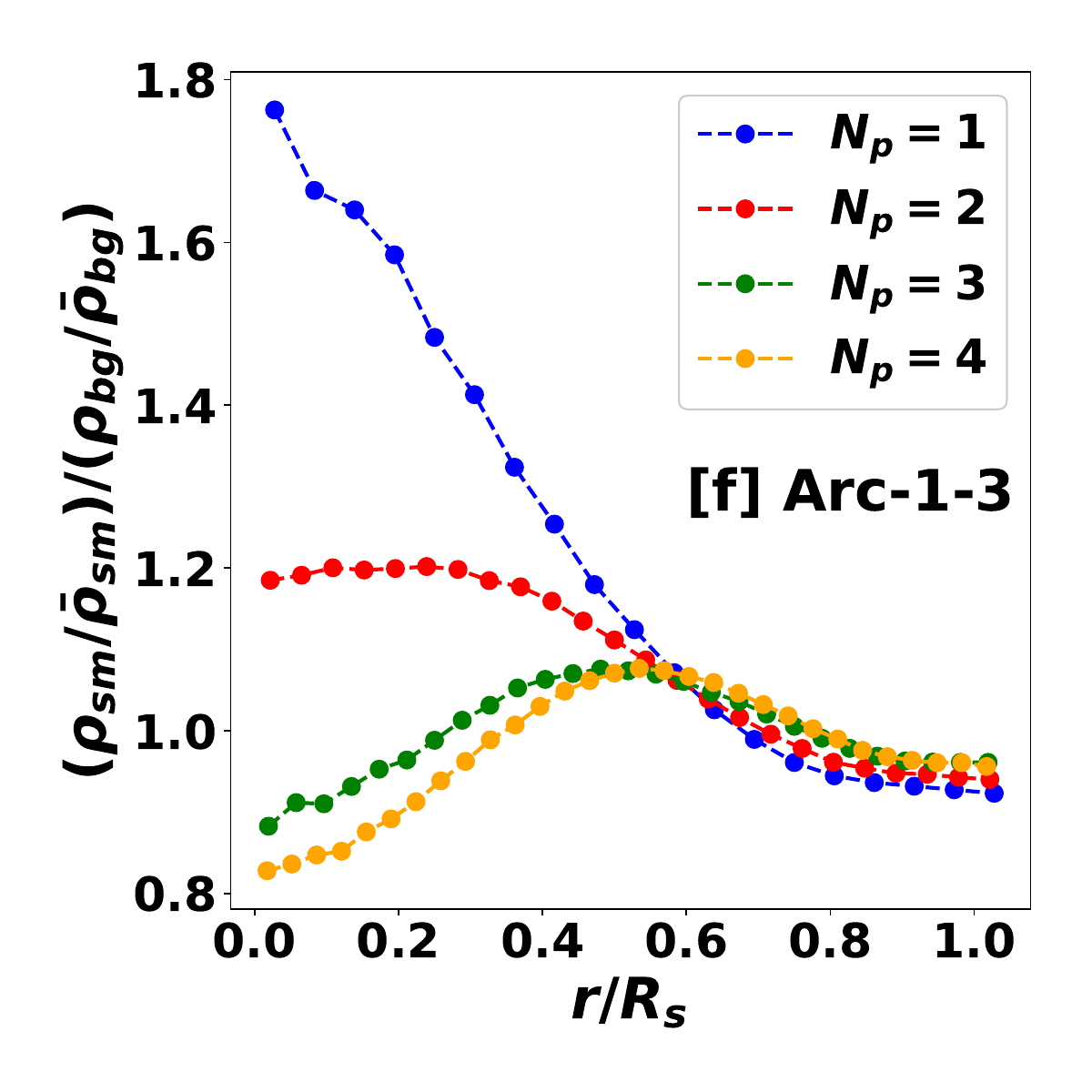}

\caption{\label{figB_1}
This figure shows normalized monomer density of the small loops and big loops when there are $N_p=2$ and $N_p=3$ polymers in the sphere for Arc-1-1 ((a)-(c)), and Arc-1-3((d)-(f)) polymers. Subfigures (c) and (f) show how the ratios of small loop monomer density to big loop monomer density changes as we go from $N_p=1$ to $N_p=4$ for Arc-1-1 and Arc-1-3 polymers respectively.
}
\end{figure*}

Manipulating the topology of polymers to modify the entropic interaction between different sections of 
a ring polymer is an interesting physical endeavour by itself. But our study is motivated to find a plausible, 
future use of these results to investigate chromosome organization within the nucleus. Eukaryotic chromosomes 
are linear polymers, with multiple internal loops which are formed by binding proteins for the purpose of gene 
regulation. The sizes of these internal loops 
can even vary due to the active process of extrusion in the interphase period of the cell cycle. Moreover, there 
can be hierarchical loops which can be formed which will depend on the number of extrusion events going on in parallel, as 
well as the position of the extrusion factors on the chain contour \cite{Alipour2012}. At the moment, for the sake 
of simplicity, we have considered the idealized situation where, firstly, we consider topologically modified 
ring polymers by neglecting the effects of free ends of linear polymers. Secondly, we consider the multiple 
modified ring polymers in our system to be identical in topology for each case study. However, even using such 
simplifying considerations, we have found that the process of entropy-maximization can cause the more compact
segments of the polymers to be pushed more towards the periphery of the sphere. This is reminiscent of the 
more-compacted heterochromatin being situated near the periphery of the nucleus. Thus, we  propose that entropic 
interactions could be one of the aspects affecting nuclear organization  in eukaryotes, and its role should 
be investigated further.

By comparing  contact maps of different polymers, we observe that 
inter-polymer contact between segments of different polymers is
reduced as we introduce internal loops within the polymers. In
hindsight, this observation is not unexpected, as we 
are introducing more internal constraints within each polymer to 
make it more compact, thereby possibly reducing mixing between 
the polymers. We are extending our current studies to 
systematically investigate this aspect for linear polymers with
internal loops where the different polymers could have different
topologies. Whether these considerations also lead to a
preference for a particular polymer (chromosome) to be  the
neighbour of other particular polymers (other chromosomes), where
the choice depends on the relative differences in topology, is an
aspect which we are also investigating.

In summary, we believe that this paper and its adjoining previous
study will open up further exploration of the emergent properties
 obtained by modifying the internal topologies of polymers. Our
 previously published works have already demonstrated that the 
 bacterial cell can modify the topology of its chromosome during 
 its cell cycle to organize itself even as replication is in 
 progress \cite{dna1,dna2,dna3}. We hope that the current work too, will be helpful in
 identifying some of the underlying mechanisms of chromosome 
 organization in eukaryotic cells. 

\section{Author Contributions}

KR implemented the model, performed calculations and analysis. The research plan was designed by 
AC, DM and interpretation of results has extensive contributions from SP. ISS did initial calculations 
which started  off the project. KR, DM, and AC wrote the paper.

\section{Acknowledgements}
Authors acknowledge useful discussions with Arieh Zaritsky, Conrad Woldringh, J. Horbach, and Sathish Sukumaran. 
AC, with DST-SERB (IN) identification No. SQUID-1973-AC-4067, acknowledges funding by DST-SERB (IN) 
project CRG/2021/007824, funding by Biosantexc research and mobility programs funded by ENS-France
and opportunity  to host 
Prof. Jie Xiao using funds from the Fulbright-Nehru specialist program from USIEF
A.C also acknowledges discussions in meetings organized by ICTS, Bangalore, India and use of the computing facilities 
by PARAM-BRAHMA. 

\section{Appendix:}

\begin{figure}[!h]
\centering
\includegraphics[width= 0.75\columnwidth,angle=0]{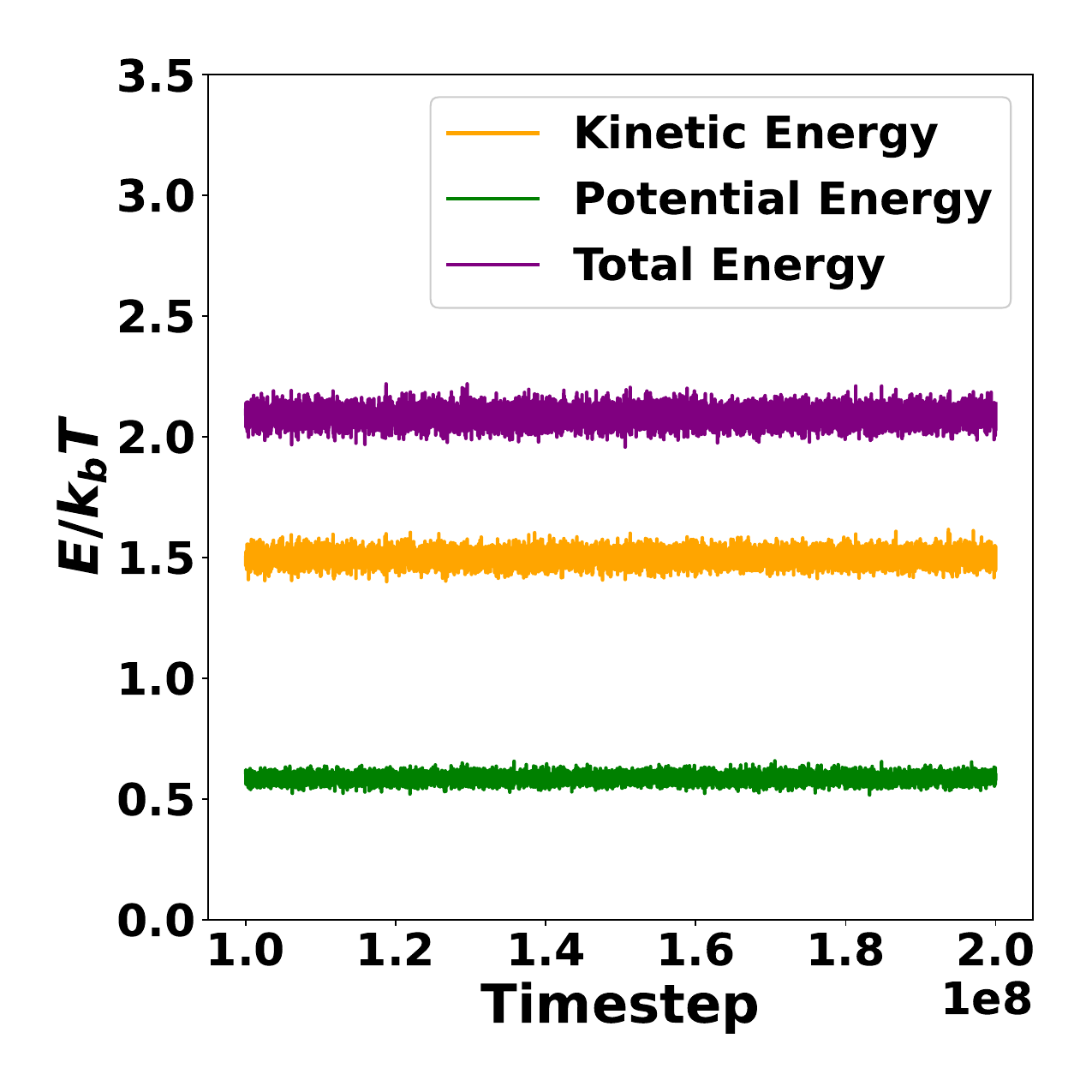}

\caption{\label{figenergy}
The figure shows the Kinetic Energy, Potential Energy, and Total Energy per monomer during a production run of $N_p=6$ Arc-1-3 polymers in a sphere.
}
\end{figure}

\subsection{Calculation of Radius of Gyration}
Estimating the radius of gyration  of a single ring polymer in good solvents (in dilute solution) remains controversial
\cite{Ravi2022}. In particular, one is interested to calculate the $f$-factor, which is the ratio 
of the square of the radius of gyration of a ring polymer and a linear polymer, {\em viz.} ${[R_g(R)/R_g(L)]}^2$.
Here $ R_g(R)$ and $R_g(L)$ is the radius of gyration of  the ring polymer and the linear polymer with 
identical contour length. For a linear polymer, an estimate of the radius of gyration is 
\begin{equation}
R_g(L) = \frac{a N^\nu}{\sqrt{(1 + 2 \nu)(2 + 2\nu)}}
\end{equation}
where $\nu =0.6$ is the Flory exponent for good solvents. 
While for a Gaussian chain, the $f$-factor is found to be  $0.5$ by analytical calculations. 
For real chains in  good solvents, various groups have estimated different values of the $f$-factor, ranging from
$0.516$ to $0.57$ \cite{Ravi2022}.  We have used $f$-factor to be $0.55$ to estimate the $R_g(R)$ of a ring-polymer
with in good solvent with excluded volume interactions.

\subsection{Cases where: $N_p=2$ and $N_p=3$}
In this section, we show the results of the simulations where we have taken either two or three polymers of the same kind within the sphere, i.e., $N_p =2$ and $N_p = 3$.
We observe that when $N_p=2$, the pattern of organization is the same as that found for $N_p=1$, where small loops are found preferentially towards the center, whereas for three polymers in the sphere, the small loops are located towards the periphery (Similar to what is found for higher numbers of polymers in the sphere). The same is found for both Arc-1-1(Fig. \ref{figB_1}(a)-(c)) and Arc-1-3(Fig. \ref{figB_1}(d)-(f)). This implies that the change in pattern of organization of loops occurs when we go from $N_p=2$ to $N_p =3$.

\subsection{Change in Energy}
To make sure that the organization we observe is solely due to maximization of entropy, we plotted the kinetic energy, potential energy(WCA+Spring energy), and total energy per monomer of the system across the entire production run in Fig. \ref{figenergy}. We find that the energy of the system remains more or less constant throughout, which means that to minimize the free energy of the system, the entropy needs to be maximised which the system achieves by preferentially localizing the differently sized loops to different parts of the sphere.

\bibliographystyle{unsrt}
\bibliography{ref.bib}

\end{document}